\documentclass[preprint2]{exoplanet}
\usepackage{times}

\newcommand{\refs}{\par\noindent\hangindent=1pc\hangafter=1}
\begin{document}

\def\e{{\rm E}}
\def\rel{{\rm rel}}

\title{\textbf{\LARGE Exoplanetary Microlensing}}

\author {\textbf{\large B. Scott Gaudi}}
\affil{\small\em The Ohio State University}

\begin{abstract}
\begin{list}{ } {\rightmargin 0.5in}
\baselineskip = 11pt
\parindent=1pc
{\small 
Gravitational microlensing occurs when a foreground star happens to
pass very close to our line of sight to a more distant background
star.  The foreground star acts as a lens, splitting the light from
the background source star into two images, which are typically
unresolved.  However, these images of the source are also magnified,
by an amount that depends on the angular separation between the lens
and source.  The relative motion between the lens and source therefore
results in a time-variable magnification of the source: a microlensing
event.  If the foreground star happens to host a planet with projected
separation near the paths of these images, the planet will also act as
a lens, further perturbing the images and resulting in a
characteristic, short-lived signature of the planet.  This chapter
provides an introduction to the discovery and characterization of
exoplanets with gravitational microlensing.  The theoretical
foundation of the method is reviewed, focusing in particular on the
phenomenology of planetary microlensing perturbations.  The strengths
and weaknesses of the microlensing technique are discussed, highlighting
the fact that it is sensitive to low-mass planetary companions to
stars throughout the Galactic disk and foreground bulge, and that its
sensitivity peaks for planet separations just beyond the snow line.
An overview of the practice of microlensing planet searches is given, with
a discussion of some of the challenges with detecting and analyzing
planetary perturbations.  The chapter concludes with a review of the
results that have been obtained to date, and a discussion of the near
and long-term prospects for microlensing planet surveys.  Ultimately,
microlensing is potentially sensitive to multiple-planet systems
containing analogs of all the solar system planets except Mercury, as
well as to free floating planets, and will provide a crucial test of
planet formation theories by determining the demographics of planets
throughout the Galaxy.
 \\~\\~\\~}
 
\end{list}
\end{abstract}

\section{INTRODUCTION}

Gravitational lensing generally refers to the bending of light rays of
a background light source by a foreground mass.  Gravitational
microlensing, on the other hand, traditionally refers to the special case when multiple images are created
but have separations of less than a few milliarcseconds, and hence are unresolved
with current capabilities.  Although the idea of the gravitational deflection of
light by massive bodies well predates General Relativity, and can be traced
as far back as Sir Isaac Newton\footnote{See {\em Schneider et al.}
1992 for a thorough recounting of the history of gravitational
lensing.}, the concept of gravitational microlensing appears to be
attributable to Einstein himself. In 1936 Einstein published a paper
in which he derived the equations of microlensing by a foreground star
closely aligned to a background star ({\em Einstein}, 1936). Indeed it
seems that Einstein had been thinking about this idea as far back as
1912 ({\em Renn et al.}, 1997), and perhaps had even hoped to use the
phenomenon to explain the appearance of Nova Geminorum 1912 ({\em
Sauer}, 2008).  However, by 1936 he had dismissed the practical
significance of the microlensing effect, concluding that ``there is no
great chance of observing this phenomenon'' ({\em Einstein}, 1936). 

Indeed, there is no ``great chance'' of observing gravitational
microlensing.  The optical depth to gravitational microlensing, i.e.,
the probability that any given star is being appreciably lensed at any
given time, is of order $10^{-6}$ toward the Galactic bulge, and is
generally similar or smaller for other lines of sight\footnote{The
phenomenon of gravitational lensing of multiply-imaged quasars by
stars in the foreground galaxy that is creating the multiple images of the quasar 
is also referred to as microlensing.  In
this instance the optical depth to microlensing can be of order unity.  However, in
this chapter we are concerned only with gravitational microlensing of
stars within our Galaxy or in nearby galaxies, where the optical depths
to microlensing are always small.}.  Thus at least partly due to this low probability,
the idea of gravitational microlensing lay mostly dormant for five
decades after Einstein's 1936 paper (with some notable exceptions,
e.g., {\em Liebes}, 1964; {\em Refsdal}, 1964).  It was not until the
seminal paper by {\em Paczynski} (1986) that the idea of gravitational
microlensing was resurrected, and then finally put into
practice with the initiation of several observational searches for
microlensing events toward the Large and Small Magellenic Clouds
and Galactic bulge in the early 1990s ({\em
Alcock et al.} 1993; {\em Aubourg et al.} 1993; {\em Udalski et al.}, 1993).
The roster of detected microlensing events now numbers in the thousands.
These events have been discovered by many different collaborations, toward several
lines of sight including the Magellenic Clouds ({\em Alcock et al.}, 1997, 2000;
{\em Palanque-Delabrouille et al.}, 1998), and M31 ({\em Paulin-Henriksson et al.} 2002;
{\em de Jong et al.}, 2004; {\em Uglesich et al.}, 2004; {\em Calchi Novati et al.}, 2005),
with the vast majority found toward the Galactic bulge ({\em Udalski et al.}, 2000; {\em Sumi et al.}, 2003;
{\em Thomas et al.}, 2005; {\em Hamadache et al.}, 2006) or nearby fields in the
Galactic plane ({\em Derue et al.}, 2000).

\begin{figure*}
\epsscale{1.8}
\plotone{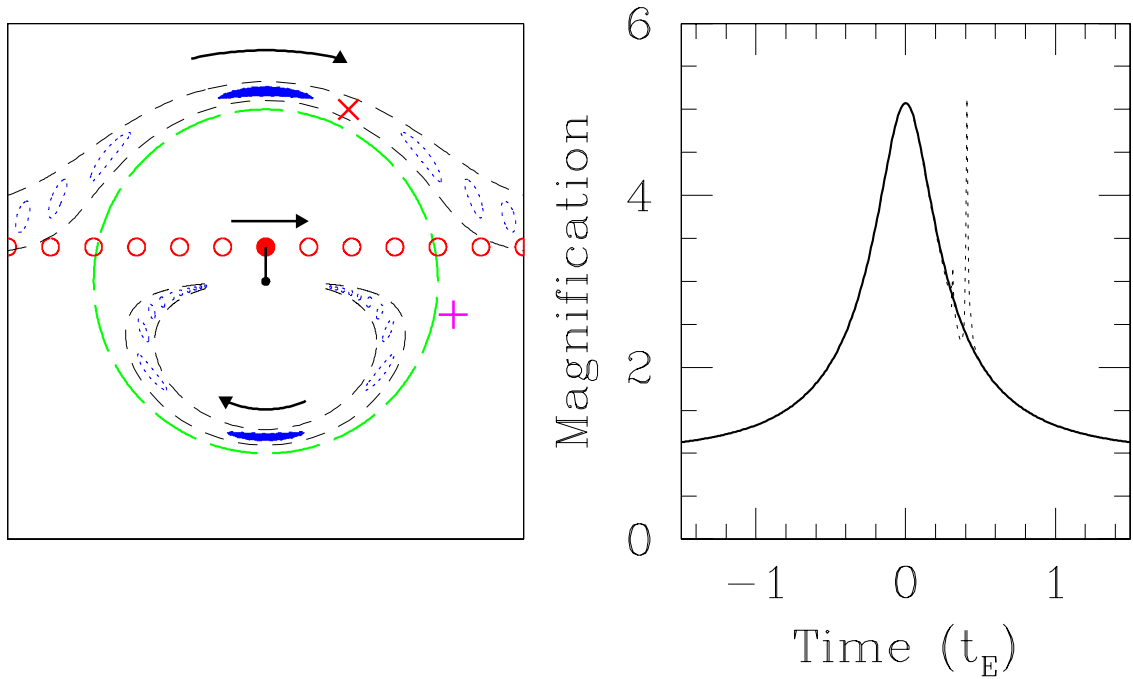}
\caption{\small 
Left: The images (dotted ovals) are shown for several
different positions of the source (solid circles), along with the primary lens
(dot) and Einstein ring (long dashed circle).  If the
primary lens has a planet near the path of one of the images,
i.e. within the short-dashed lines, then the planet
will perturb the light from the source, creating a
deviation to the single lens light curve.
Right: The magnification as a function of
time is shown for the case of a single lens (solid) and accompanying planet
(dotted) located at the position of the X in the left panel.  
If the planet was located at the + instead, then there would be no detectable perturbation, and
the resulting light curve would be essentially identical to the solid curve.
}\label{fig:cartoon}
\end{figure*}

Even before the first microlensing events were detected, it was
suggested by {\em Mao \& Paczynski} (1991) that gravitational
microlensing might also be used to discover planetary companions to
the primary microlens stars.  The basic idea is
illustrated in Figure \ref{fig:cartoon}.  As the foreground star passes close to the
line of sight to the background source, it splits the source into two
images, which sweep out curved trajectories on the sky as the
foreground lens star passes close to the line of sight to the
source. These images typically have separations of order one
milliarcsecond, and so are unresolved.  However, the total area of
these images is also larger than the area of the source, and as a
result the background star also exhibits a time-variable
magnification, which is referred to as a microlensing event.  If the
foreground star happens to host a planet with projected separation
near the paths of one of these two images, the planet will further
perturb the image, resulting in a characteristic, short-lived
signature of the planet.

{\em Gould \& Loeb} (1992) considered this novel method of detecting
planets in detail, and laid out the practical requirements for an observational
search for planets with microlensing.  In particular, they advocated a ``two tier''
approach.  First, microlensing events are discovered and alerted real-time by
a single survey telescope that monitors many square degrees of the Galactic bulge
on a roughly nightly basis.  Second, these ongoing events are then densely monitored 
with many smaller telescopes to discover the short-lived signatures of planetary
perturbations.  

The search for planets with microlensing began in earnest in 1995 with
the formation of several follow-up collaborations dedicated to
searching for planetary deviations in ongoing events ({\em Albrow et
al.}, 1998; {\em Rhie et al.} 2000).  These initial searches adopted
the basic approach advocated by {\em Gould \& Loeb} (1993): monitoring
ongoing events alerted by several survey collaborations (e.g., {\em
Udalski et al.}, 1994; {\em Alcock et al.} 1996), using networks of
small telescopes spread throughout the southern hemisphere.  Although
microlensing planet searches have matured considerably since their
initiation, this basic approach is still used to this day, with the
important modification that current follow-up collaborations tend to
focus on high-magnification events, which individually have higher
sensitivity to planets ({\em Griest \& Safizadeh}, 1998), as discussed
in detail below.

From 1995-2001, no convincing planet detections were made, primarily
because the relatively small number of events being alerted each year
($\sim 50-100$) by the survey collaborations meant that there were 
only a few events ongoing at any given time, and often these were poorly suited for
follow-up.  Although interesting upper limits were placed on the
frequency of Jovian planets ({\em Gaudi et al.}
2002, {\em Snodgrass et al.}, 2004), perhaps the most important result
during this period was the development of both the theory and practice of
the microlensing method, which resulted in its transformation from a
theoretical abstraction to a viable, practical method of searching for
planets.

In 2001, the OGLE collaboration ({\em Udalski}, 2003) upgraded to a
new camera with a 16 times larger field of view and so were able to
monitor a larger area of the bulge with a higher cadence.  As a
result, in 2002 OGLE began alerting nearly an order of magnitude more
events per year than previous to the upgrade.  These improvements in
the alert rate and cadence, combined with improved cooperation and
coordination between the survey and follow-up collaborations, led to
the first discovery of an extrasolar planet with microlensing in 2003
({\em Bond et al.} 2004).  The MOA collaboration upgraded to a 1.8m
telescope and 2 deg$^2$ camera in 2004 ({\em Sako et al.}, 2008), and
in 2007 the OGLE and MOA collaborations started alerting $\sim 850$
events per year, thus ushering in the ``golden age'' of microlensing
planet searches.

The first detections of exoplanets with microlensing
({\em Bond et al.}, 2004; {\em Udalski et
al.}, 2005; {\em Beaulieu et al.}, 2006; {\em Gould et al.}, 2006;
{\em Gaudi et al.}, 2008a; {\em Bennett et al.}, 2008; {\em Dong et
al.}, 2009; {\em Janczak et al.}, 2009; {\em Sumi et al.}, 2009) have
provided important lessons about the kinds
of information that can be extracted from observed events.  When
microlensing planet surveys were first being developed, it was thought
that the primary virtue of the method would be solely its ability to
provide statistics on large-separation and low-mass planets.  
Individual detections would be of little interest because
the only routinely measured property of the planets would be the
planet/star mass ratio, and the nature of the host star in any given
system would remain unknown because the microlensing event itself provides
little information about the properties of the host star, and follow-up
reconnaissance would be difficult or impossible due to the large
distances of the detected systems from Earth.  In fact, experience with actual
events has shown that much more information can typically
be gleaned from a combination of a detailed analysis of the light
curve, and follow-up, high-resolution imaging. As a result, in most
cases it is possible to infer the mass of the primary lens and the basic physical properties of
planetary system, including the planet mass and physical separation.
In some special cases it is possible to infer considerably more information.

The theoretical foundation of microlensing is highly technical, the
practical challenges associated with searching for planets with
microlensing are great, and the analysis of observed microlensing
datasets is complicated and time-consuming. Furthermore, although it
is possible to learn far more about individual systems than was
originally thought, extracting this information is difficult, and it
is certainly the case that the systems will never be as well
studied as planets systems found by other methods in the solar
neighborhood.  

Give the intrinsic limitations and difficulties of the microlensing method,
and might therefore wonder: ``why bother?''  The
primary motivation for  going to all this trouble is that microlensing
is sensitive to planets in a region of parameter space that is
difficult or impossible to probe with other methods, namely low-mass
planets beyond the `snow line', the point in the protoplanetary disk
beyond which ices can exist.  The snow line plays a crucial role in
the currently-favored model of planet formation.  Thus, even the first
few microlensing planet detections have provided important constraints on
planet formation theories.  The second generation of
microlensing surveys will potentially be sensitive to multiple-planet
systems containing analogs of all the solar system planets except
Mercury, as well as to free floating planets.  Ultimately, when
combined with the results from other complementary surveys,
microlensing surveys can potentially yield a complete picture of the
demographics of essentially {\it all} planets with masses greater than that of Mars.  Thus
it is well worth the effort, even given the drawbacks.

\begin{figure*}
\epsscale{2.1}
\plotone{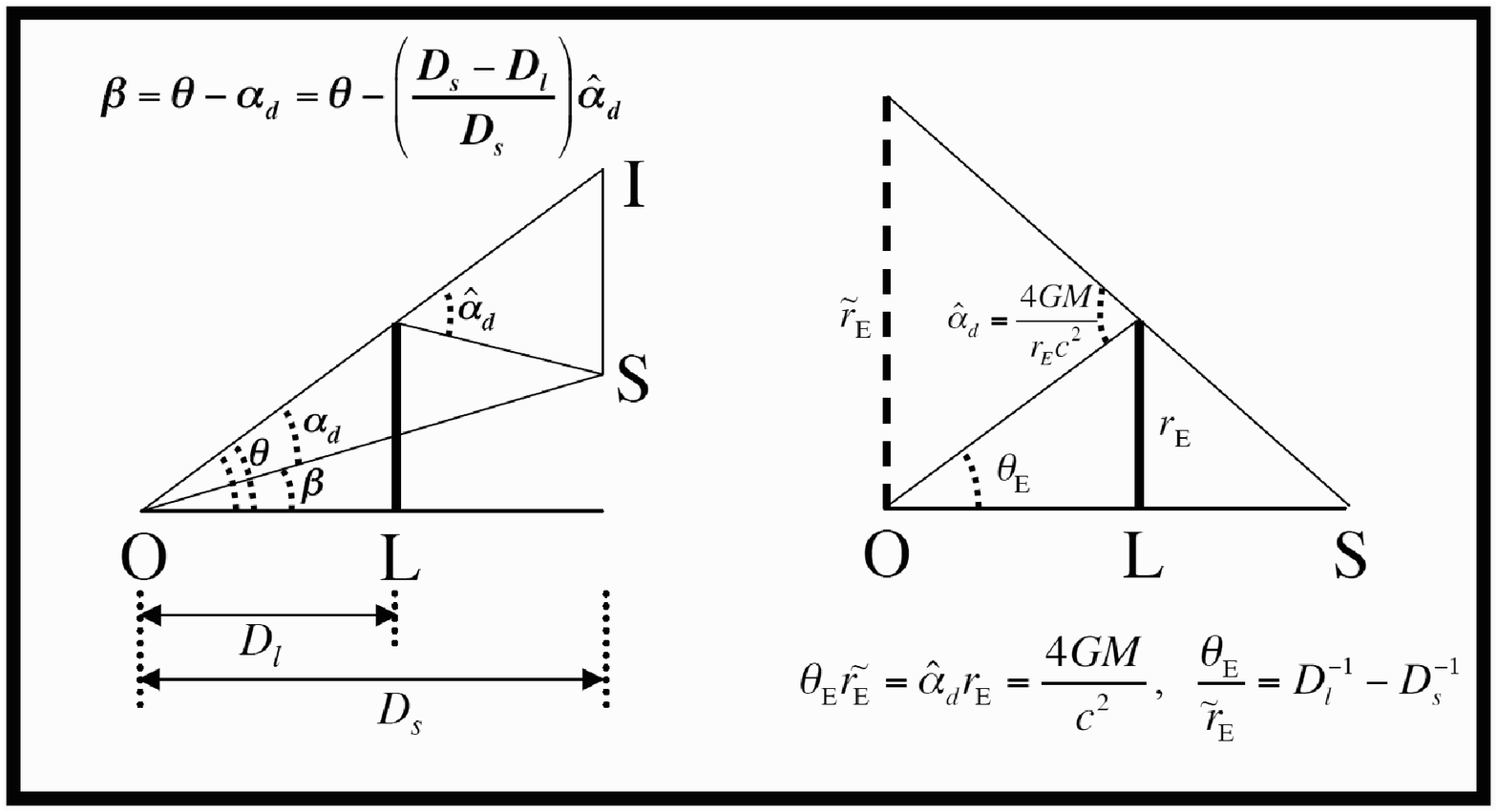}
\caption{\small 
Left:  The lens (L) at a distance $D_l$
from the observer (O) deflects light from the source
(S) at distance $D_s$ by the Einstein bending angle $\mbox{\boldmath$\hat\alpha_d$}$.
The angular positions of the images $\mbox{\boldmath$\theta$}$ and unlensed source $\mbox{\boldmath$\beta$}$
are related by the lens equation, $\mbox{\boldmath$\beta$}= \mbox{\boldmath$\theta$}- \mbox{\boldmath$\alpha_d$}
= \mbox{\boldmath$\theta$}-(D_s-D_l)/D_s\mbox{\boldmath$\hat\alpha_d$}$.  For a point lens,
$\hat\alpha_d = 4GM/(c^2D_l\theta)$.  Right:
Relation of higher-order observables, the angular ($\theta_\e$) and projected ($\tilde r_\e$)
Einstein radii, to physical characteristics of the lensing system. Adapted
from {\em Gould}, 2000.
\label{fig:dia}
}\end{figure*}

The primary goal of this chapter is to provide a general
introduction to gravitational microlensing and searches for planets
using this method.  Currently, the single biggest obstacle to the
progress of microlensing searches for planets is simply a lack of
human power. There are already more observed binary and planetary events
than can be modeled by the handful of researchers in the world with
sufficient expertise to do so.  This situation is likely to get worse
when next-generation microlensing planet surveys come on line.
However, because the theoretical foundation of microlensing is quite
technical and the practical implementation of microlensing planet
searches is fairly complex, it can be very difficult and time
consuming for the uninitiated to gain sufficient familiarity with the
field to make meaningful contributions.  This difficulty is
exacerbated by the fact that the explanations of the relevant concepts
are scattered over dozens of journal articles.  The aim of this
chapter is to partially remedy this situation by providing a reasonably
comprehensive review, which a beginner can use to gain a least a basic
familiarity with the many concepts, tools, and techniques that are
required to actively participate and contribute to current
microlensing planet searches.  Readers are encouraged to peruse reviews
by {\em Paczynski} 1996, {\em Sackett} 1999, {\em Mao} 2008,
and {\em Bennett} 2009a for additional background material. 

Section \ref{sec:found} provides an overview of the theory of gravitational microlensing searches for
planets.  This is the heart of the chapter, and is fairly long and
detailed. Much of this section may not be of
interest to all readers, and some of the material may be too technical
for a beginner to the subject.  Therefore, the first subsection
(\ref{sec:basic}) provides a primer on the basic properties and
features of microlensing.  Beginners, or casual readers who are only interested in
a basic introduction to the method itself and its features, can read only
this section (paying particular attention to Figures \ref{fig:cartoon}, \ref{fig:dia},
and \ref{fig:ped}), and then skip to Section
\ref{sec:features}. Section \ref{sec:practice} discusses the 
practical implementation of planetary microlensing, while
Section \ref{sec:features} reviews the basic advantages and drawbacks of the
method.  Section \ref{sec:results} provides a summary
of the results to date, as well as a brief discussion of their
implications.  Section \ref{sec:future} discusses future
prospects for microlensing planet searches, and in particular
the expected yields of a next-generation ground-based planet
search, and a space-based mission.

\section{FOUNDATIONAL CONCEPTS AND EQUATIONS}\label{sec:found}

\subsection{Basic Microlensing}\label{sec:basic}

This section provides a general overview of the basic equations,
scales, and phenomenology of microlensing by a point mass, and a brief
introduction to how microlensing can be used to find planets, and how
such planet searches work in practice.  It is meant to be
self-contained, and therefore the casual reader who is not interested
in a detailed discourse on the theory, phenomenology, and practice of planetary
microlensing can simply read this section and then skip to Section
\ref{sec:features} without significant loss of continuity.

A microlensing event occurs when a foreground ``lens'' happen to pass
very close to our line of sight to a more distant background
``source.''  Microlensing is a relatively improbable phenomenon, and
so in order to maximize the event rate, microlensing survey are
typically carried about toward dense stellar fields.  In particular,
the majority of microlensing planet surveys are carried out toward the
Galactic center.  Therefore, for our purposes, the lens is typically a
main-sequence star or stellar remnant in the foreground Galactic disk
or bulge, whereas the source is a main-sequence star or giant
typically in the bulge.

The left panel of Figure \ref{fig:dia} shows the basic geometry of
microlensing.  Light from the source at a distance $D_s$ is deflected
by an angle $\mbox{\boldmath$\hat\alpha_d$}$ by the lens at a distance
$D_l$. For a point lens, $\hat\alpha_d= 4GM/(c^2D_l\theta)$, where $M$
is the mass of the lens, and $\theta$ is the angular separation of the
images of the source and the lens on the 
sky\footnote{This form for the bending angle can be derived heuristically by
assuming that a photon passing by an object of mass $M$ at a distance
$b\equiv D_l \theta $ will experience an impulse given by the Newtonian acceleration
$GM/b^2$ over a time $2b/c$, thereby inducing a velocity perpendicular to the
original trajectory of $\delta v = (GM/b^2)(2b/c)= 2GM/(bc)$.   The deflection
is then $\delta v/c = 2GM/(bc^2)$.  The additional factor of two cannot be derived
classically, and arises from General Relativity (see, e.g., {\em Schneider et al.} 1992).}.
The relation between
$\theta$, and the angular separation $\beta$ between the lens and source in the
absence of lensing, is called the {\it lens equation}, and is
given trivially by $\beta = \theta - \alpha_d$.  From basic geometry
and using the small-angle approximation, $\hat \alpha_d (D_s-D_l)=
\alpha_d D_s$.  Therefore, for a point lens,
\begin{equation}
\beta = \theta - \frac{4GM}{c^2\theta}\frac{D_s-D_l}{D_sD_l}.
\label{eqn:pmlens}
\end{equation}

\begin{figure*}
\epsscale{2.0}
\plottwo{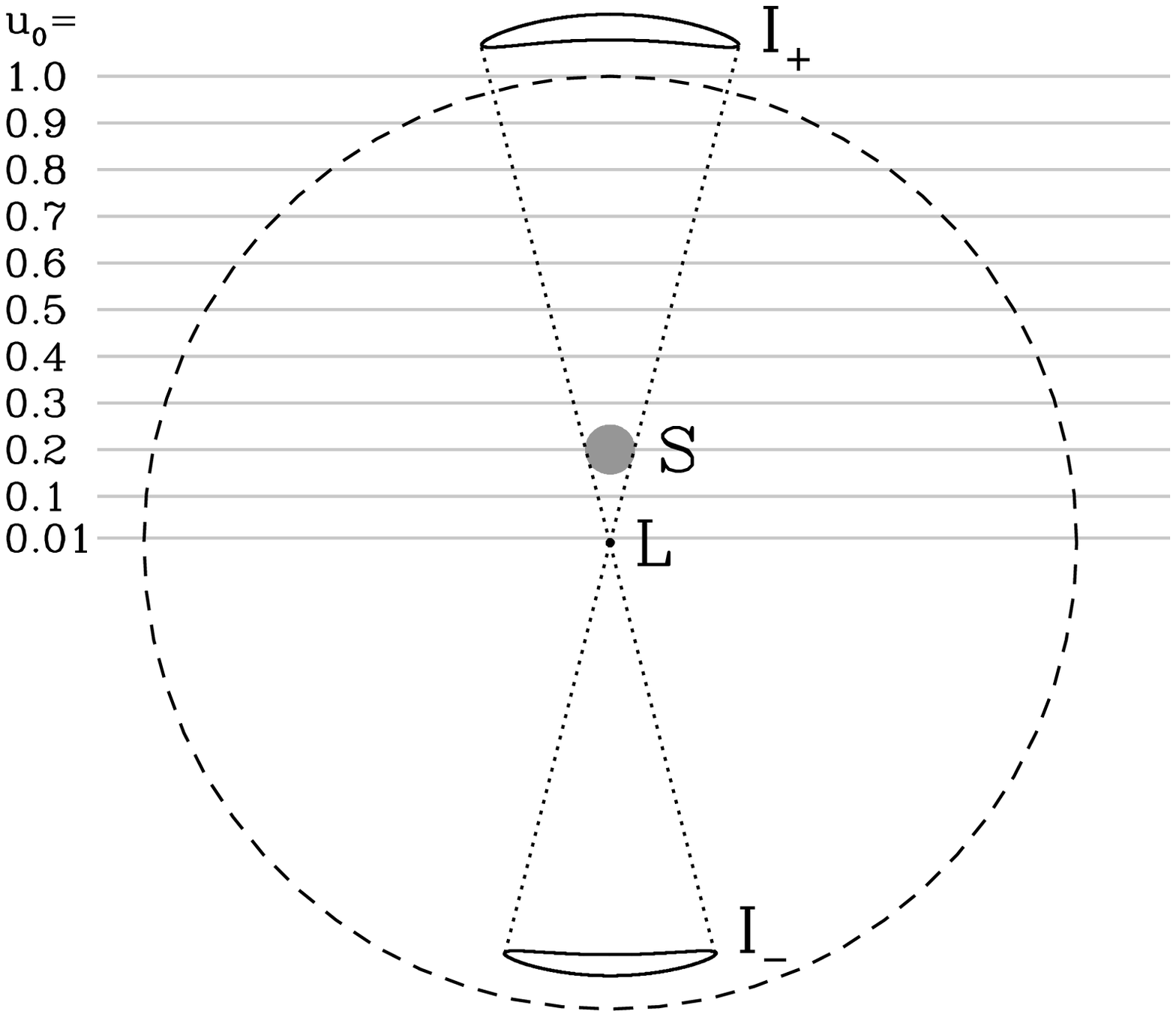}{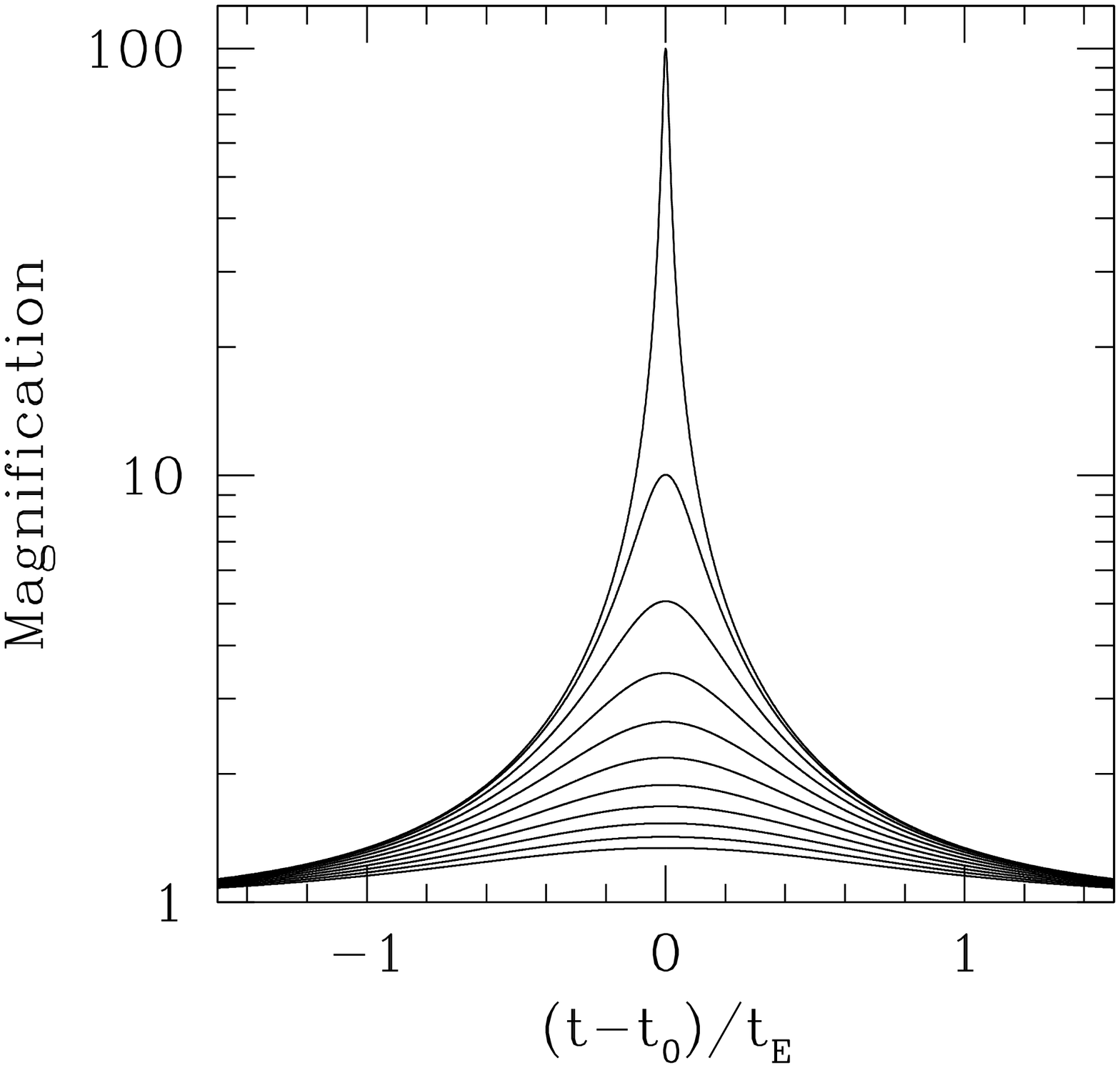}
\caption{\small Basic point-mass microlensing.
(Left) All angles are normalized by the angular Einstein ring radius $\theta_\e$, shown as a dashed circle
of radius $\theta_\e$.  The source (S) is located at
an angular separation of $u=0.2$ from the lens (L). Two images are created, one image outside 
the Einstein ring (I$_+$), on the same side of the lens as the source with position from the lens of $y_+=0.5(\sqrt{u^2+4}+u)$, and one
inside the Einstein ring, on the opposite side of the lens as the source with position from the lens
of $y_-=-0.5(\sqrt{u^2+4}-u)$.  The images are compressed radially but elongated tangentially.  Since surface
brightness is conserved, the magnification of each image is just the ratio of its area to the area of the source.
Since the images are typically unresolved, only the total magnification of the two images is measured,
which depends only on $u$.  (Right) Magnification as a function of time (light curves), for the
ten trajectories shown in the left panel with impact parameters $u_0=0.01,0.1,0.2,...,1.0$.  Time is relative
to the time $t_0$ of the peak of the event (when $u=u_0$), and in units of the angular Einstein crossing time $t_\e$. 
Higher magnification implies more elongated images, which leads to increased sensitivity to planetary companions. 
Adapted from {\em Paczynski} 1996.
}\label{fig:ped}
\end{figure*}

The left panel of Figure \ref{fig:ped} shows the basic source and image
configurations for microlensing by a single
point mass.  From Equation \ref{eqn:pmlens}, if the lens is exactly aligned with the
source ($\beta=0$), it images the source into an ``Einstein ring'' with a radius 
$\theta_{\rm E}=\sqrt{\kappa M \pi_{\rm rel}}$,
where $M$ is the mass of the lens, $\pi_{\rm rel}={\rm AU}/D_{\rm rel}$
is the relative lens-source 
parallax, $D_{\rm rel} \equiv (D_l^{-1}-D_s^{-1})^{-1}$ is the relative lens-source
distance, and $\kappa=4G/c^2{\rm AU} \simeq 8.14~{\rm mas}~M_\odot^{-1}$.  It is also
instructive to note that $\theta_{\rm E} = \sqrt{2R_{\rm Sch}/D_{\rm rel}}$, where
$R_{Sch} \equiv 2GM/c^2$ is the Schwarzschild radius of the lens.  Quantitatively,
\begin{equation}
\theta_{\rm E} = 550~{\rm mas}~\left(\frac{M}{0.3 M_\odot}\right)^{1/2}
\left(\frac{\pi_{\rm rel}}{125~\mu{\rm as}}\right)^{1/2},
\label{eqn:thetaequant}
\end{equation}
which corresponds to a physical Einstein ring radius at the distance of the lens of,
\begin{eqnarray}
r_{\rm E} &\equiv& \theta_{\rm E}D_l\\
 &=&2.2{\rm AU}\left(\frac{M}{0.3 M_\odot}\right)^{1/2}
\left(\frac{D_s}{8~{\rm kpc}}\right)^{1/2}
\left[\frac{x(1-x)}{0.25}\right]^{1/2},
\label{eqn:requant}
\end{eqnarray}
where $x\equiv D_l/D_s$.

Normalizing
all angles on the sky by $\theta_\e$, we can define $u=\beta/\theta_\e$ and $y=\theta/\theta_\e$.
Using these definitions and the definition for $\theta_\e$, the lens equation (\ref{eqn:pmlens})
reduces to the form 
\begin{equation}
u = y - y^{-1}.
\end{equation}
This is equivalent to a quadratic equation in $y$: $y^2-uy-1=0$.   Thus in the case of imperfect
alignment ($u \ne 0$), there are two images, with positions 
\begin{equation}
y_{\pm}=\pm \frac{1}{2}(\sqrt{u^2+4}\pm u).
\label{eqn:ypmfirst}
\end{equation}
The positive, or ``major'' image is always outside the Einstein ring, whereas the negative, 
or ``minor'' image is always inside the Einstein ring\footnote{The images
formed by a gravitational lens can also be found by determining the relative time delay as a function
of the (vector) angle $\mbox{\boldmath$\theta$}$ for hypothetical light rays propagating from the source.  This time
delay function includes the effect of the difference in the geometric path length, as well the gravitational
(Shapiro) delay, as a function of $\mbox{\boldmath$\theta$}$.   The images are then located at the stationary
points (maxima, minima, or saddle points) of the time delay surface.  For a point-mass lens,
the positive solution in Equation \ref{eqn:ypmfirst} corresponds to a minimum in the time delay surface, and
so is also referred to as the ``minimum'' image.  The negative solution corresponds
to a saddle point, and so is also referred to as the ``saddle'' image.
There is formally a third image, corresponding to the maximum of the time delay surface, but this
image is located behind the lens, and is infinitely demagnified (for a true point lens).}.
As can be seen in Figure \ref{fig:ped}, the angular separation between these images at
the time of closest alignment is is $\sim 2\theta_\e$.  Thus for typical lens masses ($0.1-1~M_\odot$) and 
lens and source distances ($1-10~{\rm kpc}$),
$\theta_{\rm E}\la {\rm mas}$ and so the images are not resolved.  
However, the images are also distorted, and since surface brightness is conserved, this implies
they are also magnified.  The magnification of each image is just the ratio of the area
of the image to the area of the source. As can be seen in Figure \ref{fig:ped}, the images
are typically elongated tangentially by an amount $y_\pm/u$, but are compressed
radially by an amount ${\rm d}{y_\pm}/{\rm d}u$.  The magnifications
are then 
\begin{equation}
A_\pm = \left| \frac{y_\pm}{u} \frac{{\rm d}y_\pm}{{\rm d}u} \right| = \frac{1}{2}\left[\frac{u^2+2}{u\sqrt{u^2+4}} \pm 1\right],
\label{eqn:apmped}
\end{equation}
and thus the total magnification is
\begin{equation}
A(u)= \frac{u^2+2}{u\sqrt{u^2+4}}
\label{eqn:atot1}.
\end{equation}
Note that, as $u \rightarrow \infty$, $A_+\rightarrow 1$ and $A_- \rightarrow 0$. 
Also, for $u \ll 1$, the magnification takes the form $A \simeq 1/u$.

Because the lens, source, and observer are all in relative motion,
the angular separation between the lens and source is a function of time.
Therefore, the magnification of the source is also a function of time:
a microlensing event.  In the simplest case of uniform, rectilinear motion, the
relative lens-source motion can be parametrized by
\begin{equation}
u(t) = \left[u_0^2 + \left(\frac{t-t_0}{t_\e}\right)^2\right]^{1/2},
\label{eqn:uoft1}
\end{equation}
where $u_0$ is the minimum separation
between the lens and source (in units of $\theta_\e$), $t_0$ is the
time of maximum magnification (corresponding to when $u=u_0$), and
$t_\e$ is the time scale to cross the angular Einstein ring radius,
$t_\e \equiv \theta_\e/\mu_{\rm rel}$, where $\mu_{\rm rel}$ is the proper motion 
of the source relative to the lens.    This form for $u$
gives rise to a smooth, symmetric
microlensing event with a characteristic form (often
called a ``Paczynski curve''), as shown
in the  right panel of Figure \ref{fig:ped}. 
The typical timescales for events toward the Galactic bulge
are of order a month, 
\begin{equation}
t_{\rm E} = 19~{\rm d}\left(\frac{M}{0.3 M_\odot}\right)^{1/2}
\left(\frac{\pi_{\rm rel}}{125~\mu{\rm as}}\right)^{1/2} 
\left(\frac{\mu_{\rm rel}}{10.5~{\rm mas/yr}}\right)^{-1},
\label{eqn:tequant}
\end{equation}
but can range from less than a day to 
years.  The magnification is $A>1.34$ for $u\le 1$, and so the
magnifications are substantial and easily detectable. 

If the lens star happens to have a planetary companion, and the planet
happens to be located near the paths of one or both of the two images
created by the primary lens, the companion can create a
short-timescale perturbation on the primary microlensing event when
the image(s) sweep by the planet ({\em Mao \& Paczynski} 1991; {\em
Gould \& Loeb}, 1992).  There are two conceptually different channels
by which planets can be detected with microlensing, delineated by
the maximum magnification $A_{\rm max}$ of the primary event in which the planet is detected.

Consider the case of a relatively low-magnification
(i.e., $A_{\rm max} \la 10$ or $u_0 \ga 0.1$) primary event, such as illustrated in Figure \ref{fig:cartoon}.
Over the course of
the primary event, the two images sweep out a path on the sky, with
the path of the major (minimum) image entirely contained outside the
Einstein ring radius, and the path of the minor (saddle) image
entirely contained within the Einstein ring radius.  In the
main detection channel, the primary hosts a planet which happens to be located near the
path of one\footnote{As can be seen in Figure \ref{fig:cartoon}, for low-magnification events,
the two images are well-separated, and thus a low mass ratio planet will generally only
significantly perturb one of the two images.} of these two images for such a low-magnification event.  
As the image sweeps by the position of the planet, the planet will further perturb
the light from this image and yield a short-timescale deviation
({\em Gould \& Loeb}, 1992). The
duration of this deviation is $\sim t_{\e,p} = q^{1/2} t_{\rm E}$,
where $q=m_p/M$ is the mass ratio and $m_p$ is the planet mass, and the magnitude of the perturbation depends on how close the perturbed image
passes by the planet.
Given a range of primary event durations of $\sim 10-100~{\rm days}$, 
the duration of the perturbations range from a few hours for an
Earth-mass planet to a few days for Jovian-mass planets.   As can 
be seen in Figure \ref{fig:dia}, the location of
the perturbation relative to the peak of the primary event depends on
two parameters: $\alpha$, the angle of the projected star-planet axis with respect to the 
source trajectory\footnote{Conventionally, both the lensing
deflection angle and the source trajectory angle are denoted by
the symbol $\alpha$.  In order to maintain contact with the literature, 
this unfortunate convention is maintained here, but to avoid confusion 
the deflection angle is denoted with the subscript ``$d$''.},
and $d$, the instantaneous angular separation between the planet and host star in
units of $\theta_{\rm E}$.  

Because the orientation of the source trajectory relative to the planet
position is random, the time of this perturbation is not predictable
and the detection probability is $\sim A(t_{0,p}) \theta_{{\rm
E},p}/\theta_{\rm E}$, where $A(t_{0,p})$ is the unperturbed
magnification of the image that is being perturbed at the time
$t_{0,p}$ of the perturbation ({\em Horne et al.}, 2009), and
$\theta_{{\rm E},p} \equiv q^{1/2} \theta_\e$ is the Einstein ring
radius of the planet.  Here the factor $A$ accounts for the fact that
the area of the image plane covered by magnified images is larger by
their magnification.

Since the planet must be located near one of the two primary images in
order to yield a detectable deviation, and these images are always
located near the Einstein ring radius when the source is significantly
magnified (see Figure \ref{fig:cartoon}), the sensitivity of the
microlensing method peaks for planet-star projected separations of
$\sim r_\e$, i.e., for $d\sim 1$. However, microlensing can also
detect planets well outside the Einstein ring ($d\gg 1$), albeit with less
sensitivity.  Since the magnification of the minor image decreases
with position as $y_-^4$ (see Equations \ref{eqn:ypmfirst} and
\ref{eqn:apmped}), microlensing is generally not sensitive to planets
$d\ll 1$, i.e., close-in planets.

Given the existence of a planet with a projected separation within a
factor of $\sim 2$ of the Einstein ring radius, the detection
probabilities range from tens of percent for Jovian planets to a few
percent for Earth-mass planets ({\em Gould \& Loeb}, 1992; {\em
Bolatto \& Falco}, 1994; {\em Bennett \& Rhie}, 1996; {\em Peale},
2001). Detecting planets via the main channel requires substantial
commitment of resources because the unpredictable nature of the
perturbation requires dense, continuous sampling, and furthermore the
detection probability per event is relatively low so many events must
be monitored.

A useful feature of low-magnification planetary microlensing events
such as that shown in Figure \ref{fig:cartoon} is that it is possible
to essentially `read off' the light curve parameters from the observed
features ({\em Gould \& Loeb}, 1992; {\em Gaudi \& Gould}, 1997b).
For the primary event, the three gross observable parameters are the
time of maximum magnification $t_0$, the peak magnification $A_{\rm
max}$, and a measure of the duration of the event such as its
full-width half-maximum $t_{FWHM}$.  The latter two observables can
then be related to the Einstein timescale $t_\e$ and impact parameter
$u_0$ using Equations \ref{eqn:atot1} and \ref{eqn:uoft1}
\footnote{For the purposes of exposition, this discussion ignores
blending, which is generally important and complicates the
interpretation of observed light curves.  See Section
\ref{sec:lightcurves}.}.  For example, for small $u_0$ we have $u_0
\sim A_{\rm max}^{-1}$ and $t_\e \sim
0.5t_{FWHM}u_0^{-1}$. Unfortunately, only $t_{\rm E}$ contains any
information about the physical properties of the lens, and then only
in a degenerate combination of the lens mass, distance, and relative
lens-source proper motion. However, as discussed in detail in Section
\ref{sec:properties}, in many cases it is often possible to obtain
additional information which partially or totally breaks this
degeneracy.  In particular, in those cases where it is possible to
isolate the light from the lens itself, this measurement can be used
to constrain the lens mass ({\em Bennett et al.}, 2007).  The three
parameters that characterize the planetary perturbation are the
duration of the perturbation, the time of the perturbation $t_{0,p}$,
and the magnitude of the perturbation, $\delta_p$.  As mentioned
previously, the duration of
the perturbation is proportional to $t_{\e,p}\equiv q^{1/2}t_\e$, and
so gives the planet/star mass ratio $q$.  The time and magnitude of
the perturbation then specify the projected separation $d$ and angle
of the source trajectory with respect to the binary axis $\alpha$,
since $t_{0,p}$ and $\delta_p$ depend on the location of the planet
relative to the path of the perturbed image ({\em Gaudi \& Gould},
1997b).

The other channel by which microlensing can detect planets is in high
magnification events ({\em Griest \& Safizadeh}, 1998)\footnote{There
is no formal definition for `high'-magnification events, however
typically high-magnification refers to events with maximum
magnification $A_{\rm max} \ga 100$, corresponding to impact
parameters $u_0\la 0.01$.}.  In addition to perturbing images that
happen to pass nearby, planets will also distort the perfect circular
symmetry of the Einstein ring.  Near the peak of high-magnification
events, as the lens passes very close to the observer-source line of
sight (i.e.\ when $u\ll 1$), the two primary images are highly
elongated and sweep along the Einstein ring, thus probing this
distortion.  For very high-magnification events, these images probe
nearly the entire Einstein ring radius and so are sensitive to all
planets with separations near $r_\e$, regardless of their orientation
with respect to the source trajectory.  Thus high-magnification events
can have nearly 100\% sensitivity to planets near the Einstein ring
radius, and are very sensitive to low-mass planets ({\em Griest \&
Safizadeh}, 1998). However, these events are rare: a fraction $\sim
1/A_{\rm max}$ of events have maximum magnification $\ga A_{\rm max}$.
Fortunately, these events can often be predicted several hours to
several days ahead of peak, and furthermore the times of high
sensitivity to planets are within a full-width half-maximum of the
event peak, or roughly a day for typical high-magnification events
({\em Rattenbury et al.}, 2002).  Thus scarce observing resources can
be concentrated on these few events and only during the times of
maximum sensitivity.  Because the source stars are highly magnified,
it is also possible to use more common, smaller-aperture telescopes.

\subsection{Theory of Microlensing}\label{sec:theory}

Gravitational lensing can be thought of as the mapping 
$\mbox{\boldmath$\beta$} \rightarrow \mbox{\boldmath$\theta$}$ between the angular
position of a source $\mbox{\boldmath$\beta$}$ in the absence of lensing to the angular position(s) 
$\mbox{\boldmath$\theta$}$ of the image(s) of the source under the action of the gravitational lens.  
This mapping is given by the lens equation,
\begin{equation}
\mbox{\boldmath$\beta$} = \mbox{\boldmath$\theta$} - \mbox{\boldmath$\alpha_d(\theta)$},
\label{eqn:lenseq}
\end{equation}
where $\mbox{\boldmath$\alpha_d$}$ is deflection of the source due to the lens.

Consider a source at a distance $D_s$ and a lens
located at a distance $D_l$ from the observer.
Figure \ref{fig:dia} shows the lensing geometry.  The deflection angle $\mbox{\boldmath$\alpha_d$}$ is related
to the angle $\mbox{\boldmath$\hat \alpha_d$}$ by which the lens mass bends the light ray from the source by 
$\mbox{\boldmath$\hat \alpha_d$}(D_s-D_l)=\mbox{\boldmath$\alpha_d$}D_s$.
Assume the lens is a system of $N_L$ point masses 
each  with mass $m_i$ and position $\mbox{\boldmath$\theta$}_{m,i}$.  Further assume
that the lenses are static (or, more precisely, moving much more slowly than $c$), and
their distribution along the line of sight is small in comparison to $D_l$, $D_s$, and $D_s-D_l$.
The deflection angle is then,
\begin{equation}
\mbox{\boldmath$\alpha_d$}(\mbox{\boldmath$\theta$}) = \frac{4G}{D_{\rm rel} c^2} \sum_i^{N_l} m_i 
\frac{\mbox{\boldmath$\theta$}- \mbox{\boldmath$\theta$}_{m,i} }{|\mbox{\boldmath$\theta$}-\mbox{\boldmath$\theta$}_{m,i}|^2},
\label{eqn:bend}
\end{equation}
where $D_{\rm rel} \equiv (D_l^{-1}-D_s^{-1})^{-1}$.  See {\em Schneider et al.} (1992) and {\em Petters et al.} (2001)
for the expression for $\mbox{\boldmath$\alpha_d$}$ for a general mass distribution,
as well for the derivation of Equations \ref{eqn:lenseq} and \ref{eqn:bend} 
from the time delay function and ultimately the metric.  

It is common practice to normalize all angles to the angular Einstein ring radius,
\begin{equation}
\theta_\e \equiv \left(\kappa M \pi_{\rm rel}\right)^{1/2},
\label{eqn:thetae}
\end{equation}
where $M \equiv \sum_i^{N_l}m_i$ is the total mass of the lens,
$\pi_{\rm rel} \equiv {\rm AU}/D_{\rm rel}$ is the relative lens-source parallax,
and $\kappa=4G/(c^2{\rm AU})\simeq 8.14~{\rm mas}~M_\odot^{-1}$.
The reason for this convention is clear when considering the single lens (Section \ref{sec:point}).
The dimensionless vector source 
position is defined to be ${\bf u} \equiv \mbox{\boldmath$\beta$}/\theta_\e$, and the dimensionless vector
image positions are defined to be ${\bf y} \equiv \mbox{\boldmath$\theta$}/\theta_\e$. 

It is often convenient to write the lens equation in complex coordinates ({\em Witt}, 1990).
Defining the components of the (dimensionless) source position to be ${\bf u}=(u_1,u_2)$
and the image position(s) to be ${\bf y}=(y_1,y_2)$, the two-dimensional 
source and images positions can be expressed in complex form as, 
$\zeta=u_1 + i u_2$ and $z= y_1 + i y_2$.  The lens equation can now be rewritten
\begin{equation}
\zeta = z - \sum_i^{N_l} \frac{\epsilon_i}{{\bar z} - {\bar z}_{m,i}},
\label{eqn:clenseq}
\end{equation}
where $z_{m,i}$ is the position of mass $i$, $\epsilon \equiv m_i/M$.
The overbars denote complex conjugates, which in Equation \ref{eqn:clenseq}
arise from the identity $z/|z|^2 \equiv {\bar z}^{-1}$.  
This equation can then be solved to find the image positions $z_j$.

Lensing conserves surface brightness, but because of the mapping the
angular area of each image of the source is not necessarily equal to
angular area of the unlensed source.  Thus the flux of each image (the
area times the surface brightness) is different from the flux of the
unlensed source: the source is magnified or demagnified.  For a small
source, the magnification $A_j$ of each image $j$ is given by the
amount the source is ``stretched'' due to the lens mapping.
Mathematically, the amount of stretching is given by the inverse of the determinant of the Jacobian of the mapping
(\ref{eqn:clenseq}) evaluated at the image position,
\begin{equation}
A_j = \left. \frac{1}{{\rm det} J}\right\vert_{z=z_j},\,\, {\rm
det} J \equiv \frac{\partial(x_1,x_2)}{\partial(y_1,y_2)}=1-{\partial\zeta\over \partial\bar{z}}
{\overline{\partial\zeta}\over \partial\bar{z}}.
\label{eqn:maggen}
\end{equation}
See {\em Witt} (1990) for a derivation of the rightmost equality.  Note that these magnifications can be positive or negative, where the
sign corresponds to the parity (handedness) of the image.  By definition in
microlensing the images are unresolved, and we are interested in the
total magnification which is just the sum of the magnifications of the
individual images, $A\equiv \sum_j |A_j|$.

An interesting and critical property of gravitational lenses is that
the mapping can be singular for some source positions.  At these
source positions, ${\rm det} J = 0$.    In other words, an 
infinitesimally small displacement in the source position maps to an
infinitely large separation in the image position\footnote{This is analogous to the familiar 
distortion seen in cylindrical
map projections of the globe (such as the Mercator projection) for
latitudes far from the equator, due to the singular nature of these
mappings at the poles.}.  For point sources at these positions, the
magnification is formally infinite, and for sources near these
positions, the magnification is large. 

From the lens equation (\ref{eqn:clenseq}),  
\begin{equation}
{\partial\zeta\over \partial\bar{z}} =  \sum_i^{N_l} \frac{\epsilon_i}{({\bar z}_{m,i} - {\bar z})^2},
\end{equation}
and therefore, from equation (\ref{eqn:maggen}), the image positions where ${\rm det} J = 0$ are given by,
\begin{equation}
\left| \sum_i^{N_l} \frac{\epsilon_i}{({\bar z}_{m,i} - {\bar z})^2} \right|^2 = 1.
\label{eqn:crit}
\end{equation}
The set of all such image positions define closed {\it critical curves}.  These can be found by noting
that the sum in Equation \ref{eqn:crit} must have a modulus equal to unity.  Therefore, we can solve
for the critical image positions parametrically by solving the equation
\begin{equation}
\sum_i^{N_l} \frac{\epsilon_i}{({\bar z}_{m,i} - {\bar z})^2} = e^{i\phi},
\label{eqn:critpar}
\end{equation}
for each value of the parameter $\phi = [0,2\pi)$. 
The set of source positions corresponding to these image positions define closed curves called
{\it caustics}.  
By clearing the complex conjugates ${\bar z}$ and fractions ({\em Witt}, 1990), Equation \ref{eqn:critpar} can
be written as a complex polynomial of degree $2N_l$, there are at most $2N_l$ critical curves and caustics. 
Given their importance in planetary microlensing, caustics are discussed in considerably more 
detail below.

\subsubsection{Single Lenses}\label{sec:point}

For a single point mass ($N_l=1$), defining the origin as the position of the
lens, the lens equation reduces to,
\begin{equation}
\beta = \theta - \frac{\theta_\e^2}{\theta}.
\label{eqn:singlec}
\end{equation}
Since the lens is circularly symmetric, the images are always located along the 
line connecting the lens and source, and the vector notation has been suppressed, but
the convention that positive values of $\theta$ are for images on the same
side of the lens as the source is kept.  
If the lens is perfectly aligned with the source, then $\beta=0$,
and $\theta=\theta_\e$.  In other words, the lens images
the source into a ring of radius equal to the angular Einstein ring radius.

The dimensionless single lens equation is,
\begin{equation}
u=y-\frac{1}{y}.
\label{eqn:singles}
\end{equation}
In the case of imperfect alignment, this becomes a quadratic
function of $y$, and so there are two images of the source, with positions,
\begin{equation}
y_{\pm} = \pm \frac{1}{2}\left(\sqrt{u^2+4}\pm u\right).
\label{eqn:plimages}
\end{equation}
One of these images (the major image, or minimum) is always outside the Einstein ring radius ($y_+ \ge 1$)
on the same side of the lens as the source, and the other image (the minor image, or saddle point) is always inside the Einstein ring radius ($|y_-| \le 1$) on the opposite side of the lens as the source.
The separation between the two images is $|y_+-y_-|=(u^2+4)^{1/2}$, and thus the images 
are separated by $\sim 2\theta_\e$ when both images are significantly magnified (i.e.\ when $u\la 1$).
Since $\theta_\e$ is of order a milliarcsecond for typical lens masses, and source and lens
distances in events toward the Galactic bulge, the images are unresolved.

The magnifications of each image can be found analytically, 
\begin{equation}
A_\pm = \frac{1}{2}(A\pm 1)
\label{eqn:magind}
\end{equation}
where the total magnification is,
\begin{equation}
A(u)= \frac{u^2+2}{u\sqrt{u^2+4}}.
\label{eqn:magtot}
\end{equation}
A few properties are worth noting.  
First, $u \ll 1$, $A(u)\simeq u^{-1}$.  The magnification diverges for $u\rightarrow 0$, and the
point $u=0$ defines the caustic in the single lens case.  Second, for $u \gg 1$, $A(u) \simeq 1+2u^{-4}$,
and thus the excess magnification drops rapidly for large source-lens angular separations.

\begin{figure}[h]
\epsscale{1.0}
\plotone{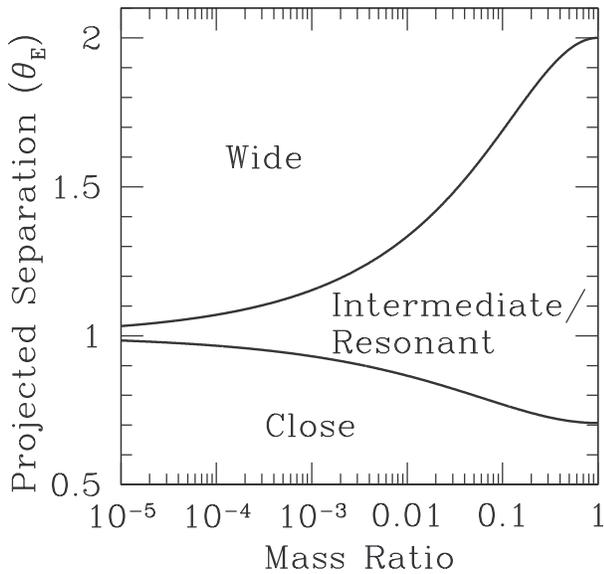}
\caption{\small 
The critical values of $d$, the projected separation in units of $\theta_{\rm E}$, at which
the caustic topology (number of caustic curves) of a binary lens changes as a function of the mass ratio $q$.
The upper curve shows $d_w$, the critical value of $d$ between the wide caustic
topology consisting of two disjoint caustics, and the intermediate or resonant caustic topology consisting
of a single caustic.  The lower curve shows $d_c$, the critical value between the resonant caustic
topology and the close caustic topology consisting of three disjoint caustics.
}\label{fig:crit}
\end{figure}

\subsubsection{Binary Lenses}\label{sec:binary}

For a two point-mass lens ($N_l=2$), the lens equation is,
\begin{equation}
\zeta = z + \frac{\epsilon_1}{{\bar z}_{m,1} - {\bar z}}+\frac{\epsilon_2}{{\bar z}_{m,2} - {\bar z}}.
\label{eqn:blenseq}
\end{equation}
This can be written as a fifth-order complex polynomial in $z$, which cannot
be solved analytically.  {\em Witt \& Mao}, 1995 provide the coefficients of the 
polynomial. It can easily be solved with standard numerical routines,
e.g., Laguerre's method ({\em Press et al.}, 1992), to yield the image
positions $z$.  It is important to note that that the solutions
to the fifth-order complex polynomial are not necessarily solutions to the lens equation
(Equation \ref{eqn:blenseq}).  Depending on the location of the source
with respect to the lens positions, two of the images can be spurious.  Thus there
are either three or five images.  

The boundaries of the three and five image regions are the caustic
curves (where ${\rm det} J =0$), and thus the number of images changes
by two when the source crosses the caustic.  A binary lens has one,
two, or three closed and non-self-intersecting caustic curves.  Which
of these three topologies is exhibited depends on the mass ratio of
the lens $q \equiv m_1/m_2$ and on the angular separation of the two
lens components in units of Einstein ring radius of binary, $d \equiv |z_{m,1}-z_{m,2}|$.  
For a given $q$, the values of $d$ for which the topology changes are given by 
({\em Schneider \& Weiss}, 1986; {\em Dominik}, 1999b),
\begin{equation}
\frac{q}{(1+q)^2} = \frac{(1-d_c)^3}{27d_c^8},\qquad d_w=\frac{(1+q^{1/3})^{3/2}}{(1+q)^{1/2}}.
\end{equation}
For $d\le d_c$, there are three caustic curves, for $d_c \le d \le d_w$, there
is one caustic curve, and for $d\ge d_w$, there are two caustic curves. 
These are often referred to as the ``close'', ``intermediate'' or ``resonant'',
and ``wide'' topologies, respectively. Figure \ref{fig:crit} plots $d_c$ and $d_w$ as a function
of $q$. For equal-mass
binaries, $q=1$, the critical values of the separation are $d_c=2^{-1/2}$ and $d_w=2$.
 
A useful property of binary lenses is that the total magnification of all
the images is always $A\ge 3$ when the source is interior to the caustic
curve, i.e., when there are five images of the source ({\em Witt \& Mao}, 1995).
Thus if the magnification of a source is observed to be less than 3 during a microlensing
event, it can be immediately concluded that either the source is exterior to the caustic,
or the source is significantly blended with an unrelated, unlensed star\footnote{There
is a third, less likely possibility that there are additional bodies in the system. In
this case, the bound that $A\ge 3$ interior to the caustic can be violated ({\em Witt \& Mao}, 1995)}, thus diluting
the magnification.

\subsubsection{Triple Lenses and Beyond}\label{sec:triple}

For a triple lens ($N_l=3$), the lens equation can be written a
tenth-order complex polynomial in $z$ ({\em Gaudi et al.}, 1998), which can be solved using the
same techniques as in the binary lens case.  {\em Rhie}, 2002 provides the coefficients of the 
polynomial.  There are a maximum of
ten images, and there are a minimum of four images, with the number
of images changing by a multiple of two when the source crosses the caustic.  
The caustics of triple lenses can exhibit quite complicated topologies, including
nested and/or self-intersecting caustic curves.  The topology depends
on five parameters: two mass ratios, two projected separations, and
the angle between projected position vectors of the companions and the
primary lens.

In general, the lens equation for a system of $N_l$ point lenses is
can be written as a complex polynomial of order $N_l^2+1$.  However, it
has been shown that the maximum number of images is $5(N_l-1)$ for
$N_l\ge 2$ ({\em Rhie}, 2001, 2003; {\em Khavinson \& Neumann}, 2006). Thus for $N_l>3$, it is always
the case that some of the roots of the polynomial are not solutions to
the lens equation.

\subsubsection{Magnification Near Caustics}\label{sec:caustics}

As mentioned previously, the caustics are the set of source positions
for which the magnification of a point source is formally infinite.
Caustic curves are characterized by multiple concave segments called
{\it folds} which meet at points called {\it cusps}.  Folds and cusps
are so named because the local lensing properties of sources close to
these caustics are equivalent to the generic fold and cusp
mapping singularities of mathematical catastrophe theory.  For a more
precise formulation of this statement and additional discussion, see {\em Petters et al.}, 2001.  
This leads to the particularly important property
of fold and cusps that their local lensing properties are universal, regardless
of the global properties of the lens.  Thus the positions and magnifications
of the critical images of sources near folds and cusps have universal scaling behaviors 
that can be described essentially analytically, and whose normalization
depend only on the local properties of the lens potential.  
This universal and local behavior of the magnification
near caustics has proven quite useful, in that it allows one to analyze light curves
near caustic crossings separate from and independent of the global lens model (see,
e.g., {\em Gaudi \& Gould}, 1999, {\em Albrow et al.}, 1999b; {\em Rhie \& Bennett}, 1999, {\em Afonso et al.}, 2001; 
{\em Dominik}, 2004a,b). 

The lensing behavior near
folds and cusps has been discussed in detail by a number of authors ({\em Schneider \& Weiss}, 1986,1992;
{\em Mao}, 1992; {\em Zakharov}, 1995,1999;
{\em Fluke \& Webster}, 1999; {\em Petters et al.}, 2001, {\em Gaudi \& Petters}, 2002a,b; 
{\em Pejcha \& Heyrovsky}, 2009).  The salient properties are briefly reviewed
here, but the reader is encouraged to consult these papers for a more in-depth discussion.  

For sources close to and interior\footnote{Here interior to a fold
caustic is defined such that the caustic curves away from the source.}
to a fold caustic, there are two highly-magnified images with nearly
equal magnification and opposite parity that are nearly equidistant
from the corresponding critical curve.  Neglecting the curvature of
the caustic and any changes in the lensing properties parallel to the caustic,
the total magnification of these divergent images is ({\em Schneider \& Weiss}, 1986),
\begin{equation}
A_{\rm div}(\Delta u_\perp)= \left(\frac{ \Delta u_\perp}{u_r}\right)^{-1/2}\Theta(\Delta u_\perp),
\label{eqn:fold}
\end{equation} 
where $\Delta u_\perp$ is the perpendicular distance to
the caustic, with $\Delta u_\perp>0$ for sources interior to the caustic, $\Theta(x)$ is the Heaviside step function,
and $u_r$ is the characteristic `strength' of the fold caustic locally, and is related to local
derivatives of the lens potential. 
As the source approaches the caustic, the images
brighten and merge, disappearing when the source
crosses the caustic.  The behaviors of the remaining (non-critical) images are
continuous as the source crosses the caustic.  Thus immediately
outside of a fold, the magnification is finite and (typically) modest.

For sources close to and interior to a cusp, there are three
highly-magnified images.  For a source on the axis of symmetry of the cusp, the
total magnification of the three images is $\propto \Delta u_c^{-1}$,
where $\Delta u_c$ is the distance of the source from the cusp. As the
source approaches the cusp, the magnification of all three images 
increases and the images merge.  Two of the three images disappear as the source exits
the cusp.  The magnification of the remaining image is continuous
as the source exits the cusp, and in particular the image remains highly magnified, also with
magnification which is $\propto \Delta u_c^{-1}$. Thus, in contrast to
folds, sources immediately exterior to cusps are highly magnified.  
As with the fold, the behaviors of the non-critical images are
continuous as the source crosses the caustic.

Caustic curves are closed, and thus for any given source trajectory
(which is simply a continuous path through the source plane), caustic
crossings come in pairs.  Generally, since the majority of the length
of a caustic is made up of fold caustics, both the caustic entry and exit
are fold crossings.  Since the magnification immediately outside of a
fold is not divergent, it is usually impossible to predict a caustic
entry beforehand.  Once one sees a fold caustic entry, a caustic exit
is guaranteed, and this is typically a fold exit.  Monitoring a
caustic exit is useful for two reasons.  First, the strong finite
source effects (see below) during the crossing can be used to provide
additional information about the lens (see Section \ref{sec:properties}), and measure
the limb-darkening of the source (e.g., {\em Albrow et al.}, 1999c; {\em Fields et al.}, 2003).
In addition, during the caustic crossing the source is highly
magnified and potentially very bright, which allows for 
otherwise impossible spectroscopic
observations to determine properties of the source star, such as
its effective temperature and atmospheric abundances ({\em Minniti et al.}, 1998; 
{\em Johnson et al.}, 2008).  Unfortunately, it is typically difficult to
predict when this exit will happen well before the crossing ({\em
Jaroszy{\'n}ski \& Mao}, 2001).

For both fold and cusp caustics, the magnification of a source of
finite size begins to deviate by more than a few percent from the
point-source approximation when the center of the source is
within several source radii of the caustic ({\em Pejcha \&
Heyrovsky}, 2009).  Formally, the magnification of a finite source can
be found simply by integrating the point-source magnification over of
source, weighting by the source surface brightness distribution.
Practically, this approach is difficult and costly to implement
precisely due to the divergent magnification near the caustic curves.
An enormous amount of effort has been put into developing robust and
efficient algorithms to compute the magnification for finite sources
({\em Dominik}, 1995, 2007; {\em Wambsganss}, 1997; {\em Gould \&
Gaucherel}, 1997; {\em Griest \& Safizadeh}, 1998; {\em Vermaak},
2000; {\em Dong et al.}, 2006; {\em Gould}, 2008; {\em Pejcha \&
Heyrovsky}, 2009; {\em Bennett}, 2009b).  The most efficient of these algorithms use a
two-pronged approach.  First, semi-analytic approximations to the
finite-source magnification derived from an expansion of the
magnification in the vicinity of the source are used where appropriate
({\em Gould}, 2008; {\em Pejcha \& Heyrovsky}, 2009).  Second, where
necessary a full numerical evaluation of the finite source
magnification is performed by integrating in the image plane.
Integrating in the image plane removes the difficulties with the
divergent behavior of the magnification near caustics, because the
surface brightness profiles of the images are smooth and continuous.
Thus one simply `shoots' rays in the image plane, and determines which
ones `land' on the source using the lens equation.  The ratio of the
total area of all the images divided by the area of the source
(appropriately weighted by the surface brightness profile of the
source) gives the total magnification.  The devilish details then lie
in the manner in which one efficiently samples the image plane 
(see e.g., {\em Rattenbury et al.}, 2002; {\em Dong et al.}, 2006; {\em Pejcha \& Heyrovsky}, 2009; {\em Bennett}, 2009b).

For planetary microlensing events, efficient routines for evaluating
the finite-source magnification are crucial, for two reasons.  First, 
nearly all planetary perturbations are
strongly affected by finite source effects.  Second, the processes of
finding the best-fit model to an observed light curve and evaluating
the model parameter uncertainties requires calculating tens of thousands of
trial model curves (or more), and the majority of the computation time is spent
calculating the finite-source magnifications.

\begin{figure*}[htp]
\epsscale{2.1}
\plotone{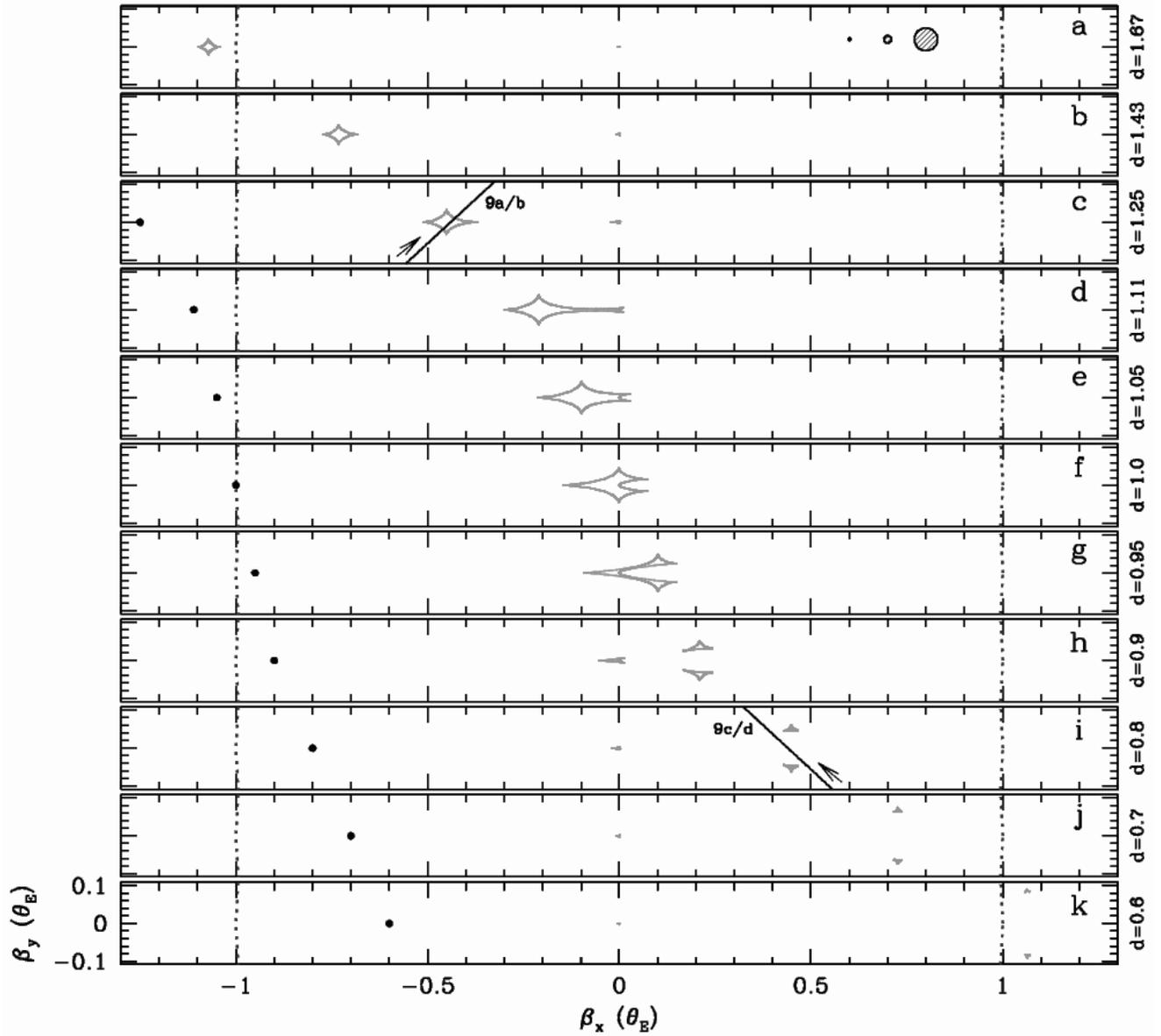}
\caption{\small 
The grey curves show the caustics for a planetary lens with mass ratio $q=0.001$, and 
various values of $d$, the projected separation in units of $\theta_\e$.  The dotted
lines show sections of the Einstein ring.  The dots show the location of the
planet. In panels c and i, an example trajectory is shown which produces a 
perturbation by the planetary caustic; the resulting light curves are shown in Figure \ref{fig:plcurves}.  In panel a,
three different representative angular source sizes in units of $\theta_\e$ are shown, $\rho_*=0.003, 0.01$,
and 0.03. For typical microlensing event parameters, these correspond to stars in the Galactic bulge with
radii of $\sim R_\odot, \sim 3R_\odot$, and $\sim 10R_\odot$, i.e., a main-sequence turn-off star, a subgiant, and a clump giant.
}\label{fig:pcaustics}
\end{figure*}

\subsection{Planetary Microlensing Phenomenology}\label{sec:phenom}

For binary lenses in which the companion mass ratio is $q\ll 1$ (i.e.,
planetary companions), the companion will cause a small perturbation to the
overall magnification structure of the lens.  Thus the majority of
source positions will give rise to magnifications that are essentially
indistinguishable from a single lens.  The source positions for which
the magnification deviates significantly from a single lens are all
generally confined to a relatively narrow region around the caustics.
Thus much of the phenomenology of planetary microlensing can be
understood by studying the structure of the caustics and the magnification
pattern near the caustics.

Recall there are three different caustic topologies for binary lenses:
close, intermediate, and wide (See Figure \ref{fig:crit}).  For $q\ll 1$, the critical values of
the separation where these caustic topologies change can be
approximated by $d_c\simeq 1-3q^{1/3}/4$ and $d_w \simeq 1+3q^{1/3}/2$
({\em Dominik}, 1999b).  Therefore, for planetary lenses, the
intermediate or resonant caustics are confined to a relatively
narrow range of separations near $d=1$, and this range shrinks as
$q^{1/3}$.  Figure \ref{fig:pcaustics} shows the caustics for a Jupiter/Sun mass ratio
of $q=0.001$, and 11 different separations $d=0.6,0.7,0.8,0.9, 0.95,
1.0, 1.05, 1.11, 1.25, 1.43, 1.67$.

\begin{figure}[htp]
\epsscale{1.0}
\plotone{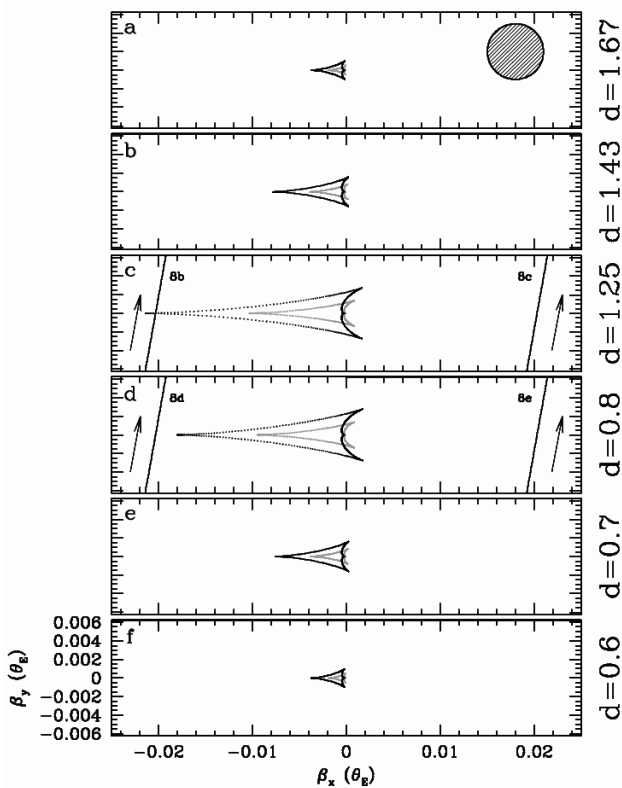}
\caption{\small 
The black curves show the central caustics for a planetary lens with $q=0.001$, and 
various values of $d$, the projected separation in units of $\theta_\e$.  The primary
lens is located at the origin, and so trajectories which probe
the central caustic correspond to events with small impact
parameter $u_0$, or events with high maximum magnification. The grey
curves show the central caustic for a mass ratio of $q=0.0005$, demonstrating
that the size of the central caustic scales as $q$.  For $q\ll 1$, the central
caustic and proximate magnification patterns are essentially identical under the transformation $d \leftrightarrow d^{-1}$. The degree of asymmetry, i.e.\ the
length to width ratio, of the central caustic depends on $d$, such that the caustic becomes more asymmetric
as $d\rightarrow 1$.  In panels c and d, example trajectories are shown which produce 
perturbations by the central caustic; the resulting light curves are shown in Figure \ref{fig:clcurves}.
In panel a, a representative angular source size in units of $\theta_\e$ of $\rho_*=0.003$ is
shown. For typical microlensing event parameters, this correspond to a star in the Galactic bulge of
radius $\sim R_\odot$, i.e., a main-sequence turn-off star.
}\label{fig:ccaustics}
\end{figure}

\begin{figure*}
\epsscale{1.8}
\plotone{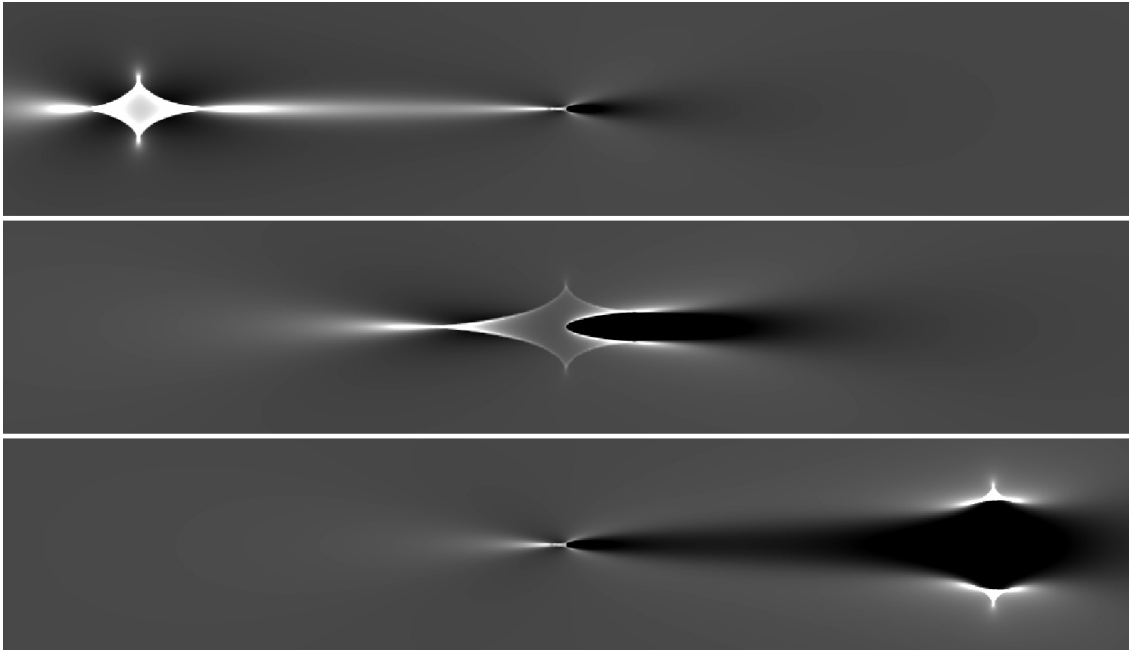}
\caption{\small 
The magnification pattern as a function of source position for a planetary companion with 
$q=0.001$ and $d=1.25$ (top panel),
$d=1.0$ (middle panel), and $d=0.8$ (bottom panel), corresponding to wide, intermediate/resonant,
and close topologies, respectively.  The greyscale shading denotes $2.5\log(1+\delta)$,
where $\delta$ is the fractional deviation from the single-lens (i.e., no planet) magnification. White
shading corresponds to regions with positive deviation from the single lens magnification,
whereas black shading corresponds to negative deviations.  For the wide and close topology,
there are two regions of large deviations, corresponding to the central caustics located
at the position of the primary (the center of each panel), and the planetary caustics.  For the intermediate/resonant
topology, there is only one large caustic, which produces relatively weak perturbations
for a large fraction of the caustic area. 
}\label{fig:magmap}
\end{figure*}

\begin{figure}[htp]
\epsscale{1.0}
\plotone{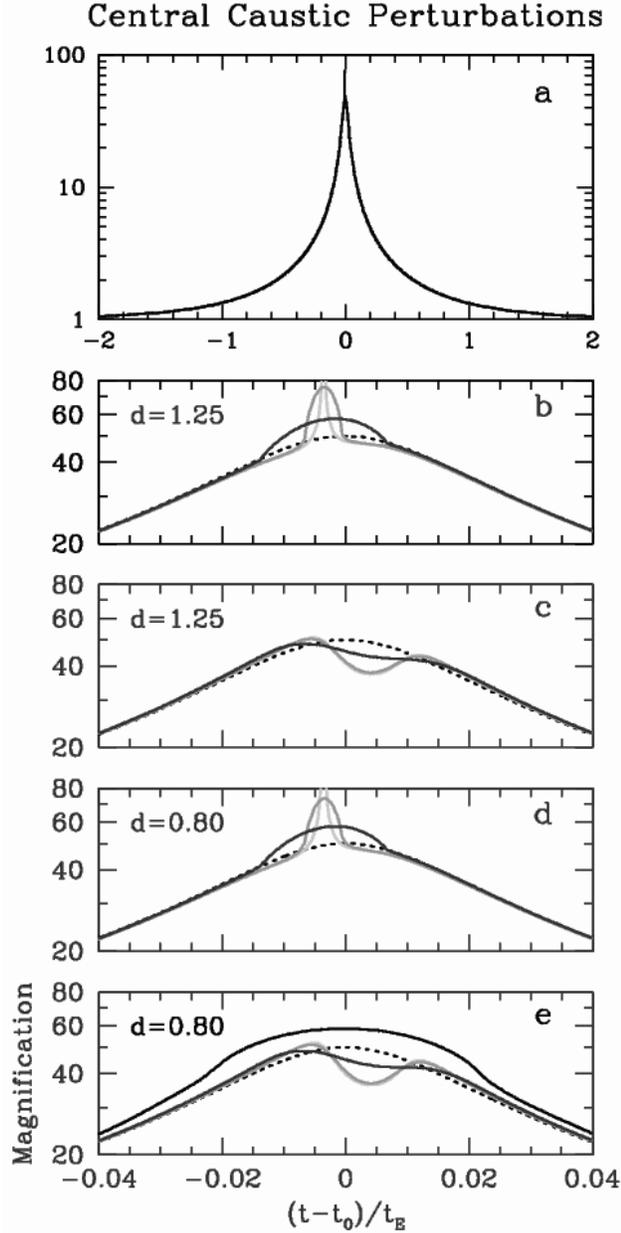}
\caption{\small 
Example light curves of planetary perturbations arising from
the source passing close to the central caustic in a high-magnification
event, for a planet/star mass ratio of $q=0.001$. Panel (a) shows
the overall light curve.  The impact parameter of the event with respect to the 
primary lens is $u_0=0.02$, corresponding to a peak magnification
of $A_{\rm max} \sim u_0^{-1}=50$.  Panels (b-e) show zooms of the light curve peak.  Two
different cases are shown, 
one case of the wide planetary companion with $d=1.25$ (b,c), and a close planetary companion with $d=0.8$ (d,e).
These two cases satisfy $d \leftrightarrow d^{-1}$ and demonstrate the close/wide degeneracy. 
The source passes close to the central
caustic; two example trajectories are shown in Figure \ref{fig:ccaustics} and the resulting
light curves including the planetary perturbations are shown in panels b-e. 
The dotted line shows the magnification with no planet, whereas the
solid lines show the planetary perturbations with source
sizes of $\rho_*=0,0.003$, and $0.01$, (lightest to darkest).  
In panel e, the light curve for $\rho_*=0.03$ is also shown.  In this case,
the primary lens transits the source, resulting in a `smoothed' peak. 
Although the planetary deviation is largely washed out, it is still detectable
with sufficiently precise photometry.
}\label{fig:clcurves}
\end{figure}

For both the close ($d<d_c$; Figure \ref{fig:pcaustics}i-k) and wide ($d>d_w$; Figure \ref{fig:pcaustics}a-c) topologies, 
one caustic is always located near the position of the primary
(the origin in Figure \ref{fig:pcaustics}).  This is known as the central caustic.
Figure \ref{fig:ccaustics} shows an expanded view of these caustics, which have a highly
asymmetric `arrow' shape, with one cusp at the arrow tip pointing
toward the planet, and three cusps at the `back end' of the caustic
pointing away from the planet.  The on-axis cusp pointing away from
the planet is generally much `weaker' than the other cusps, in the
sense that the scale of the gradient in magnification along the cusp
axis is smaller for the weaker cusp, so that at fixed distance from
the cusp, the excess magnification is smaller for the weaker cusp.
The magnification pattern near a central caustic is illustrated in Figure \ref{fig:magmap}.
The light curves (one-dimensional slices through the magnification pattern) 
for sources passing perpendicular to the binary-lens
axis close the back-end will exhibit a `U'-shaped double-peaked
deviation from the single-lens form (Figure \ref{fig:clcurves}c,d), whereas sources passing
perpendicular to the binary-lens axis close to the tip of the central
caustic will exhibit a single bump (Figure \ref{fig:clcurves}b,d).  Sources passing the caustic
parallel to the binary-lens axis 
will exhibit little deviation from the single lens form, essentially unless they
cross the caustic.  

In the limit that $q\ll 1$, and $|d-1|\gg q$, it is possible show
that, for fixed $d$, the size of the central caustic scales as $q$.
Furthermore, the overall shape of the caustic, as quantified,
e.g., by its length-to-width ratio, depends only on $d$, such that the
caustic becomes more asymmetric (the length-to-width ratio increases)
as $d\rightarrow 1$. Finally, the central caustic shape and size is
invariant under the transformation $d\rightarrow d^{-1}$.  See {\em
Chung et al.}, 2005 and Figure \ref{fig:ccaustics}.  As illustrated
in Figure \ref{fig:clcurves}, the $d \rightarrow d^{-1}$ duality
results in a degeneracy between light curves produced by
central caustic perturbations due to planetary companions, typically referred to as the close/wide
degeneracy ({\em Griest \& Safizadeh} 1999; {\em Dominik}, 1999b; {\em Albrow et al.}, 2000;
{\em An}, 2005)

\begin{figure}[htp]
\epsscale{1.0}
\plotone{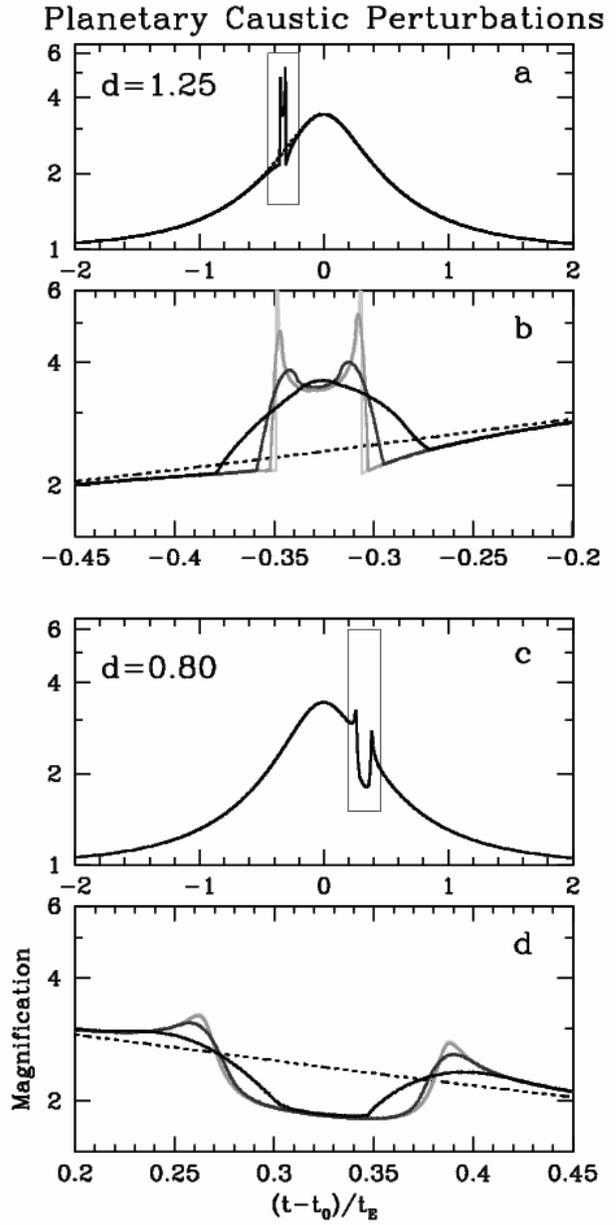}
\caption{\small 
Example light curves of planetary perturbations arising from
the source passing close to the planetary caustic for a planet/star mass ratio of
$q=0.001$.    Panels (a,c) show the overall light curves, whereas
panels (b,d) show zooms of the planetary deviation.
Two cases are shown,
one case of the wide planetary companion with $d=1.25$ (a,b), and a close planetary companion with $d=0.8$ (c,d).
In both cases, the impact parameter of the event with respect to the 
primary lens is $u_0=0.3$.  
The trajectories for the light curves displayed are shown in Figure \ref{fig:pcaustics}. 
The  dotted line shows the magnification with no planet, whereas the
solid lines show the planetary perturbations with source sizes
of $\rho_*=0, 0.003$, $0.01$, and $0.03$ (lightest to darkest).\vspace{25mm}
}\label{fig:plcurves}
\end{figure}

For the close $(d<d_c)$ topology (Figure \ref{fig:pcaustics}h-j), there are three caustics, the
central caustic and two larger, triangular-shaped caustics with three
cusps.  The latter caustics are referred to as the planetary caustics,
and are centered on the planet/star axis at angular separation of $u_c
\simeq |d-d^{-1}|$ from the primary lens, on the opposite side of the
primary from the planet\footnote{It is possible to derive the location
of the planetary caustic(s) for both the close and wide topologies by noting
that, when the source crosses the planetary caustic(s), the planet is perturbing one of the two images 
of the source created
by the primary lens (the major image in the case of the wide topology, and
the minor image in the case of the close topology). 
For a source position $u$, the locations of the two primary 
images $y_\pm (u)$ are given by Equation \ref{eqn:plimages}.  The
planet must therefore be located at $d\sim y_{\pm}(u)$ to significantly perturb
the image, and thus the center of caustic $u_c$ is located
at the solution of the inversion of Equation
\ref{eqn:plimages}, i.e., $u(y_\pm=d)$, which yields $u_c=d-d^{-1}$.}.
The caustics are symmetrically displaced
perpendicular to the planet/star axis, with the separation between the
caustics increasing with decreasing $d$ and so increasing $u_c$.  As
illustrated in Figure \ref{fig:magmap}, the magnification pattern near these caustics
is characterized by small regions surrounding the caustics where a
source exhibits a positive deviation from the single lens
magnification, and a large region between the caustics where a source
exhibits a negative deviation from the single lens magnification.
Figure \ref{fig:plcurves}c,d shows a representative light curve from a source passing
near the planetary caustics of a close planetary lens with $d=0.8$ and
$q=0.001$.

For the wide $(d>d_w)$ topology (Figure \ref{fig:pcaustics}h-j), there are two caustics, the
central caustic and a single planetary caustic with four cusps.  As for the
close planetary caustics, the wide planetary caustic is centered on
the planet/star axis at an angular separation of $u_c \simeq
|d-d^{-1}|$ from the primary lens, but in this case on the same side
of the primary from the planet.  The caustic is an asteroid shape,
with the length along the planet/star axis being generally longer than the width.
The asymmetry (i.e., length-to-width ratio) of the planetary caustic
increases as $d\rightarrow 1$.  The magnification pattern near the
wide planetary caustic is characterized by large positive deviations
interior to the caustic, and lobes of positive deviation extending
outward along the axes of the four cusps, particularly along the
planet/star axis in the direction of the primary.  There are
relatively small regions of slight negative deviation from the
single-lens magnification immediately outside the fold caustic between the
cusps.  Figure \ref{fig:plcurves}a,b shows a representative light curve from a source
passing through the planetary caustic of a wide planetary lens with
$d=1.25$ and $q=0.001$. 

In the wide  $(d>d_w)$ and close $(d<d_c)$ cases, the planet is essentially perturbing one
of the two images created by the primary lens, and the other image (on
the other side of the Einstein ring) is essentially unaffected.  In
this case, and in the limit that $q \rightarrow 0$, the lensing
behavior near the planet and the perturbed primary image is equivalent
to a single lens with pure external shear ({\em Gould \& Loeb}, 1992;
{\em Dominik}, 1999; {\em Gaudi \& Gould}, 1997a), i.e., a Chang-Refsdal lens ({\em Chang \&
Refsdal}, 1979).  The Chang-Refsdal lens has been studied extensively (e.g.,
{\em Chang \& Refsdal}, 1984; {\em An}, 2005; {\em An \& Evans},
2006), and its properties are well-understood.  The lens equation is,
\begin{equation}
\zeta = z - \frac{1}{\bar z}-\gamma {\bar z},
\label{eqn:crlenseq}
\end{equation}
where $\gamma$ is the shear and the origin is taken to be the location 
of the planet.  This is equivalent to a fourth-order
complex polynomial in $z$, which can be solved analytically, or numerically in the same manner as
the binary-lens case (see Section \ref{sec:theory}).  There are 2 or 4 images, depending
on the source position.  The correspondence between the Chang-Refsdal
lens and the wide/close planetary case is achieved by setting
$\gamma=d^{-2}$ and choosing the origin of the binary lens to be
$d-d^{-1}$ from the primary lens, and including in the Chang-Refsdal
approximation the magnification of the
unperturbed image created by the primary on the other side of the
Einstein ring.  Note that $\gamma >1$ corresponds to the planetary
caustics for the close topology, whereas $\gamma<1$ corresponds
to the wide topology.

The properties of the caustics of a Chang-Refsdal lens (and thus the
planetary caustics of wide/close planetary lenses) can be studied
analytically.  Based on expressions from {\em Bozza}, 2000, {\em Han},
2006 has studied the scaling of the planetary caustics.  The overall
size of the wide $(d>1)$ planetary caustic is approximately $\propto
q^{1/2}d^{-2}$, and the length-to-width ratio is $\sim 1+d^{-2}$.
The shape is independent of $q$ in this approximation.  The overall
size of the close $(d<1)$ planetary caustics are approximately
$\propto q^{1/2}d^{3}$, and their shape is also independent of
$q$. The vertical separation between the two planetary caustics in the
close topology is $\propto q^{1/2}d^{-1}$.

For the resonant case ($d_c\le d \le d_w$) there
exists a single, relatively large caustic with six cusps.  For fixed
$q$, the resonant caustic is larger than either the central or planetary caustics.
The large size of these caustics results in a large
cross-section and thus an enhanced detection probability.  Indeed, in the first
planet detected by microlensing, the source crossed a resonant caustic.  The large
size also means that the light curve deviation can last a significant fraction
of the duration of the event. 
However, resonant caustics are also `weak' in the sense that for a large fraction of the area
interior to or immediate outside the caustic,
the excess magnification relative to a single lens is small (see Figure \ref{fig:magmap}). 
The exceptions to this are source positions in the vicinity the cusp located on the planet/star axis 
pointing toward the planet, and source positions near the `back end' of the caustic
near the position of the primary, which are characterized by large negative deviations relative
to the single-lens magnification.  The precise shape of the resonant caustic depends
sensitively on $d$, and thus
small changes in the value of $d$ lead to large changes in the caustic morphology, as can
be seen in Figure \ref{fig:pcaustics}.  As a result, the effects of orbital motion, which result in a change
in $d$ over the course of the event, are expected to be more
important for resonant caustic perturbations.  The size of resonant caustics scales 
as $q^{1/3}$, in contrast to planetary caustics, which scale as $q^{1/2}$, and central
caustics, which scale as $q$.  

\section{PRACTICE OF MICROLENSING}\label{sec:practice}

\subsection{Light Curves and Fitting}\label{sec:lightcurves}

The apparent relative motion between the lens and the source gives
rise to a time-variable magnification of the source: a microlensing event. 
Is is often (but not always, see Section \ref{sec:higher}) a good
approximation that the source, lens, and observer are in uniform,
rectilinear motion, in which case the angular separation between the
lens and source as a function of time can be written as, 
\begin{equation}
u(t)= \left( \tau^2 +u_0^2\right)^{1/2},
\label{eqn:uoft}
\end{equation}
where $\tau \equiv (t-t_0)/t_\e$, 
$t_0$ is the time of closest alignment, which is also the time of maximum magnification,
$u_0$ is the impact parameter of the event, and $t_\e$ is the Einstein ring crossing time,
\begin{equation}
t_\e \equiv \frac{\theta_\e}{\mu_{\rm rel}},
\end{equation}
where $\mu_{\rm rel}$ is the relative lens-source proper motion. 
Figure \ref{fig:ped} shows the magnification as a function of time for 
a microlensing event due to a single lens, with impact parameters of $u_0=0.01,0.1,0.2,...,0.9,1.0$, which serve to illustrate
the variety of light curve shapes.  

\begin{figure}[htp]
\epsscale{0.56}
\plotone{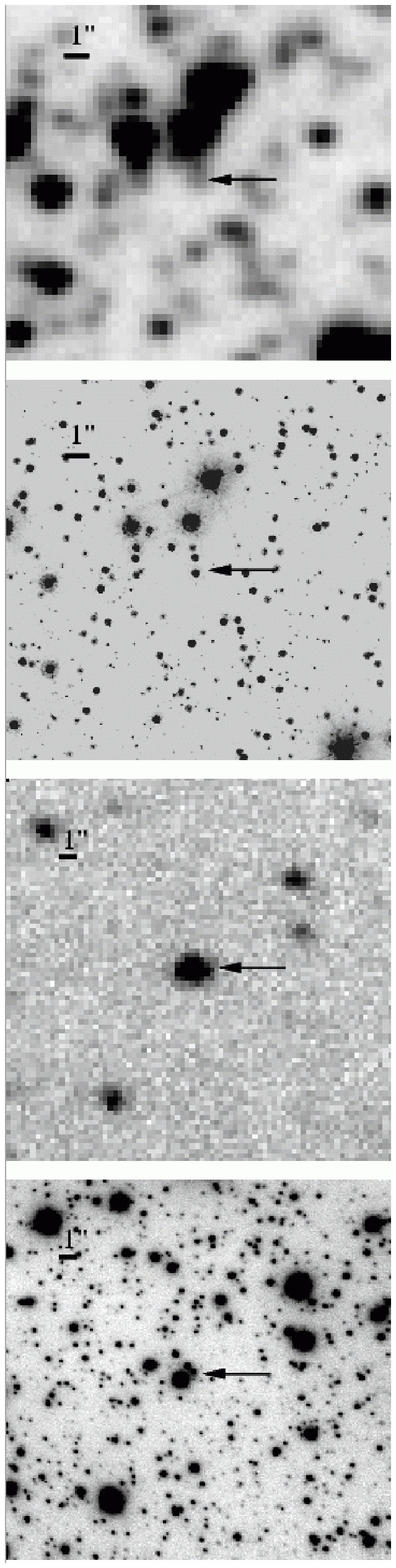}
\caption{\small 
Top panel:  A 15''$\times$ 15'' $I$-band image of the field
of planetary microlensing event OGLE-2005-BLG-071 obtained the OGLE 1.3m Warsaw telescope
at Las Campanas Observatory in Chile.  Second panel from top: Same
field as the top panel, but taken with the {\it Advanced Camera for Surveys}
instrument on the {\it Hubble Space Telescope} in the F814W filter.
Third panel from top:  22''$\times$ 22'' $H$-band image of the field
of planetary microlensing event MOA-2008-BLG-310 obtained with the CTIO/SMARTS2 1.3m telescope
at Cerro Tololo InterAmerican Observatory in Chile.  Bottom panel:
Same field and filter as the panel above, but taken with the NACO instrument
on VLT.  In all panels, the arrow indicates the microlensing target.
}\label{fig:blending}
\end{figure}

Of course, what is observed is not the magnification, but the flux of a photometered
source as a function of time, which is given by,
\begin{equation}
F(t)= F_s A(t) + F_b.
\label{eqn:foft}
\end{equation}
Here $F_s$ is the flux of the microlensing star, and $F_b$ is the flux of any 
unresolved light (or ``blended light'') that is not being lensed.
The latter can include light from a companion to the source, light from
unrelated nearby stars, light from a companion to the lens, and (most
interestingly) light from the lens itself.  Microlensing experiments
are typically carried out toward crowded fields in order to maximize the event rate,
and therefore one often finds unrelated stars blended with the microlensed source
for typical ground-based resolutions of $\sim 1''$.  Even in the most crowded
bulge fields, most unrelated background
stars are resolved at the resolution of the {\it Hubble Space Telescope (HST)}. Figure \ref{fig:blending}
shows the fields of two events as observed from the ground with typical seeing, with 
{\it HST}, and with ground-based adaptive optics (AO).

The observed flux as a function of time for a microlensing event due
to a single lens can be fit by five parameters: $t_0, u_0, t_\e$,
$F_s$, and $F_b$.  It is important to note that several of these
parameters tend to be highly degenerate.  There are only four gross
observable properties of a single-lens curve: $t_0$, the overall timescale of the event (i.e., 
$t_{FWHM}$), and the peak and baseline
fluxes.  Thus $u_0, t_\e, F_s$, and $F_b$ tend to be highly
correlated, and are only differentiated by relatively subtle
differences between light curves with the same values of the gross
observables but different values of $u_0, t_\e, F_s$, and $F_b$ ({\em
Wozniak \& Paczynski}, 1997; {\em Han}, 1999; {\em Dominik}, 2009).
As a result of these degeneracies, when fitting to data it is often
useful to employ an alternate parametrization of the single-lens model
that is more directly tied to these gross observables, in order to avoid
strong covariances between the model parameters.  

In practice, several different observatories using several different
filter bandpasses typically contribute data to any given observed
microlensing event. Since the flux of the source and blend will vary
depending the specific bandpass, and furthermore different
observatories may have different resolutions and thus different
amounts of blended light, one must allow for a different source and
blend flux for each filter/observatory combination.  Thus the total
number of parameters for a generic model fit to an observed dataset is
$N_{nl}+2\times N_O$, where $N_{nl}$ is the number of (non-linear) parameters
required to specify the magnification as a function of time, and $N_O$
is the total number of independent datasets.  Since the observed flux
is a linear function of $F_s$ and $F_b$, for a given set of $N_{nl}$
parameters which specify $A(t)$, the set of source and
blend fluxes can be found trivially using a linear least-squares fit.
Thus in searching for the best-fit model, one typically uses a hybrid
method in which the best-fit non-linear parameters (e.g., $t_0, u_0,
t_\e$ for a single lens) are varied using, e.g., a downhill-simplex,
Markov Chain Monte Carlo (MCMC), or grid-search method, and the specific best-fit
$F_s$ and $F_b$ values for each trial set of non-linear parameters are
determined via linear least squares.

The addition of a second lens component increases the number of model
parameters by (at least) three, and the complexity enormously.  As
discussed in Section \ref{sec:phenom}, the magnification pattern (magnification as a
function of source position) for a binary lens is described two
parameters, $d$, the separation of lens components in units of the
Einstein ring of the lens, and $q$, the mass ratio of the lens.  Four
additional parameters specify the trajectory of the source through the
magnification patters as a function of time: $t_0, u_0, t_\e$, and
$\alpha$.  The first three are analogous to the single lens case:
$u_0$ is the impact parameter of the trajectory from the origin of the
lens in units of $\theta_\e$, $t_0$ is the time when $u=u_0$, and
$t_\e$ is the Einstein ring crossing time.  Finally, $\alpha$
specifies the angle of the trajectory relative to the binary-lens
axis.  Thus $6+2\times N_O$ parameters are required to specific the
light curve arising from a generic, static binary lens.  There are
many possibles choices for the origin of the lens, as well as the mass
used for the normalization of $\theta_\e$.  The optimal choice depends
on the particular properties of the lens being considered, i.e., wide
stellar binary, close stellar binary, or planetary system ({\em
Dominik}, 1999b).  For planetary systems, one typically chooses the
location of the primary for the origin and normalizes $\theta_\e$ the
total mass of the system.

Light curves arising from binary lenses exhibit an astonishingly
diverse and complex phenomenology.  While this diversity makes for a
rich field of study, it complicates the interpretation of observed
light curves mightily, for several reasons.  First, other than a few
important exceptions (i.e., for planetary caustic perturbations, see Section \ref{sec:basic}), the salient features of binary and planetary
light curves have no direct relationship the canonical parameters of
the underlying model.  Thus it is often difficult to choose initial
guesses for the fit parameters, and even if a trial solution is found,
it is difficult to be sure that all possible minima have been located.
Second, small changes in the values of the canonical parameters can
lead to dramatic changes the resulting light curve.  In particular,
the sharp changes in the magnification that occur when the source
passes close to or crosses a caustic can make any goodness-of-fit
statistic such as $\chi^2$ very sensitive to small changes in the
underlying parameters.  This, combined with the shear size of
parameter space, makes brute-force searches difficult
and time-consuming.

The complicated and highly corrugated shape of the $\chi^2$ surface
also causes many of the usual minimization routines (i.e.\ downhill
simplex) to fail to find the global or even local minimum.  These
difficulties are compounded by the fact that the magnification of a
binary lens is non-analytic and time-consuming to calculate when
finite source effects are important.

Aside from their diverse phenomenology, binary lens light curves also
have the important property that they can be highly degenerate, in the
sense that two very different underlying lens models can produce very
similar light curves.  These degeneracies can be accidental, in the
sense that that with relatively poor quality data or incomplete
coverage of the diagnostic light curve features, otherwise
distinguishable light curves can provide equally good statistical fits
to a given dataset ({\em Dominik \& Hirshfeld}, 1996; {\em Dominik},
1999; {\em Albrow et al.}, 1999b).  This is particularly problematic
for caustic-crossing binary-lens light curves in which only one (i.e.,
the second) caustic crossing is observed.  In this instance, it is
typically the case that very different models can fit a given
dataset\footnote{An important corollary is that, given an
observed first caustic crossing, it is very difficult to predict the
time of the second caustic crossing well in advance.  See {\em Albrow
et al.}, 1999b and {\em Jaroszy{\'n}ski, \& Mao}, 2001 for further
discussion.}.   The more
insidious degeneracies, however, are those that arise from
mathematical symmetries in the lens equation itself ({\em Dominik},
1999b).  For example, in the limit of a very widely separated ($d_w\gg
1$) or very close ($d_c\ll 1$) binary lenses, Taylor expansion reveals
that the lens equations in the two cases
are identical (up to an overall coordinate translation) to order $d_c^2$ or $d_w^{-2}$ for $d_w \leftrightarrow
d_c^{-1}(1+q)^{1/2}$ ({\em Dominik}, 1999b; {\em Albrow et al.}, 2002;
{\em An}, 2005). Thus the magnifications are also identical to this order in these two cases. 
Note that for $q_c,q_w \ll 1$, this degeneracy is
simply $d_c \leftrightarrow d_w^{-1}$.  This degeneracy was discussed in the context
of planetary lenses in Section \ref{sec:phenom}.  It is not known
if there exist analogous degeneracies for more complex (i.e., triple)
lenses.  These mathematical degeneracies are insidious because, if the
model is deep within the limits where the degeneracies manifest
themselves, it is essentially impossible, even in principle, to
distinguish between the degenerate models with the photometric data
alone (but see {\em Gould \& Han}, 2000 for a way to resolve the
close/wide degeneracy with astrometry).  Both classes of degeneracies
complicate the fitting of observed light curves.  First, the existence
of these degeneracies implies that simply finding a fit does not imply
that the fit is unique.  Second, these degeneracies may complicate the
physical interpretation of observed light curves, because the
degenerate models generally imply very different lens properties.

A number of authors have developed routines to locate fits to observed
binary lens light curves.  {\em Mao \& DiStefano}, 1999 compiled a
large library of point-source binary lens light curves, and classified
these according to their salient features, such as, e.g., the number
of peaks and time between peaks.  They then match the features in the
observed light curves to those in the library to find trial solutions
for minimization routines.  This approach effectively works by
establishing a mapping between the light curve features and the
canonical parameters.  {\em DiStefano \& Perna}, 1997 adopted a
conceptually similar approach, where they decomposed the observed
light curve into a linear combination of basis functions, then
compared the resulting coefficients of this fit to those found for 
a library of events, again to identify promising trial solutions.
Similarly, {\em Vermaak}, 2003 used artificial neural networks to
identify promising regions of parameter space.  While these methods
can in principle be applied to any binary-lens light curve, because they
are not intrinsically systematic, it is difficult to be sure that all possible fits
have been identified.
 
{\em Albrow et al.}, 1999b and {\em Cassan}, 2008 developed algorithms to
find a complete set of solutions for binary-lens light curves where the
source crosses a caustic.  These algorithms are robust in the sense
that they will identify all possible fits to such light curves, and
thus uncover all possible degenerate solutions (e.g., {\em Afonso et
al.}, 2000).  Unfortunately, they are also fairly user-intensive, and are
obviously only applicable to a limited subset of events.

Currently, the most robust and efficient approaches to fitting
observed binary-lens light curves use some variant of a hybrid approach,
for example a grid
search over those non-linear parameters that are not simply related to
the light curve features are thus are poorly-behaved (such as the mass
ratio $q$, projected separation $d$, and/or the angle of the trajectory
$\alpha$), combined with a downhill-simplex, steepest-descent, or MCMC fit to
the remaining non-linear parameters that are more directly related to
the observed light curve features.  The linear parameters (such as the
source and blend flux) are trivially fit for each trial solution.
Often when there is some prior gross knowledge of the approximate
model parameters (i.e, a wide or close planetary or binary lens) a
judicious choice of parametrization guided by the inherent
magnification properties of the underlying lens geometry (i.e., the
caustic structure of the lens) can both speed up the fit and improve
the robustness of the fitting.  Once trial solutions are identified,
they can be more carefully explored using, e.g., Markov Chain Monte Carlo techniques.
See, e.g., {\em Gould et al.}, 2006, {\em Dong et al.}, 2007, and {\em Bennett}, 2009. 
Essentially all of these approaches must be
modified (although sometimes trivially) to include higher-order
effects when these provide significant perturbations to the observed
light curve (see Section \ref{sec:higher}).  

In general, no robust, practical,
universal, and efficient algorithm exists for fitting an arbitrary
binary lens light curve in an automated way that is not highly
user-intensive. For higher-multiplicity (i.e., triple) lenses, the
algorithms are generally even less well-developed. As mentioned
in the introduction, this is likely the current single biggest impediment to the 
progress of microlensing planet searches. Thus there is a
urgent and growing need for the development of (more automated) analysis software.

\subsection{Higher-Order Light Curve Effects}\label{sec:higher}

As reviewed in Section \ref{sec:lightcurves}, the simplest model of the observed light curve arising from
an isolated lens can be
described by $3+2\times N_O$ parameters: the three parameters that
describe the magnification as a function of time ($u_0,t_0, t_{\rm
E}$), and a blend and source flux for each of the $N_0$
observatory/filter combinations.  This model, which accurately describes
the vast majority of observed microlensing light curves, 
is derived under a number of assumptions, some of which may 
break down under certain circumstances.  When a light curve deviates from this
classic `Paczynski' or point-source single-lens form, it is generally classified as `anomalous'. 
Not surprisingly, it is frequently the case that a robust measurement of these `anomalies' allows one to infer additional
properties of the primary, planet, or both (see Section \ref{sec:properties}).

The most obvious and most common anomaly is when the lens is not isolated. 
Each additional lens component requires three additional parameters\footnote{If the additional component is very far away or very close to the primary lens,
the effects of the additional component can be described by only two additional parameters:
either the shear due the wide companion or the quadrupole moment of the lens for the close companion, and the angle of the trajectory
with respect to the projected binary axis.}  (e.g., mass ratio, projected
separation, and angle between the additional component and the
primary), and so the light curve arising from a lens consisting of
$N_l$ masses requires a total of $3\times N_l+2\times N_O$ parameters.
Roughly $\sim 10\%$ of all microlensing lightcurves are observed to be due to binaries. In fact, the true fraction
of binary lenses is likely considerably higher, since the effects of a binary companion are only
apparent when $d$ is of order unity. 

\noindent$\bullet$ {\bf Finite Source Effects}

The second most common anomaly occurs when the assumption of a point
source breaks down ({\em Gould}, 1994; {\em Witt \& Mao}, 1994; {\em
Witt}, 1995).  As already discussed in Section \ref{sec:caustics}, this assumption
breaks down when the curvature of the magnification pattern is
significant over the angular size of the source, where `significant'
depends on the photometric precision with which the light curve is
measured.  Generally, the center of the source must pass within a few
angular source radii $\theta_*$ of a caustic in order for finite
source effects to be important.  The magnitude of finite source
effects are parametrized by the angular size of the source in units
of the angular Einstein ring radius, $\rho_*\equiv
\theta_*/\theta_{\rm E}$.  For typical main-sequence sources in the
Galactic bulge, $\rho_*$ is of order $10^{-3}$, whereas for typical giant
sources, $\rho_*$ is of order $10^{-2}$.  In addition, the
amount of (filter-dependent) limb-darkening can also have an important
effect on the precise shape of the portions of the light curve affected
by finite-source effects.

\noindent$\bullet$ {\bf Parallax}

Another common anomaly occurs when the assumptions of co-spatial
and/or non-accelerating observers break down.  For example, the usual
expression for $u$, the angular separation between the lens and source
in units of $\theta_E$, presented in Equation \ref{eqn:uoft} assumes that the
observer (as well as the lens and source) are moving with constant
velocity.  The assumption that the values of $t_0$ and $u_0$ are the
same for all observers implies that they are all co-spatial.  When
either of these assumptions break down, this is generally termed
(microlens) parallax.  This occurs in one of three ways.
First, when the duration of the microlensing event is a significant
fraction of a year, the acceleration of the Earth leads to a
significant non-uniform and/or non-rectilinear trajectory of the lens
relative to the source, which leads to deviations in the observed
light curve ({\em Gould}, 1992).  This is referred to as `orbital
parallax'.  Second, for sources very close to a caustic, small
parallax displacements due to the differences in the perspective leads
to differences in the magnifications seen by observers located at
different observatories for fixed time ({\em Holz \& Wald}, 1996).
This is commonly known as `terrestrial parallax'.  
Finally, if the observers are separated by a significant fraction of an
AU, for example if the event is simultaneously observed from the Earth and a satellite in a solar orbit,
then the differences in the magnification are large even for moderate magnification ({\em Refsdal}, 1966;
{\em Gould}, 1994).  This is known as `satellite parallax'. 
In all three cases, the magnitude of the effect depends on the size of
relevant length scale that gives rise to the different or changing
perspective (i.e., the projected separation between the observatories,
the projected size of the Earth's orbit, or the projected separation between
the satellite and Earth) relative to ${\tilde
r}_{\rm E}\equiv D_{\rm rel}\theta_{\rm E}$, the angular Einstein ring radius projected on
the plane of the observer (See Figure \ref{fig:dia}).

\noindent$\bullet$ {\bf Xallarap}

Analogously to parallax, if the source undergoes significant
acceleration over the course of the event due to a binary companion,
this will give rise to deviations from the canonical light curve form
({\em Dominik}, 1998).  This effect is commonly known as `xallarap',
i.e., parallax spelled backwards, to highlight the symmetry between
this effect and orbital parallax.  In fact, it is always possible to
{\it exactly} mimic the effects of orbital parallax with a binary
source with the appropriate orbital parameters ({\em Smith et al.}
2003). For sufficiently precise data, such that the parallax/xallarap
parameters are well constrained, these two scenarios can be
distinguished because it would be {\it a priori} unlikely for a binary
source to have exactly the correct parameters to mimic the effects of
orbital parallax ({\em Poindexter et al.}, 2005).  The magnitude of
xallarap effect depends on the semimajor axis of the binary
relative to ${\hat r}_{\rm E} \equiv D_s\theta_{\rm E}$, the Einstein ring radius projected on
the source plane.

\noindent$\bullet$ {\bf Orbital Motion}

For binary or higher multiplicity lenses, relative orbital motion of
the components can also give rise to deviations from the light curve
expected under the usual static lens assumption ({\em Dominik}, 1998).
For the general case of a binary lens and a Keplerian orbit, an
additional five parameters beyond the usual
static-lens parameters are needed to specify the light curve ({\em
Dominik}, 1998). Typically, however, only two additional parameters
can be measured: the two components of the relative projected velocity
of the lenses.  These two components can be parametrized by the rate
of the change of the binary projected binary axis ${\dot d}$ and the
angular rotation rate of the binary $\omega$.  The effect of the
latter is simply to rotate that magnification pattern of the lens on
the sky, whereas the former results in a change in the lens mapping
itself, and thus a change in the caustic structure and magnification
pattern.

\noindent$\bullet$ {\bf Binary Sources}

If the source is a binary and the lens happens to pass sufficiently
close to both sources, the light curve can exhibit a deviation from the
generic single-lens, single source form ({\em Griest \& Hu}, 1992).
For static binaries, the morphology of the deviation depends on the
flux ratio of the sources, their projected angular separation in units
of $\theta_{\rm E}$, and the impact parameter of the lens from the
source companion.  By convention, such `binary source' effects are
conceptually distinguished from xallarap effects by the nature of the
deviation: if the deviation is caused by the source companion being
significantly magnified by the lens, is called a binary source effect,
if it caused by the acceleration due to the companion, it is called
xallarap.  Of course, depending on the parameters of the source and
lens, light curves can exhibit only binary source effects, only
xallarap, or both xallarap and binary-source effects.

\noindent$\bullet$ {\bf Other Miscellaneous Effects}

A number of additional high-order effects have been discussed in the
literature, for example the effects of the finite physical size of the
lens ({\em Bromley}, 1996; {\em Agol}, 2002).  In most cases,
these higher-order effects are expected to unobservable and/or extremely rare.

\subsection{Properties of the Detected Systems}\label{sec:properties}

For microlensing events due to single, isolated lenses, the parameters
that can routinely be measured are the time of maximum magnification
$t_0$, the impact parameter of the event in units of the Einstein ring
radius $u_0$, and the Einstein timescale $t_\e$, along with a source
flux $F_s$ and blend flux $F_b$ for each observatory/filter
combination.  Of these parameters, $u_0$ and $t_0$ are simply
geometrical parameters and contain no physical information about the
lens.  The Einstein timescale $t_\e$ is a degenerate combination of
the lens mass $M$, the relative lens-source parallax $\pi_{\rm rel}$,
and the relative lens-source proper motion $\mu_{\rm rel}$.  Therefore
it is not possible to uniquely determine the mass and distance to the
lens from a measurement of $t_\e$ alone.  The blend flux $F_b$
contains light from any source that is blended with the source,
including light from the lens if it is luminous.  Unfortunately, for
the typical targets toward the Galactic bulge, ground-based images
typically contain light from other, unrelated sources, and it is not
possible to isolate the light from the lens (see Figure \ref{fig:blending}).  Therefore, in the vast
majority of microlensing events, the mass, distance, and proper motion
of the lens are unknown.

As discussed below, for binary and planetary microlensing events it is
routinely possible to infer the mass ratio $q$, and $d$, the instantaneous
projected separation between the planet and star in units of
$\theta_\e$.  However, the mass of the planet is typically not
known without a constraint on the primary mass. Furthermore, a
measurement of $d$ alone provides very little information about the
orbit, since $\theta_\e$ and the inclination, phase, and ellipticity
of the orbit are all unknown {\it a priori}.
For these reasons, when microlensing planet searches were first
initiated it was typically believed that detailed information about
individual systems would be very limited for planets detected in
microlensing light curves.  This apparent deficiency was exacerbated by the
perception that the host stars would typically be
too distant and faint for follow-up observations. 

Fortunately, in reality much more
information can typically be gleaned from a combination of a detailed
analysis of the light curve and follow-up, high-resolution
imaging, using the methods outlined below.

\noindent $\bullet${\bf Mass Ratio and Projected Separation}

First, the requirements for accurately measuring the minimum three
additional parameters needed to describe the light curves of binary
and planetary microlensing events are discussed.  These
three additional parameters are the aforementioned $q$, $d$, and $\alpha$.

For planetary caustic perturbations, these parameters can essentially
be `read off' of the observed light curve.  In this case, the gross
properties of the planetary perturbation can be characterized by three
observable quantities: the time of the planetary perturbation
$t_{0,p}$, the timescale of the planetary perturbation $t_{\e,p}$, and the
magnitude of the perturbation $\delta_p$.  These quantities then
simply and completely specify the underlying parameters $q,d,\alpha$,
up to a two-fold discrete degeneracy in $d$, corresponding to whether
the planet is perturbing the major or minor image, i.e, $d
\leftrightarrow d^{-1}$.  Since these two situations result in very
different types of perturbations (see Figure \ref{fig:plcurves}), this discrete
degeneracy is easily resolved ({\em Gaudi \& Gould}, 1997b).  If finite
source effects are important but not dominant, then there also exists
a continuous degeneracy between $q$ and $\rho_*$, stemming from the
fact that in this regime, both determine the width of the perturbation
$t_p$ ({\em Gaudi \& Gould}, 1997b).  However, this degeneracy is
easily broken by good coverage and reasonably accurate photometry in
the wings of the perturbation.  Finally, there is also a degeneracy
between major image ($d>1$) planetary caustic perturbations and a
certain class of binary-source events, namely those with extreme
flux ratio between the two sources.  Specifically, it is always
possible to find a binary-source light curve that can exactly
reproduce the observables $t_{0,p}, t_{\e,p}$, and $\delta_p$ ({\em Gaudi},
1998).  This degeneracy is also easily broken by good coverage and
accurate photometry ({\em Gaudi}, 1998; {\em Gaudi \& Han}, 2004; {\em Beaulieu et al.}, 2006).

For central or resonant caustic perturbations in high-magnification events,
extracting the parameters $q,d,\alpha$ is typically more
complicated due to the fact that there is no simple, general relationship
between the salient features of the light curve perturbation and these
these parameters.  Thus fitting these perturbations typically requires
a more sophisticated approach, as discussed in Section \ref{sec:lightcurves}.  In addition,
there are a number of degeneracies that plague central caustic
perturbations.  First, as discussed in \ref{sec:phenom} and illustrated in
Figures \ref{fig:ccaustics} and \ref{fig:clcurves}, there is a close/wide duality such that the central
caustic shape and associated magnification pattern are highly
degenerate under the transformation $d \leftrightarrow d^{-1}$ ({\em
Griest \& Safizadeh} 1999; {\em Dominik}, 1999b).  This degeneracy
becomes more severe for very close/very wide planets, and in some
cases it is essentially impossible to distinguish between the two
solutions, even with extremely accurate photometry and dense coverage
of the perturbation (e.g., {\em Dong et al.}, 2009a).  There also
exists a degeneracy between central caustic planetary perturbations,
and perturbations due to very close or very wide binary lenses. Very
close binaries have small, asteroid-shaped caustic located at the
center-of-mass of the system, whereas very wide binaries have small,
asteroid-shaped caustics near the positions of each of the lenses.
The gross features of central caustic perturbations can be reproduced
by a source passing by the asteroid-shape caustic produced by a
close/wide binary lens ({\em Dominik}, 1999, {\em Albrow et al.},
2000; {\em An}, 2005).  Since close/wide binary lenses are themselves
degenerate under the transformation $d_w \leftrightarrow
d_c^{-1}(1+q)^{1/2}$, there is a four-fold degeneracy for
perturbations near the peak of high-magnification events.  Fortunately,
the degeneracy between planetary central caustic perturbations and
close/wide binary lens perturbations can be resolved with good coverage
of the perturbation and accurate photometry ({\em Albrow et al.},
2000; {\em Han \& Gaudi}, 2008; {\em Han}, 2009).

\noindent $\bullet${\bf Einstein Ring Radius} 

The requirement
for detecting a planet via microlensing is generally that the source must
pass reasonably close to the caustics produced by the planetary companion.
However, this is also basically the condition for finite source size effects to be important.  
Thus for most planetary microlensing events, it is possible
to infer the angular size of the source in units of the angular Einstein
ring radius, $\rho_* \equiv \theta_*/\theta_\e$.  

The angular size of the source can be estimated by its de-reddened
color and magnitude using empirical color-surface brightness relations
determined from angular size measurements of nearby stars ({\em van
Belle}, 1999; {\em Kervella et al.}, 1994).  The source flux $F_s$ is
most easily determined by a fit to the microlensing light curve of the
form $F(t)=F_s A(t)+F_b$, as the variable magnification of the source
allows one to `deblend' the source and blend flux.  Determining the
color of the source in this manner requires measurements in two
passbands, and thus while observations are typically focused on a
single passband (typically a far-red visible passband such as $R$ or
$I$), it is important to acquire a few points in a second filter to
determine the source color.  The extinction toward the source can be
approximately determined by comparison to nearby red giant clump stars
({\em Yoo et al.}, 2004), which have a known and essentially constant
luminosity and intrinsic color (e.g., {\em Paczynski \& Stanek},
1998).  An error in the extinction affects both the inferred color and
magnitude of the source, fortunately these have opposite and nearly
equal effects on the inferred value of $\theta_*$ ({\em Albrow et
al.}, 1999a).

Thus for events in which finite source effects are robustly
detected, it is possible to measure $\theta_{\rm E}$ ({\em Gould}, 1994).  
This partially breaks the timescale degeneracy, since
\begin{equation}
\frac{M}{D_{\rm rel}}=\frac{c^2}{4G} \theta_\e^2.
\label{eqn:massdrel}
\end{equation}
The distance to the source is typically known approximately from
its color and magnitude (and furthermore the overwhelming majority of sources
are in the bulge), and so 
a measurement of $\theta_\e$ essentially provides a mass-distance relation
for the lens. 

\noindent $\bullet${\bf Light from the Lens}

Although the majority of the lenses that give rise to microlensing
events are distant and low-mass main-sequence stars, most are
nevertheless usually sufficiently bright that their flux can be
measured to relative precision of $\la 10\%$ with moderate-aperture (1-2m) telescopes and
reasonable ($10^2-10^4~{\rm s}$) exposure times, provided that the
light from the lens can be isolated.  The fit to the
microlensing light curve gives the flux of the source $F_s$, and the
blend flux $F_b$.  The latter contains the flux from any stars that
are not being lensed but are unresolved on the image, i.e., blended
with the source star.  Generally, this blend flux can be decomposed
into several contributions,
\begin{equation}
F_b = F_l + F_{l,c} + F_{s,c} + \sum_j F_{u,j},
\label{eqn:fblend}
\end{equation}
where $F_l$ is the flux from the lens, $F_{l,c}$ is the flux from any
(blended) companions to the lens, $F_{s,c}$ is the flux from any
(blended) companions to the source, and $F_{u,j}$ is the flux from
each unrelated nearby star $j$ that is blended with the source.  As
illustrated in Figure \ref{fig:blending}, at typical ground-based resolutions of $1''$
and in the extremely crowded target fields toward the bulge where
microlensing surveys are carried out, it is often the case that there
are several unrelated stars blended with the source star.  Therefore,
the lens light cannot be uniquely identified based on such data
alone.  At the higher resolutions of $0.05-0.1''$ available from
{\it HST} or ground-based AO
imaging, essentially all stars unrelated to the source our lens are
resolved.  Since the source and lens must be aligned to $\la \theta_\e
\sim 1~{\rm mas}$ for a microlensing event to occur and the typical
relative lens-source proper motions are $\mu_{\rm rel} \sim 5-10~{\rm
mas~yr^{-1}}$ for microlensing events towards the bulge, the lens and
source will be blended in images taken within $\sim 10$ years of the
event, even at the resolution of {\it HST}.  However, because the
microlensing fit gives $F_s$, the lens flux can be determined by
subtracting this flux from the combined unresolved lens+source flux in
the high-resolution image, assuming no blended companions to the lens
or source ({\em Bennett et al.}, 2007).

There are several potential complications to this procedure to
determine the lens flux.  First, $F_s$ determined from the
microlensing fit will generally not be absolutely calibrated.  Thus
the high-resolution photometric data must be `photometrically aligned'
to the microlens dataset. Typically this is done by matching stars
common to both sets of images.  The accuracy of this alignment is
usually limited to $\sim 1\%$ due to the small number of common,
isolated stars available ({\em Dong et al.}, 2009b). It may also be the high-resolution images
are taken in a different filter than that for which the source flux $F_s$
is determined, necessitating a (model-dependent) color transformation
and introducing additional uncertainties (e.g., {\em Bennett et al.}, 2008).  
Note that it is possible to
avoid this procedure entirely if the high-resolution images can be
taken at two different epochs with substantially different source star
magnifications.  However, since this means at least one epoch must be
taken when the source is significantly magnified during the event,
this requires target-of-opportunity observations.  Finally, any light
in excess of the source detected in the high-resolution images may be
attributed to close physical companions to the lens or source.  In
some favorable cases (i.e., high magnification events), it is possible
to exclude these scenarios by the (lack of) second order effects the
companion would produce in the observed light curve.  For example, a
binary companion to the lens produces a caustic which would
be detectable in sufficiently high-magnification events, whereas a
sufficiently close companion to the source would give rise to xallarap
effects which would be detectable in long timescale events ({\em Dong et al.}, 2009b).

A measurement of the flux of the lens in a single passband, along with
a model for extinction as a function of distance and a mass-luminosity
relationship, gives a mass-distance relationship for the lens ({\em Bennett et al.}, 2007).  A
second measurement of the flux in a different passband can provide a
unique mass and distance to the lens, subject to the uncertainties in
the intrinsic color as a function of mass and the dust extinction
properties as a function of wavelength and distance. 

\noindent $\bullet${\bf Proper Motion} 

For typical values of $\mu_{\rm rel} \sim 5-10~{\rm mas/yr}$ for microlensing events toward the Galactic bulge,
after a few years, the lens and source will be displaced by $\sim 0.01$ arcseconds.  
For luminous lenses, and using space telescope or AO imaging, it is possible
measure the relative lens-source proper motion, either by measuring
the elongation of the PSF or by measuring the difference in the
centroid in several filters if the lens and source have significantly
different colors ({\em Bennett et al.}, 2007).  The proper motion can be combined 
with the timescale to give the angular Einstein ring radius, $\theta_{\rm E}=\mu_{\rm rel} t_{\rm E}$.

\noindent $\bullet${\bf Microlens Parallax}

For some classes of events, 
it is possible to obtain additional information about the lens
by measuring the microlensing parallax, $\mbox{\boldmath$\pi$}_\e$, a vector with 
magnitude $|\mbox{\boldmath$\pi$}_\e| = \rm AU/\tilde r_\e$, and direction
of the relative lens-source proper motion.  Recall
$\tilde r_\e \equiv D_{\rm rel}\theta_\e$ is the Einstein ring radius
projected onto the observer plane.  As discussed
in Section \ref{sec:higher} microlens parallax effects arise in one of three varieties:
orbital parallax due to the acceleration of the Earth during the
event, terrestrial parallax in high-magnification events 
observed by non-cospatial observers, and satellite parallax for events
observed from the ground and a satellite in solar orbit. 

Orbital parallax deviations are generally only significant for events
with timescales that are a significant fraction of a year, and so long
as compared to the median timescale of $\sim 20~{\rm days}$.
Furthermore, the deviations due to orbital parallax are subject to an
array of degeneracies ({\em Gould et al.}, 1994; {\em Smith et al.},
2003; {\em Gould}, 2004), which can hamper the ability to extract
unique microlens parallax parameters.  The severity of these
degeneracies depend on the particular parameters of the event in
question, but for most events it is the case that the only
robustly-measured effect in the light curve is an overall asymmetry,
which only yields one projection of $\mbox{\boldmath$\pi$}_\e$, namely
that in the direction perpendicular to the instantaneous Earth-Sun
acceleration vector at the time of the event ({\em Gould et al.}
1994). If the direction of the relative lens-source proper motion vector
$\mbox{\boldmath$\mu$}_{\rm rel}$
can be independently determined from the proper motion of a luminous lens, then it is possible
to determine the full $\mbox{\boldmath$\pi$}_\e$ vector,
since $\mbox{\boldmath$\pi$}_\e$ is parallel to $\mbox{\boldmath$\mu$}_{\rm rel}$.

Orbital parallax measurements made from two observatories are also
subject to several degeneracies, which have been studied by several
authors ({\em Refsdal}, 1966; {\em Gould}, 1994,1995; {\em Boutreux \&
Gould}, 1996; {\em Gaudi \& Gould}, 1997a). These can be resolved
in a number of ways ({\em Gould}, 1995; {\em Gould}, 1999; {\em Dong et al.}, 2007),
including observing from a third observatory, which allows one
to uniquely `triangulate' the parallax effects ({\em Gould}, 1994).  
Similarly, terrestrial parallax measurements from only two observatories
are subject to degeneracies which can be resolved with simultaneous observations
from a third observatory that is not co-linear with the other two.   

A measurement of the microlens parallax allows one to partially break
the timescale degeneracy and provides a mass-distance relation for the lens,
\begin{equation}
M D_{\rm rel}=\frac{c^2}{4G} \tilde
r_{\rm E}^2.
\label{eqn:drelmass}
\end{equation}

\noindent $\bullet${\bf Orbital Motion of the Planet} 

In at least two
cases, the orbital motion of the planet during the microlensing event
has been detected ({\em Dong et al.}, 2009a; {\em Bennett et al.}, 2009).  The effects of orbital motion generally allow the
measurement of the two components of the projected velocity of the
planet relative to the primary star. If an external measurement of the
mass of the lens is available, and under the assumption of a circular
orbit, these two components of the projected velocity completely
specify the full orbit of the planet (including inclination), up to a
two-fold degeneracy ({\em Dong et al.}, 2009a). In some cases, higher-order effects of
orbital motion can be used to break this degeneracy and even constrain
the ellipticity of the orbit ({\em Bennett et al.}, 2009).

\noindent $\bullet${\bf Bayesian Analysis}

In the cases when only $t_{\rm E}$, $q$, and $d$ can be measured, constraints on the 
mass and distance to the lens (and so mass and semimajor axis of the planet),
must rely on a Bayesian analysis which incorporates priors on the distribution
of microlens masses, distances and velocities (e.g., {\em Dominik}, 2006; {\em Dong et al.}, 2006).

\noindent $\bullet${\bf Complete Solutions}

In many cases, several of these pieces of information can be measured
in the same event, often providing complete or even redundant measurements
of the mass, distance, and transverse velocity of the event.  For example,
a measurement of $\theta_{\rm E}$ from finite source effects, when combined
with a measurement of $r_{\rm E}$ from microlens parallax, yields
the lens mass,
\begin{equation}
M=\left(\frac{c^2}{4G}\right)\tilde r_{\rm E} \theta_{\rm E},
\label{eqn:masssol}
\end{equation}
distance
\begin{equation}
D_{l}^{-1} = \frac{\theta_{\rm E}}{\tilde r_{\rm E}} + D_s^{-1},
\label{eqn:dlsol}
\end{equation}
and transverse velocity ({\em Gould}, 1996).

\subsection{Practical Aspects of Current Microlensing Searches}\label{sec:practical}

The microlensing event rate toward the Galactic bulge is ${\cal
O}(10^{-6})$ events per star per year ({\em Paczynski}, 1991).  In a typical field toward the
Galactic bulge, the surface density of stars is $\sim 10^{7}$ stars
per deg$^2$ to $I\sim 20$ ({\em Holtzman et al.}, 1998).  Thus to
detect $\sim 100$ events per year, $\sim 10$ deg$^2$ of the Bulge must
be monitored.  Until relatively recently, large-format CCD cameras
typically had fields of view of $\sim 0.25$ deg$^2$, and thus $\sim
40$ pointings were required and so fields could only be monitored once
or twice per night.  While this cadence is sufficient to detect the
primary microlensing events, it is insufficient to detect and
characterize planetary perturbations, which last a few days or less.

As a result, microlensing planet searches have operated using a
two-stage process.  The Optical Gravitational Lens Experiment (OGLE,
{\em Udalski}, 2003) and the Microlensing Observations in Astrophysics
(MOA, {\em Sako et al.}, 2008) collaborations monitor several tens of
square degrees of the Galactic bulge, reducing their data real-time in
order to alert microlensing event in progress.  A subset of these
alerted events are then monitored by several follow-up collaborations,
including the Probing Lensing Anomalies NETwork (PLANET, {\em Albrow
et al.}, 2008), RoboNet ({\em Tsapras et al.}, 2009), Microlensing
Network for the Detection of Small Terrestrial Exoplanets (MiNDSTEp,
{\em Dominik et al.}, 2008), and Microlensing Follow Up Network
($\mu$FUN, {\em Yoo et al.}, 2004) collaborations.  Since only individual
microlensing events are monitored, these teams can achieve the sampling and
photometric accuracy necessary to detect planetary perturbations.  In
fact, the line between the `alert' and `follow-up' collaborations is
now somewhat blurry, both because the MOA and OGLE collaborations
monitor some fields with sufficient cadence to detect planetary
perturbations, and because there is a high level of communication
between the collaborations, such that the observing strategies are
often altered real time based on available information about ongoing
events.

There are two conceptually different channels by which planets can be
detected with microlensing, corresponding to whether the planet is
detected through perturbations due to the central caustic (as shown in
Figure \ref{fig:clcurves}), or perturbation due to the planetary
caustic(s) (as shown in Figure \ref{fig:plcurves}).  Because, for any
given planetary system, the planetary caustics are always larger than
the central caustics, the majority of planetary perturbations are
caused by planetary caustics.  Thus searching for planets via the
influence of the planetary caustics is termed the `main channel'.
However, detecting planets via the main channel requires substantial
commitment of resources because the unpredictable nature of the
perturbation requires dense, continuous sampling, and furthermore the
detection probability per event is relatively low so many events must
be monitored.  Detecting planets via their central caustic
perturbations requires monitoring high magnification primary events
near the peak of the event. In this case, the trade-off is that
although high-magnification events are rare (a fraction $\sim 1/A_{\rm
max}$ of events have maximum magnification $\ga A_{\rm max}$), they are individually
very sensitive to planets.  The primary challenge lies with this
channel lies with identifying high-magnification events real time.

Which approach is taken depends on the resources that the individual
collaborations have available.  The PLANET collaboration has
substantial access to 0.6-1.5m telescopes located in South Africa,
Perth, and Tasmania.  With these resources, they are able to monitor
dozens of events per season, and so are able to search for planets via
the main channel.  This tactic led to the detection of the first cool
rocky/icy exoplanet OGLE-2005-BLG-390Lb ({\em Beaulieu et al.}, 2006).
On the other hand, the $\mu$FUN collaboration use a single 1m
telescope in Chile to monitor promising alerted events in order try to
identify high-magnification events substantially before peak.  When
likely high-magnification events are identified, the other telescopes in
the collaboration are then engaged to obtain continuous coverage of
the light curve during the high-magnification peak. High-magnification
events often reach peak magnitudes of $I \la 15$, and thus can
be monitored with relatively small-apertures (0.3-0.4m).  This allows
amateur astronomers to contribute to the photometric follow-up. Indeed
over half of the members of the $\mu$FUN collaboration are amateurs.

\section{FEATURES OF THE MICROLENSING METHOD}\label{sec:features}

The unique way in which microlensing finds planets leads to some
useful features, as well as some (mostly surmountable) drawbacks.
Most of the features of the microlensing method can be understood
simply as a result of the fact that planet detection relies on the
direct perturbation of images by the gravitational field of the
planet, rather than on light from the planet, or the indirect effect
of the planet on the parent star.

\noindent$\bullet$ {\bf Peak Sensitivity Beyond the Snow Line}

The peak sensitivity of microlensing is for planet-star separations of 
$\sim r_\e$, which corresponds to equilibrium temperatures of
\begin{equation}
T_{\rm eq} = 278~{\rm K} \left(\frac{L}{L_\odot}\right)^{1/4} \left(\frac{r_\e}{{\rm AU}}\right)^{-1/2}
\sim 70~{\rm K} \left(\frac{M}{0.5M_\odot}\right),
\label{eqn:Teq}
\end{equation}
where the rightmost expression assumes $L/L_\odot = (M/M_\odot)^5$, $D_l=4~{\rm kpc}$, and $D_s=8~{\rm kpc}$.
Thus microlensing is most sensitive to planets in the regions beyond the `snow line,' the 
point in the protoplanetary disk
exterior to which the temperature is less than the condensation temperature of water
in a vacuum ({\em Lecar et al.}, 2006; {\em Kennedy et al.}, 2007; {\em Kennedy \& Kenyon}, 2008).
Giant planets are thought to form in the region immediately beyond the snow line, where the 
surface density of solids is highest ({\rm Lissauer}, 1987).

Is microlensing sensitive to habitable planets?  For assumptions above, and further assuming the habitable zone is centered
on $a_{HZ}={\rm AU}(L/L_\odot)^{1/2}$, the projected separation of a planet in
the habitable zone in units of $\theta_\e$ is $d_{HZ} \sim 0.25 (M/M_\odot)^2$.  Thus for typical
hosts of $M\la M_\odot$, the habitable zone is well inside the Einstein ring radius ({\em Di Stefano}, 1999).
Since microlensing is much less sensitive to planets
with separations much smaller than the Einstein ring radius as these can only 
perturb highly demagnified images, it is much less
sensitive to planets in the habitable zones of their parent stars ({\em Park et al.}, 2006)
for typical events.
However, it is important to note that this is primarily a statistical statement about
the typical sizes of the angular Einstein ring radii of microlensing events, rather
than a statement about any intrinsic limitations of the microlensing method. 
A fraction of events have substantially smaller Einstein ring radii,
and for these events there is significant sensitivity to habitable planets. 
In particular, lenses closer to the observer have smaller Einstein ring radii and so microlensing
events from nearby stars will have more sensitivity to habitable planets ({\em DiStefano \&
Night}, 2008; {\em Gaudi et al.} 2008b).   Indeed,
a space-based microlensing planet search mission (described in Section \ref{sec:future}) will have significant
sensitivity to habitable planets, primarily due to the large number of events being
searched for planets ({\em Bennett et al.}, 2008). 

\noindent$\bullet$ {\bf Sensitivity to Low-mass Planets}

The amplitudes of the perturbations caused by planets are typically large,
$\ga 10\%$,  Furthermore,
although the durations of the perturbations get shorter with planet
mass (as $\sqrt{m_p}$)
and the probability of detection decreases (also roughly as $\sqrt{m_p}$), 
the amplitude of the perturbations are independent of the planet mass.
This holds until the `zone of influence' of the planet, which has a
size $\sim \theta_{\e,p}$, is smaller than the angular size of the source
$\theta_*$.  When this happens, the perturbation is `smoothed'
over the source size, as demonstrated in Figures \ref{fig:clcurves} and \ref{fig:plcurves}.  For typical parameters, $\theta_{\e,p}\sim \mu{\rm
as}(m_p/M_\oplus)^{1/2}$, and for a star in the bulge,
$\theta_*\sim \mu{\rm as} (R_*/R_\odot)$.
This `finite source' suppression essentially precludes the detection
of planets with mass $\la 5~M_\oplus$ for clump giant sources in the bulge with $R_* \sim 10R_\odot$ ({\em Bennett \& Rhie}, 1996).
For main-sequence sources ($R\sim R_\odot$), finite source effects become
important for planets with the mass of the Earth, but does
not completely suppress the perturbations and render then undetectable until masses 
of $\sim 0.02~M_\oplus \sim 2~M_{\rm Moon}$ for main-sequence sources ({\em Bennett \& Rhie}, 1996; {\em Han et al.}, 2005).  Thus
microlensing is sensitive to Mars mass planets and even planets a few times the mass
of the Moon, for sufficiently small source sizes.

\noindent$\bullet$ {\bf Sensitivity to Long-Period and Free-Floating Planets}

Since microlensing can `instantaneously' detect planets without waiting
for a full orbital period, it is immediately sensitive to planets with very long
periods.  Although the probability of detecting a planet decreases
for planets with separations larger than the Einstein ring radius
because the magnifications of the images decline, it does not drop to
zero.  For events in which the primary star is also detected, the detection
probability for very wide planets with $d\gg 1$ is $\sim (\theta_{\e,p}/\theta_\e) d^{-1} = q^{1/2} d^{-1}$
({\em  Di Stefano \& Scalzo}, 1999b). 
Indeed since microlensing is directly sensitive to the planet
mass, planets can be detected even without a primary microlensing event ({\em  Di Stefano \& Scalzo}, 1999a).  
Even free-floating planets that
are not bound to any host star are detectable in this way ({\em Han et al.}, 2005). 
Microlensing is the only method that can detect old, free-floating
planets.  A significant population of free-floating planets planets is a
generic prediction of most planet formation models, particular those
that invoke strong dynamical interactions to explain the observed
eccentricity distribution of planets ({\em Goldreich et al.}, 2004; {\em Juric \& Tremaine}, 2008;
{\em Ford \& Rasio}, 2004).

\begin{figure}[htp]
\epsscale{1.0}
\plotone{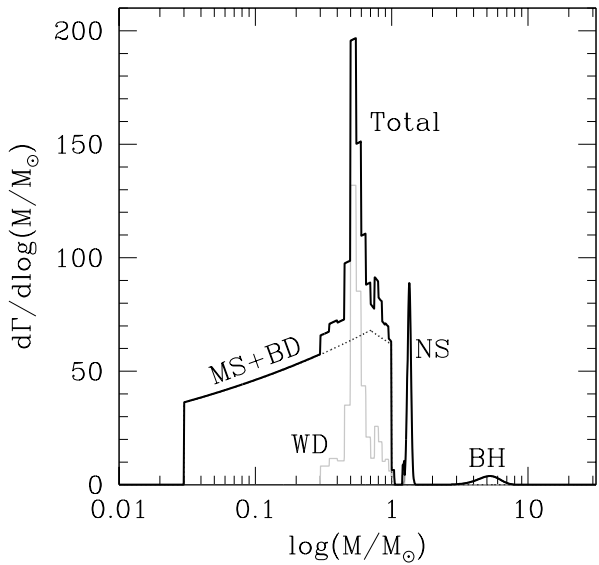}
\caption{\small 
The rate of microlensing events toward the Galactic bulge as a function
of the mass of the lens, for main sequence (MS) stars and brown dwarfs (BD, $0.03M_\odot < M < 1 M_\odot$) (bold
dashed curve) and white dwarfs (WD), neutron stars (NS), and black hole (BH) remnants (solid curves). The total is shown by
a bold solid curve. From {\em Gould}, 2000.
}\label{fig:massdist}
\end{figure}

\noindent $\bullet$ {\bf Sensitivity to Planets Orbiting a Wide Range
of Host Stars} 

The hosts probed by microlensing are
simply representative of the population of massive objects along the
line of sight to the bulge sources, weighted by the lensing
probability.  Figure \ref{fig:massdist} shows a model by {\em Gould}, 2000 for the microlensing event
rate toward the Galactic bulge as a function of the host star
mass for all lenses, and also broken down by the type of
host, i.e., star, brown dwarf, or remnant.  The sensitivity of microlensing is weakly
dependent on the host star mass, and has essentially no dependence on
the host star luminosity.  Thus microlensing is about equally
sensitive to planets orbiting stars all along the main sequence, from
brown dwarfs to the main-sequence turn-off, as well as 
planets orbiting white dwarfs, neutron stars, and black
holes.

\noindent $\bullet$ {\bf Sensitivity to Planets Throughout the Galaxy}

Because microlensing does not rely on light from the planet
or host star, planets can be detecting orbiting stars with distances
of several kiloparsecs.  The microlensing event rate depends on the intrinsic
lensing probability, which peaks for lens distances about
halfway to the typical sources in the Galactic bulge, and the number density distribution of lenses
along the line-of-sight toward the bulge, which peaks at the bulge.  The event rate remains substantial
for lens distances in the range $D_l\sim 1-8~{\rm kpc}$.  Roughly
40\% and 60\% of microlensing events toward the bulge are expected to be
due to lenses in the disk and bulge, respectively ({\em Kiraga \& Paczynski}, 1994).
Specialized surveys may be sensitive to planets with $D_l \la 1~{\rm kpc}$ 
({\em Di Stefano}, 2008; {\em Gaudi et al.}, 2008),
as well as planets in M31 ({\em Covone et al.}, 2000; {\em Chung et al.}, 2006;
{\em Ingrosso et al.}, 2009).

\noindent $\bullet$ {\bf Sensitivity to Multiple-Planet Systems}

For low-magnification events, multiple planets in the same system
can be detected only if the source crosses the planetary caustics
of both planets, or equivalently only if both planets happen to have projected
positions sufficiently close to the paths of the two images created by
the primary lens.  The probability of this is simply the product of the
individual probabilities, which is typically ${\cal{O}}(1\%)$ or less  ({\em Han \& Park}, 2002).  In
high-magnification events, however,
individual planets are detected with near-unity
probability regardless of the orientation of the planet with respect to
the source trajectory ({\em Griest \& Safizadeh}, 1998). This immediately implies all planets
sufficiently close to the Einstein ring radius will be revealed in
such events ({\em Gaudi et al.}, 1998).  This, along with the fact that high-magnification events
are potentially sensitive to very low-mass planets, makes such events excellent probes
of planetary systems. 

\noindent $\bullet$ {\bf Sensitivity to Moons of Exoplanets}

Because microlensing is potentially sensitive to planets with mass
as low as that of a few times the mass of the Moon ({\em Bennett \& Rhie}, 1996), it is 
potentially sensitive to large moons of exoplanets.  A system with a star of mass $M$, planet of mass $m_p$, and moon
of mass $m_m$ corresponds to a triple
lens with the hierarchy $m_m \ll m_p \ll M$. {\em Bennett \& Rhie}, 2002, {\em Han \& Han}, 2002,
{\em Han}, 2008, and {\em Liebig \& Wambsganss}, 2009 have all considered the detectability of moons of various masses, 
orbiting planets of various masses.  Generally, these studies find that massive terrestrial $(m_m \sim M_\oplus)$ moons (should
they exist) are readily detectable.  Less massive moons, with masses similar to our Moon,
are considerably more difficult to detect, but may be detectable in next-generation microlensing planet
searches, particularly those from space, in some favorable circumstances (i.e., for sources
with small dimensionless source size $\rho_*$). 
Analogous to planetary microlensing, moons with projected separation
much smaller than the Einstein radius of the planet $r_{\e,p}\equiv \theta_{\e,p}D_l$ are difficult to detect ({\em Han}, 2008).
Contrary to planetary microlensing, however, the signal of the moon does not 
diminish for angular separations much greater than $r_{\e,p}$ ({\em Han}, 2008),
although of course it becomes less likely that both the planetary and moon signal will be detected simultaneously.

A minimum requirement for a moon to be stable is that its semimajor axis
must be less the Hill radius of the planet, which is given by 
\begin{equation}
r_H = a_p \left(\frac{m_p}{3M}\right)^{1/3},
\label{eqn:rhill}
\end{equation}
where $a_p$ is the semimajor axis of the planet.  The ratio of $r_H$ to the Einstein ring
radius of the planet is,
\begin{equation}
\frac{r_H}{r_{\e,p}} \sim d q^{-1/6}.
\label{eqn:rhillre}
\end{equation}
Thus for $q\la 10^{-3}$, detectable moons also (fortuitously) happen to be stable.

\section{RECENT HIGHLIGHTS}\label{sec:results}

\begin{figure*}[htp]
\epsscale{1.9}
\plottwo{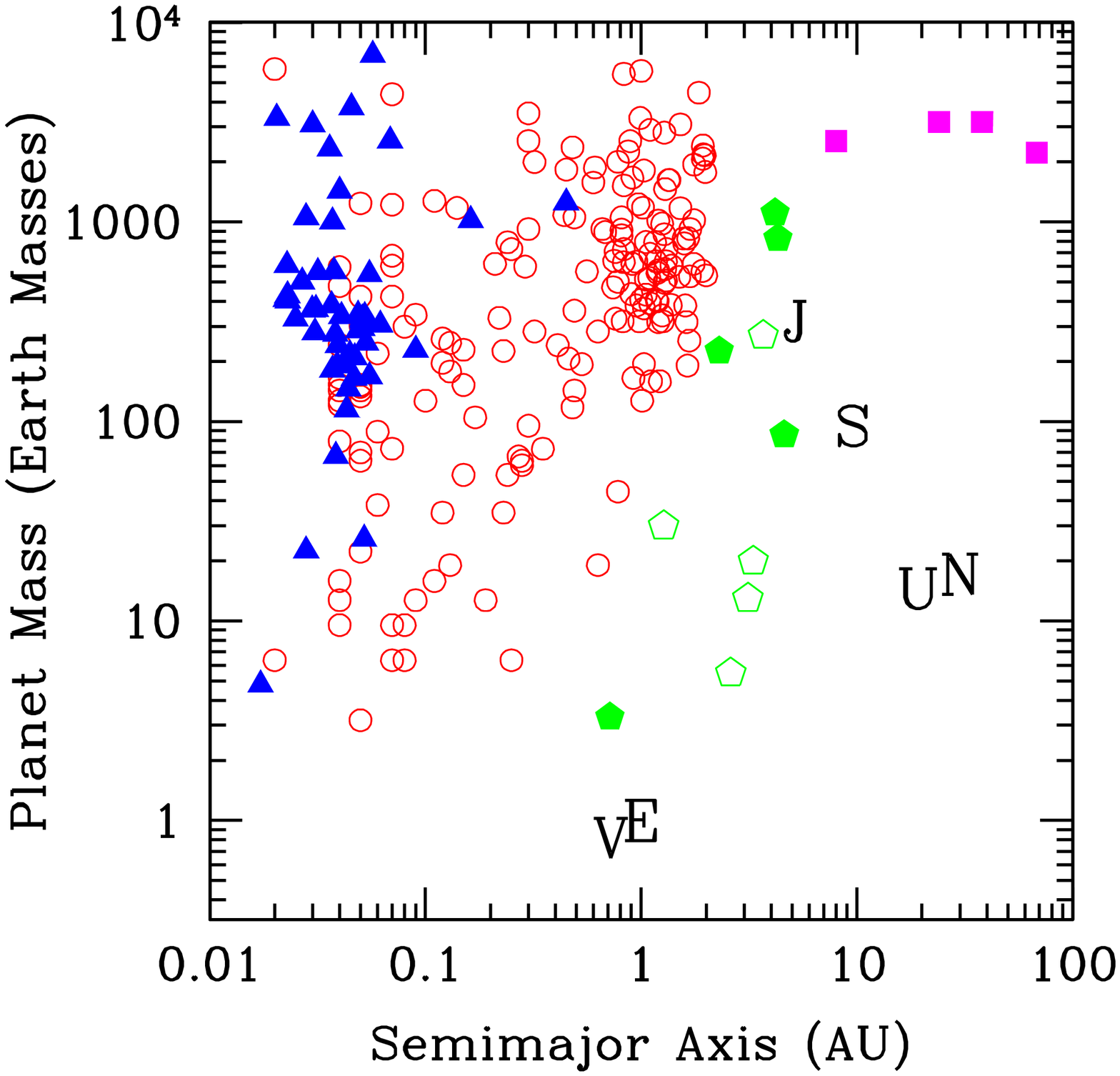}{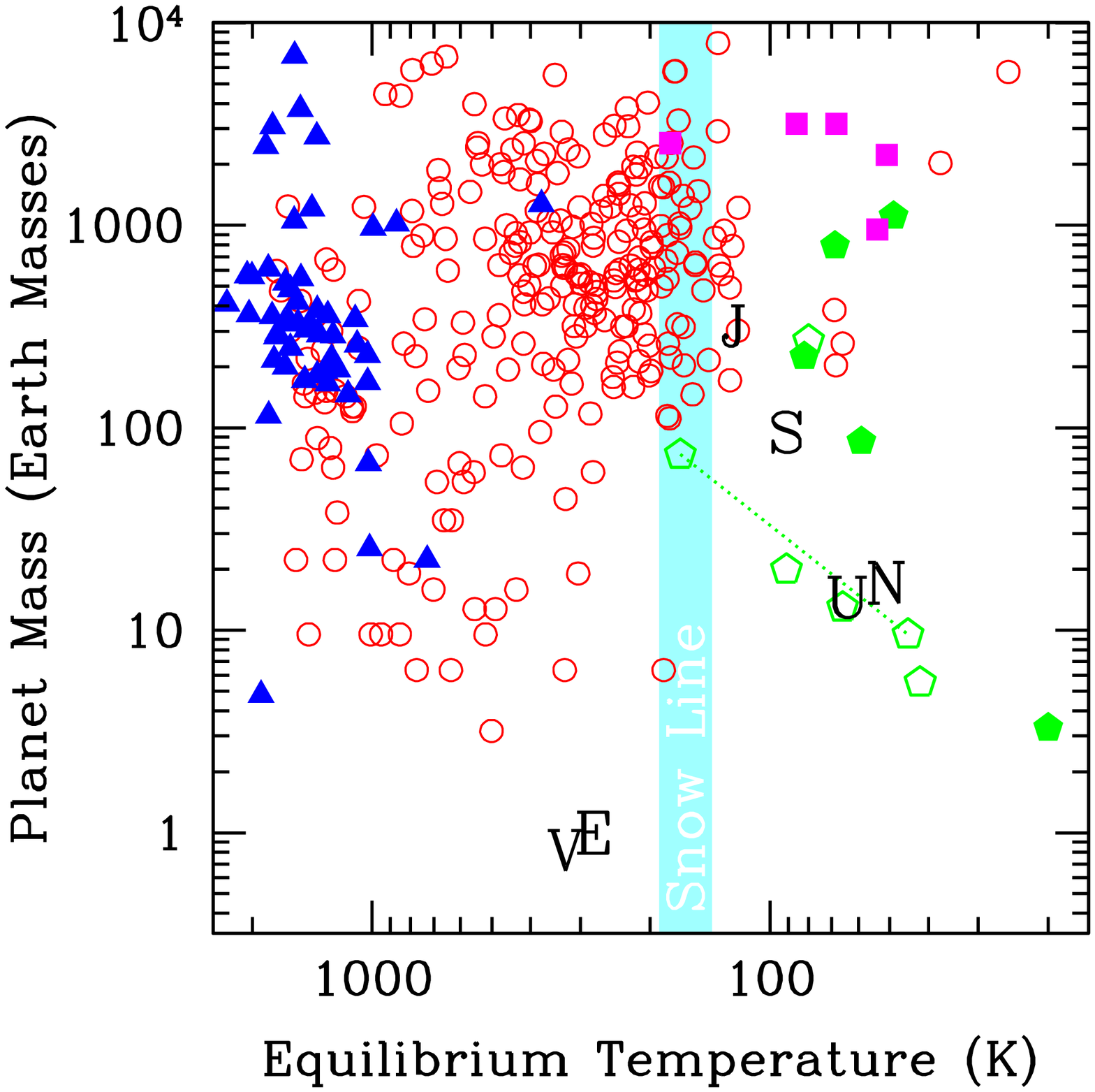}
\caption{\small
(Left) Mass versus semi-major axis for known exoplanets. (Right) Mass versus equilibrium temperature for known exoplanets.
In both panels, radial velocity detections are indicated by red circles, transiting planets
are indicated by blue triangles, microlensing detections are indicated
by green pentagons, and direct detections are indicated by magenta squares.
The letters indicate the locations of the Solar System planets. The solid pentagons are those microlensing detections for which the primary
mass and distance are measured, and thus planet mass, semimajor axis, and equilibrium
temperature are well-constrained.  The open pentagons indicate those microlensing detections
for which only partial information about the primary properties is available,
and thus the planet properties are relatively uncertain.  An extreme example
is shown in right panel, where two points connected
by a dotted line are plotted for event MOA-2008-BLG-310Lb ({\em Janczak et al.} 2009),
corresponding to the extrema of the allowed range of planet properties. In the right panel,
the light blue stripe roughly indicates the location of the 
snow line.  These figures also demonstrate the complementarity of the various planet detection
techniques: transit and radial velocity surveys are generally sensitive to planets
interior to the snow line, whereas microlensing and direct imaging surveys are sensitive to more distant
planets beyond the snow line.}
\label{fig:observed}
\end{figure*}

To date, ten detections of planets with microlensing have been
announced, in nine systems ({\em Bond et al.}, 2004; {\em Udalski et
al.}, 2005; {\em Beaulieu et al.}, 2006; {\em Gould et al.}, 2006;
{\em Gaudi et al.}, 2008a; {\em Bennett et al.}, 2008; {\em Dong et
al.}, 2009; {\em Janczak et al.}, 2009; {\em Sumi et al.}, 2009).  In
addition, there are another seven events with clear, robust planetary
signatures that await complete analysis and/or publication.  The
masses, separations, and equilibrium temperatures of the ten announced
planets are shown in Figure \ref{fig:observed}. We can expect to a
handful of detections per year at the current rate.

The right panel of this Figure \ref{fig:observed} demonstrates that
the first microlensing planet detections are probing a region of
parameter space that has not been previously explored by any method,
namely planets beyond snow line.  As a result, although the total
number of planets found by microlensing to date is small in comparison
to the sample of planets revealed by the radial velocity and transit
methods, these discoveries have already provided important empirical
constraints on planet formation theories.  In particular, the
detection of four cold, low mass ($5-20~M_\oplus$) planets amongst the
sample of microlensing detections indicates that these planets are
common ({\em Beaulieu et al.}, 2006; {\em Gould et al.}, 2006, {\em
Sumi et al.}, 2009).  The detection of a Jupiter/Saturn analog also
suggests that solar system analogs are probably not rare ({\em Gaudi
et al.}, 2008).  Finally, the detection of a low-mass planetary
companion to a brown-dwarf star suggests that such objects can form
planetary systems similar to those around solar-type main-sequence
stars ({\em Bennett et al.} 2008).

Another important lesson learned from these first few detections is
that it is possible to obtain substantially more information about the
planetary systems than previously thought.  In all ten cases, finite source
effects have been detectable and so it has been possible to measure
$\theta_{\rm E}$, which yields a mass-distance relation for the
primary (see Section \ref{sec:properties}).  Furthermore, for four systems, additional constraints allow
for a complete solution for the primary mass and distance, and so
planet mass. 

\subsection{Individual Detections}
\label{sec:ind}

The first two planets found by
microlensing, OGLE-2003-BLG-235/MOA-2003-BLG-53Lb ({\em Bond et al.}, 2004), and
OGLE-2005-BLG-071Lb ({\em Udalski et al.}, 2005), are Jovian-mass objects with separations of $\sim
2-4~{\rm AU}$. While the masses and separations of these planets
are similar to many of the planets discovered via radial velocity
surveys, their host stars are generally less massive and so the
planets have substantially lower equilibrium temperatures of $\sim 50-70$~K,
similar to Saturn and Uranus.

The third and fourth planets discovered by microlensing are
significantly lower mass, and indeed inhabit a region of
parameter space that was previously unexplored by any
method.  OGLE-2005-BLG-390Lb is a very low-mass planet with a
planet/star mass ratio of only $\sim 8 \times 10^{-5}$ ({\em Beaulieu et al.}, 2006).  A Bayesian
analysis combined with a measurement of $\theta_{\rm E}$ from finite
source effects indicates that the planet likely orbits a low-mass M
dwarf with $M=0.22_{-0.11}^{+0.21}M_\odot$, and thus has a mass of
only $5.5_{-2.7}^{+5.5}M_\oplus$.  Its separation is
$2.6_{-0.6}^{+1.5}~{\rm AU}$, and so has a cool equilibrium
temperature of $\sim 50~{\rm K}$.  OGLE-2005-BLG-169Lb is another
low-mass planet with a mass ratio of $8\times 10^{-5}$ ({\em Gould et al.}, 2006), essentially
identical to that of OGLE-2005-BLG-390Lb.  A Bayesian
analysis indicates a primary mass of $0.52_{-0.22}^{+0.19}M_\odot$,
and so a planet of mass $\sim 14_{-6}^{+5} M_\oplus$, a separation
of $3.3_{-0.9}^{+1.9}~{\rm AU}$, and an equilibrium temperature of
$\sim 70~{\rm K}$ ({\em Bennett et al.}, 2007).  In terms of its mass and equilibrium temperature,
OGLE-2005-BLG-169Lb is very similar to Uranus.  OGLE-2005-BLG-169Lb
was discovered in a high-magnification $A_{\rm max}\sim 800$ event; as argued
above, such events have significant sensitivity to multiple
planets.  There is no
indication of any additional planetary perturbations in this event,
which excludes Jupiter-mass planets with separations between 0.5-15~AU,
and Saturn-mass planets with separations between 0.8-9.5~AU.  Thus
it appears that this planetary system
is likely dominated by the detected Neptune-mass companion.  

The fifth and sixth planets discovered by microlensing were detected
is what is arguably one of the most information-rich and complex microlensing
events ever analyzed, OGLE-2006-BLG-109 ({\em Gaudi et al.}, 2008; {\em Bennett et al.}, 2009).  The light curve exhibited
five distinct features.  Four of these features are attributable to the
source crossing the resonant caustic of a Saturn-mass planet, and
include a short cusp crossing, followed by a pair of caustic
crossings, and finally a cusp approach, all spanning roughly two weeks.
The fifth feature cannot be explained by the Saturn-mass planet
caustic, but rather is due to the source approaching the tip of the
central caustic of a inner, Jupiter-mass planet.  

The OGLE-2006-BLG-109Lb planetary system bears a remarkable similarity
to a scaled version of Jupiter and Saturn.  By combining all the
available information, it is possible to infer that the primary is an
M dwarf with a mass of $M=0.51^{+0.05}_{-0.04}~M_\odot$ at distance of
$D_l=1.51^{+0.11}_{-0.12}~{\rm kpc}$.  The two planets have masses of $0.73\pm
0.06~M_{\rm Jup}$ and $0.27\pm 0.02~M_{\rm Jup}$ and separations of
$2.3 \pm 0.5~{\rm AU}$ and $4.5^{+2.1}_{-1.0}~{\rm AU}$, respectively.  Thus
the mass and separation ratios of the two planets are very similar to
Jupiter and Saturn.  Although the two planets in the
OGLE-2006-BLG-109Lb system orbit at about half the distance of their
Jupiter/Saturn analogs, their equilibrium temperatures are also similar
(although somewhat cooler) to Jupiter and Saturn, because of the lower
primary mass and luminosity. 
 
The seventh planet detected by microlensing is also a very low-mass,
cool planet ({\em Bennett et al.}, 2008).  MOA-2007-BLG-192Lb has a
mass ratio of $q= 2 \times 10^{-4}$, somewhat smaller than that of
Saturn and the sun.  However, this event exhibits orbital parallax
and the best-fit model also shows evidence for finite source effects,
allowing a measurement of the primary mass.  In this case, the
inferred primary mass is quite low:
$M=0.060_{-0.021}^{+0.028}~M_\odot$, implying a sub-stellar
brown-dwarf host, and very low mass for the planet of
$3.3_{-1.6}^{+4.9}~M_\oplus$.  Remarkably, this planet was detected
from data taken entirely by the MOA and OGLE survey collaborations,
without the benefit of additional coverage from the follow-up
collaborations.  This event therefore serves as an early indication of
how future microlensing planet surveys will operate, as detailed in
the next section.  Unfortunately, in this case, coverage of the
planetary perturbation was sparse, and as a result the limits on the
mass ratio and primary mass are relatively weak.  Fortunately, the inferred primary
mass can be confirmed and made more precise with follow-up VLT and/or
{\it HST} observations.  If confirmed, this discovery will demonstrate that
brown dwarfs can form planetary systems similar to those around
solar-type main-sequence stars.

The eighth and ninth microlensing planets, MOA-2007-BLG-400Lb ({\em
Dong et al.}, 2008b) and MOA-2008-BLG-310Lb ({\em Janczak et al.},
2009) were detected in events that bear some remarkable similarities.
Both were high-magnification events ($A_{\rm max}\sim 628$ and $A_{\rm
max} \sim 400$, respectively) in which the primary lens transited the
source, resulting in a dramatic smoothing of the peak of the event
(see panel e of Figure \ref{fig:clcurves} for an example). By eye, single lens
models provide relatively good matches to the data. Nevertheless, weak
but broad and significant residuals to the single-lens model are
apparent in both cases, which are well-fit by perturbations due to the
central caustic of a planetary companion.  In both cases, the inferred
caustic size is significantly smaller than the source size.  Also in
both cases, precise constraints on the mass ratio are possible despite
the weak perturbation amplitudes.  MOA-2007-BLG-400Lb is a cool,
Jovian-mass planet with mass ratio of $q=0.0025 \pm 0.0004$, and
MOA-2008-BLG-310Lb is sub-Saturn mass planet with $q=(3.3 \pm 0.3)
\times 10^{-4}$.  The angular Einstein ring radius of the primary lens 2008-BLG-310-310L
as inferred from finite source effects is quite
small ($\theta_\e = 0.162 \pm 0.015~{\rm mas}$), implying that if
the primary is a star ($M\ge 0.08~M_\odot$), it must have $D_l > 6.0~{\rm kpc}$
and so be in the bulge ({\em Janczak et al.}, 2009).  This is the best candidate yet for a bulge
planet.  Unfortunately, it is not possible to definitely determine if the system is in the
bulge with currently-available information, and it is possible that 
the primary is a low-mass star or brown dwarf in the foreground Galactic disk.  This ambiguity
can be resolved with follow-up high resolution
imaging taken immediately, followed by additional observations taken in $\sim 10$ years, at which point the 
source and lens will have separated by $\sim 50~{\rm mas}$.  

The tenth planet, OGLE-2007-BLG-368Lb, is another cold
Neptune ({\em Sumi et al.,} 2009), making it the fourth
low-mass ($\la 20~M_\oplus$) planet discovered with microlensing.  This
planet was detected in a relatively
low-magnification event with a maximum magnification of only $\sim 13$.
In this case, the source crossed between the two, triangular-shaped planetary caustics of
a close $(d<1)$ topology planetary lens, and thus the light curve exhibited a large ($\sim 20\%$) dip, characteristic
of the planetary perturbations due to such configurations (see Figure \ref{fig:plcurves}c,d).

\subsection{Frequency of Cold Planets}
\label{sec:freq}

The microlensing detection sensitivity declines with planet mass as
$\sim m_p^{1/2}$, and thus the presence of two low-mass planets
amongst the first four detections was an indication that the frequency
of cold super Earths/Neptunes ($5-15~M_\oplus$) is substantially
higher than that of cold Jovian-class planets.  {\em Gould et al.}
2006 performed a quantitative analysis that accounted for the
detection sensitivities and Poisson statistics and demonstrated that,
at 90\% confidence, $38_{-22}^{+31}\%$ of stars host cold super
Earths/Neptunes Neptunes with separations in the range $1.6-4.3~{\rm
AU}$.  An updated analysis taking into account the planets that have
since been detected, as well as those events that did not yield
planets, revises this number downward to $\sim 20\pm 10\%$ (D.\
Bennett, private communication).  Thus, such planets are common, which
is ostensibly a confirmation of the core accretion model of planet
formation, which predicts that there should exist many more `failed
Jupiters' than bona-fide Jovian-mass planets at such separations,
particularly around low-mass primaries ({\em Laughlin et al.}, 2004;
{\em Ida \& Lin}, 2005).

By adopting a simple model for the scaling of the detection efficiency
with $q$, {\em Sumi et al.} 2009 used the distribution of mass ratios
of the ten microlensing planets discovered to date to derive an
intrinsic mass (ratio) function of exoplanets beyond the snow line of
${\rm d}N/{\rm d}q \propto q^{-1.7\pm 0.2}$.  This implies that cold
Neptunes/super Earths ($q \sim 5 \times 10^{-5}$) are $7^{+6}_{-3}$
times more common than cold Jupiters ($q\sim 10^{-3}$), reinforcing
the conclusions of {\em Gould et el.} 2006.

\subsection{Properties of the Planetary Systems}
\label{sec:propobs}

As already mentioned, for all of the planet detections, it has been
possible to obtain additional information to improve the constraints
on the properties of the primaries and planets.  For example, for
OGLE-2003-BLG-235/MOA-2003-BLG-53Lb, follow-up imaging with {\it HST}
yielded a detection of light from the lens, which constrains the mass
of the primary and planet to $\sim 15\%$,
$M=0.63_{-0.09}^{+0.07}M_\odot$ and $m_p=2.6_{-0.6}^{+0.8}~M_{\rm
Jup}$ ({\em Bennett et al.}, 2006).  However, it is the analysis of
events OGLE-2006-BLG-109 ({\em Bennett et al.}, 2009) and
OGLE-2005-BLG-071 ({\em Dong et al.}, 2009a) which demonstrate most strikingly the detailed
information that can be obtained for
planets detected via microlensing.

For microlensing event OGLE-2006-BLG-109, besides the basic signatures of the two planets,
the light curve displays finite source effects and orbital
parallax, which allow for a complete solution to the lens system and
so measurements of the mass and distance to the primary, as well as a
the masses of the planets ({\em Gaudi et al.}, 2008a; {\em Bennett et al.}, 2009).  This primary mass measurement is corroborated by a Keck
AO $H$-band image of the target that yields a measurement of the
flux of the lens, which in this case turns out to be brighter than the
source.  The lens flux is consistent with the mass inferred entirely
from the microlensing observables.  In addition, the orbital motion of
the outer Saturn-mass planet was detected.  This was possible because the
source crossed or approached the resonant caustic due to this planet at four
distinct times.  Furthermore, as discussed in Section \ref{sec:phenom},
the exact shape and size of resonant caustics depend sensitively on $d$,
which changes over the two weeks of the planetary deviations.  In fact,
four of the six orbital parameters of the Saturn-mass planet are
well-measured, and a fifth parameter is weakly constrained.  This information,
combined with an assumption of coplanarity with the Jupiter-mass planet and 
the requirement of stability, enables a constraint on the orbital eccentricity of the Saturn-mass
planet of $e=0.15^{+0.17}_{-0.10}$, and inclination of the system of $i=64^{+4}_{-7}$ degrees
({\em Bennett et al.}, 2009).

In the case of OGLE-2005-BLG-071Lb, {\it HST} photometry, when combined with
information on finite source effects and
microlens parallax from the light curve, allows for a measurement of the primary
mass and distance, and so planet mass and projected separation.  This
leads to the conclusion that the companion is a massive planet with 
$m_p=3.8 \pm 0.4~M_{\rm Jup}$ and projected separation $r_\perp = 3.6 \pm 0.2~{\rm AU}$, orbiting 
an M-dwarf primary with a mass of $M=0.46 \pm 0.04 M_\odot$ and a distance of
$D_l=3.2 \pm 0.4~{\rm kpc}$ ({\em Dong et al.}, 2009a).  Furthermore, the primary
has thick-disk kinematics with a projected 
velocity relative to the Local Standard of Rest of  $v=103\pm 15~{\rm km~s^{-1}}$, suggesting that
it may be metal-poor.  Thus, OGLE-2005-BLG-071Lb
may be a massive Jovian planet orbiting a metal-poor, thick-disk
M-dwarf.  The existence of such a planet may pose a challenge for core-accretion models
of planet formation (e.g., {\em Laughlin et al.}, 2004; {\em Ida \& Lin}, 2004,2005).  

Interestingly, all four microlensing planet hosts for which it has
been possible to measure the distance to the system lie in the
foreground disk. As mentioned above, MOA-2008-BLG-310Lb is the most
promising candidate for a bulge planet, but in this case the primary
could still be a foreground disk brown dwarf.  In contrast, roughly
60\% of all microlensing events toward the Galactic bulge are due to
lenses in the bulge ({\em Kiraga \& Paczynski}, 1994).  The lack of
confirmed microlensing bulge planets could be due to the selection effect
that longer events, which are more likely to arise from disk lenses, are
preferentially monitored by the follow-up collaborations, or it could
reflect a difference between the planet populations in the disk and
bulge.  Regardless, with a larger sample of
planets with well-constrained distances, and a more careful accounting
of selection effects, it should be possible to compare the
demographics of planets in the Galactic disk and bulge.

\begin{figure*}
\epsscale{2.2}
\plottwo{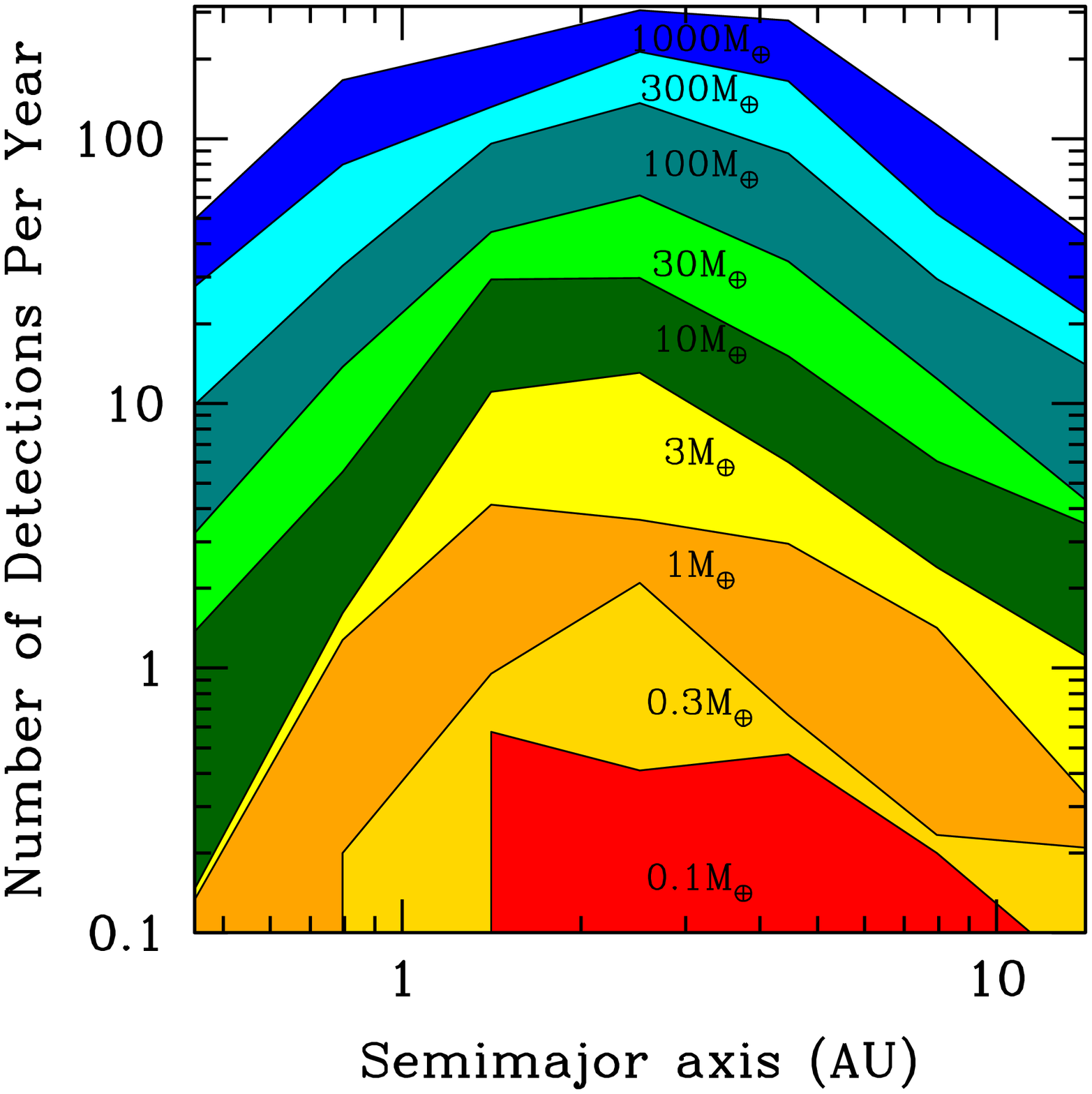}{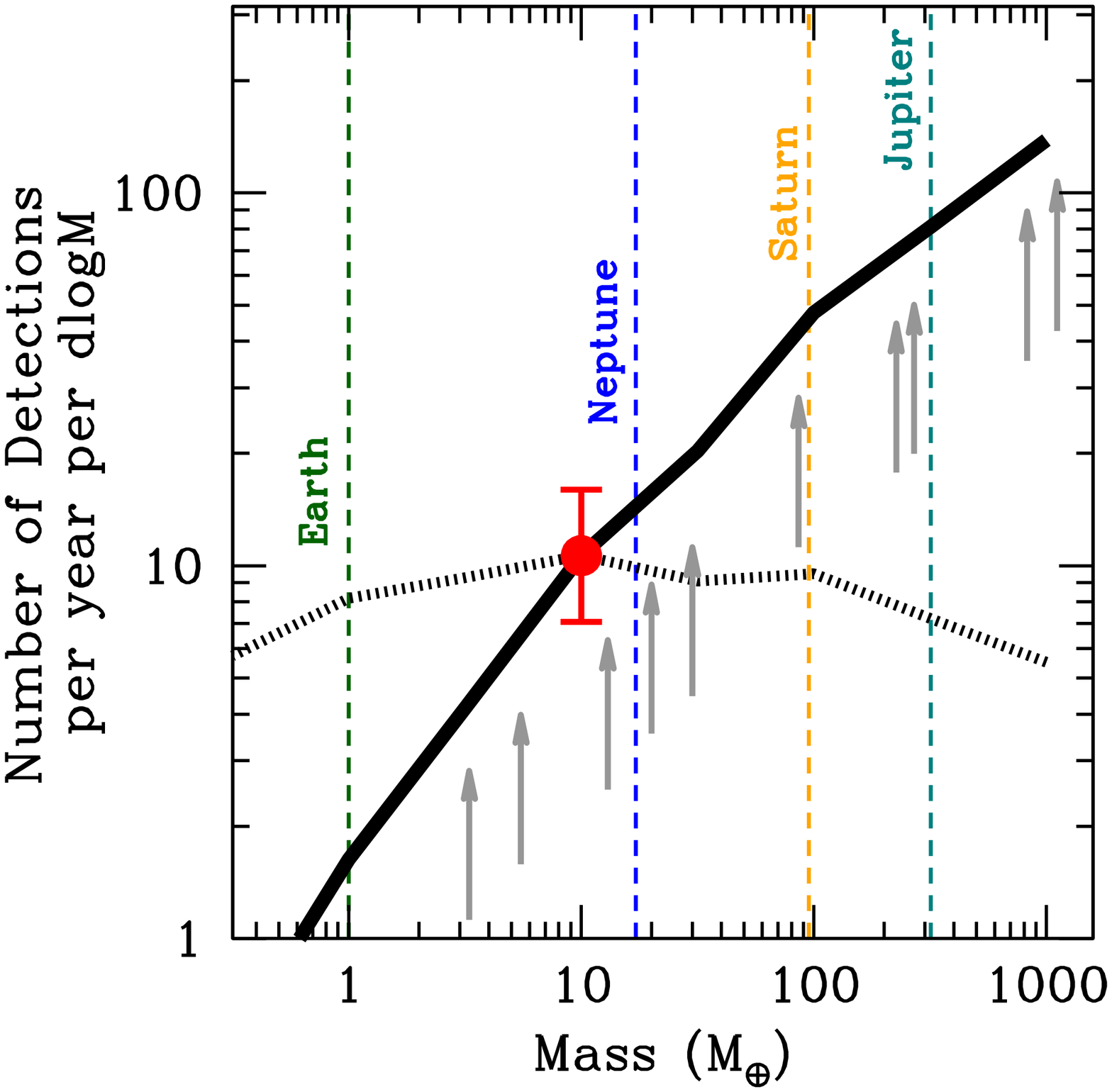}
\caption{\small
Expectations from the next generation of ground-based microlensing surveys,
including MOA-II, OGLE-IV, and a KMTNet telescope in South Africa.
These results represent the average of two independent simulations,
which include very different input assumptions 
but differ in their predictions by only $\sim 0.3$ dex.
(Left)  Number of planets detected per year as a function
of semimajor axis for various masses, assuming every star
has a planet of the given mass and semimajor axis. 
(Right) Number of planets detected per year
as a function of planet mass, normalized by the number of $\sim 10M_\oplus$ found to date by microlensing
(indicated by the red dot). The solid line is the prediction assuming an equal number
of planets per logarithmic mass interval, and the dotted curve assumes
that the number of planets per log mass scales as $m_p^{-0.7}$ ({\em Sumi et al.}, 2009). 
In both cases, planets are
assumed to be distributed
uniformly in $\log(a)$ between 0.4-20~AU. Arrows indicate the locations of
the ten published exoplanets.
}\label{fig:nextgen}
\end{figure*}

\section{FUTURE PROSPECTS}\label{sec:future}

In the four seasons of 2003-2006 there were five planetary events,
containing six detected planets. With the MOA upgrade in 2006 to the
MOA-II phase, the rate of planet detections has increased
substantially.  From the 2007, 2008, and 2009 bulge seasons, there
were four, three, and four secure planetary events, respectively
(seven of these await publication).  Thus even maintaining the current
rate, we can expect of order a dozen new planet detections over the
next several years.  In fact, as described below, we can expect the rate of planet
detections to increase substantially, as microlensing planet searches
transition toward the next generation of surveys.  Besides finding
more planets, these surveys will also have improved sensitivity to
lower-mass, terrestrial planets. Thus we can expect to have robust
constraints on the frequency of Earth-mass planets beyond the snow
line within the next decade.  A space-based survey would determine
determine the demographics of planets with mass greater than that of Mars and
semimajor axis $\ga 0.5$AU, determine the frequency of free-floating,
Earth-mass planets, and determine the frequency of terrestrial planets in the
outer habitable zones of solar-type stars in the Galactic disk and
bulge.

The transition to the next generation
of ground-based surveys is enabled by the advent of large-format cameras with fields-of-view (FOV) of several
square degrees.  With such large FOVs, it becomes possible to monitor tens of
millions of stars every $10-20$ minutes, and so discover thousands of
microlensing events per year. Furthermore, these events are then
simultaneously monitored with the cadence required to detect
perturbations due to very low-mass ($\sim M_\oplus$) planets.  Thus
next-generation searches will operate in a very different mode than
the current alert/follow-up model.  In order to obtain round-the-clock
coverage and so catch all of the perturbations, several such
telescopes would be needed, located on 3-4 continents roughly evenly
spread in longitude.

In fact, the transition to the `next-generation' is happening already.  
The MOA-II telescope in New Zealand (1.8m and 2 deg$^2$ FOV)
already represents one leg of such a survey. The OGLE team has recently
upgraded to the OGLE-IV phase with a 1.4 deg$^2$ camera, which will represent the second
leg in Chile when it becomes operational in 2010.  Although the OGLE telescope
has a smaller FOV camera and a smaller aperture (1.3m) than the
MOA telescope, these are mostly compensated by the
better site quality in Chile.  Finally, the 
Korean Microlensing Telescope Network (KMTNet) is
a project with plans to build three 1.6m telescopes with 4 deg$^2$ FOV
cameras, one each in South Africa, South America, and Australia.  With
the completion of the South African leg of this network (planned
2012), a next-generation survey would effectively be in place.  In addition, astronomers from Germany and China
are considering initiatives to secure funding to build 1-2m class
telescopes with wide FOV cameras in southern Africa or Antarctica.

Detailed simulations of such a next-generation microlensing survey
have been performed by several groups ({\em Gaudi et al.,
unpublished}; {\em Bennett}, 2004).  These simulations include models
for the Galactic population of lenses and sources that match all
constraints ({\em Han \& Gould}, 1995, 2003), and account for
real-world effects such as weather, variable seeing, moon and sky
background, and crowded fields.  They reach similar conclusions.  Such
a survey would increase the planet detection rate at fixed mass by at
least an order of magnitude over current surveys.  Figure
\ref{fig:nextgen} shows the predictions of these simulations for the
detection rate of planets of various masses and separations using a
survey including MOA-II, OGLE-IV, and a Korean telescope in South
Africa.  In particular, if Earth-mass planets with semimajor axes of
several AU are common around main-sequence stars, a next generation
microlensing survey should detect several such planets per year.  This
survey would also be sensitive to free-floating planets, and would
detect them at a rate of hundreds per year if every star has ejected
Jupiter-mass planet.

Ultimately, however, the true potential of microlensing cannot be
realized from the ground.  Weather, seeing, crowded fields, and
systematic errors all conspire to make the detection of
planets with mass less than Earth effectively impossible from the
ground ({\em Bennett}, 2004).  As outlined
in {\em Bennett \& Rhie}, 2002, a space-based microlensing survey offers several advantages:
the main-sequence bulge sources needed to detect sub-Earth
mass planets are resolved from space, the events can be monitored
continuously, and it is possible to observe the moderately reddened
source stars in the near infrared to improve the photon collection
rate.  Furthermore, the high spatial resolution afforded by space
allows unambiguous identification of light from the primary
(lens) stars and so measurements of the primary and planet 
masses ({\em Bennett et al.}, 2007).

The expectations from a Discovery-class space-based microlensing
survey are impressive.  Such a survey would be sensitive to all
planets with mass $\ga 0.1M_\oplus$ and separations $a \ga 0.5~{\rm
AU}$, including free-floating planets ({\em Bennett \& Rhie}, 2002;
{\em Bennett et al.}, 2009).  This range includes analogs to all the
solar system planets except Mercury. If every main-sequence star has
an Earth-mass planet in the range $1-2.5~{\rm AU}$, the survey would
detect $\sim 500$ such planets within its mission lifetime.  The
survey would also detect a comparable number of habitable Earth-mass
planets as the {\it Kepler} mission ({\em Borucki et al.}, 2003), and so would provide an important
independent measurement of $\eta_{\oplus}$.  When combined with
complementary surveys (such as {\it Kepler}), a space-based
microlensing planet survey would determine
the demographics  of both bound and free-floating planets with masses greater than that
of Mars orbiting stars with masses less than that of the Sun.

The basic requirements for a space-based microlensing planet survey
are relatively modest: an aperture of at least 1m, a large FOV camera
(at least $\sim$0.5 deg$^2$) with optical or near-IR detectors,
reasonable image quality of better than $\sim 0.25$'', and an orbit
for which the Galactic bulge is continuously visible. The {\it
Microlensing Planet Finder (MPF)} is an example of a space-based
microlensing survey that can accomplish these objectives, essentially
entirely with proven technology, and at a cost of $\sim$\$300 million
excluding launch vehicle ({\em Bennett et al.}, 2009).
Interestingly, the requirements for a space-based microlensing planet
survey are very similar to, or less stringent than, the requirements
for a number of the proposed dark energy missions, in particular those
that focus on weak lensing measurements.  Thus, it may be attractive
to consider a combined dark energy/planet finding mission that could
be accomplished at a substantial savings compared to doing each
mission separately ({\em Gould}, 2009).

\bigskip
\textbf{ Acknowledgments.} I would like to thank Subo Dong for preparing Figure \ref{fig:blending},
Andrzej Udalski for reading over the
manuscript, and two anonymous referees for catching some important
errors, and comments that greatly improved the presentation.

\bigskip

\centerline\textbf{ REFERENCES}
\bigskip
\parskip=0pt
{\small
\baselineskip=11pt

\refs Afonso, C., Alard, C., Albert, J.~N., Andersen, J., Ansari, R., 
Aubourg, {\'E}., Bareyre, P., Bauer, F., Beaulieu, J.~P., Bouquet, A., 
Char, S., Charlot, X., Couchot, F., Coutures, C., Derue, F., Ferlet, R., 
Glicenstein, J.~F., Goldman, B., Gould, A., Graff, D., Gros, M., 
Haissinski, J., Hamilton, J.~C., Hardin, D., de Kat, J., Kim, A., Lasserre, 
T., Lesquoy, {\'E}., Loup, C., Magneville, C., Marquette, J.~B., Maurice, 
{\'E}., Milsztajn, A., Moniez, M., Palanque-Delabrouille, N., Perdereau, 
O., Pr{\'e}vot, L., Regnault, N., Rich, J., Spiro, M., Vidal-Madjar, A., 
Vigroux, L., Zylberajch, S., Alcock, C., Allsman, R.~A., Alves, D., 
Axelrod, T.~S., Becker, A.~C., Cook, K.~H., Drake, A.~J., Freeman, K.~C., 
Griest, K., King, L.~J., Lehner, M.~J., Marshall, S.~L., Minniti, D., 
Peterson, B.~A., Pratt, M.~R., Quinn, P.~J., Rodgers, A.~W., Stetson, 
P.~B., Stubbs, C.~W., Sutherland, W., Tomaney, A., Vandehei, T., Rhie, 
S.~H., Bennett, D.~P., Fragile, P.~C., Johnson, B.~R., Quinn, J., Udalski, 
A., Kubiak, M., Szyma{\'n}ski, M., Pietrzy{\'n}ski, G., Wo{\'z}niak, P., 
Zebru{\'n}, K., Albrow, M.~D., Caldwell, J.~A.~R., DePoy, D.~L., Dominik, 
M., Gaudi, B.~S., Greenhill, J., Hill, K., Kane, S., Martin, R., Menzies, 
J., Naber, R.~M., Pogge, R.~W., Pollard, K.~R., Sackett, P.~D., Sahu, 
K.~C., Vermaak, P., Watson, R., 
\& Williams, A.\ (2000) Combined Analysis of the Binary Lens Caustic-crossing Event MACHO 98-SMC-1.  {\em ApJ,  532}, 340-352

\refs Afonso, C., Albert, J.~N., Andersen, J., Ansari, R., Aubourg, {\'E}., 
Bareyre, P., Bauer, F., Blanc, G., Bouquet, A., Char, S., Charlot, X., 
Couchot, F., Coutures, C., Derue, F., Ferlet, R., Fouqu{\'e}, P., 
Glicenstein, J.~F., Goldman, B., Gould, A., Graff, D., Gros, M., 
Haissinski, J., Hamilton, J.~C., Hardin, D., de Kat, J., Kim, A., Lasserre, 
T., LeGuillou, L., Lesquoy, {\'E}., Loup, C., Magneville, C., Mansoux, B., 
Marquette, J.~B., Maurice, {\'E}., Milsztajn, A., Moniez, M., 
Palanque-Delabrouille, N., Perdereau, O., Pr{\'e}vot, L., Regnault, N., 
Rich, J., Spiro, M., Vidal-Madjar, A., Vigroux, L., Zylberajch, S., 
\& The EROS collaboration (2001) Photometric constraints on microlens spectroscopy of EROS-BLG-2000-5.  {\em A\&A,  378}, 1014-1023 

\refs Albrow, M., Beaulieu, J.-P., Birch, P., Caldwell, J.~A.~R., Kane, S., 
Martin, R., Menzies, J., Naber, R.~M., Pel, J.-W., Pollard, K., Sackett, 
P.~D., Sahu, K.~C., Vreeswijk, P., Williams, A., Zwaan, M.~A., 
\& The PLANET Collaboration (1998) The 1995 Pilot Campaign of PLANET: Searching for Microlensing Anomalies through Precise, Rapid, Round-the-Clock Monitoring.  {\em ApJ,  509}, 687-702

\refs Albrow, M.~D., Beaulieu, J.-P., Caldwell, J.~A.~R., Depoy, D.~L., 
Dominik, M., Gaudi, B.~S., Gould, A., Greenhill, J., Hill, K., Kane, S., 
Martin, R., Menzies, J., Naber, R.~M., Pollard, K.~R., Sackett, P.~D., 
Sahu, K.~C., Vermaak, P., Watson, R., Williams, A., 
\& Pogge, R.~W.\ (1999a) The Relative Lens-Source Proper Motion in MACHO 98-SMC-1.  {\em ApJ,  512}, 672-677 

\refs Albrow, M.~D., Beaulieu, J.-P., Caldwell, J.~A.~R., Depoy, D.~L., 
Dominik, M., Gaudi, B.~S., Gould, A., Greenhill, J., Hill, K., Kane, S., 
Martin, R., Menzies, J., Naber, R.~M., Pogge, R.~W., Pollard, K.~R., 
Sackett, P.~D., Sahu, K.~C., Vermaak, P., Watson, R., Williams, A., 
\& The PLANET Collaboration (1999b) A Complete Set of Solutions for Caustic Crossing Binary Microlensing Events.  {\em ApJ,  522}, 1022-1036 

\refs Albrow, M.~D., Beaulieu, J.-P., Caldwell, J.~A.~R., Dominik, M., 
Greenhill, J., Hill, K., Kane, S., Martin, R., Menzies, J., Naber, R.~M., 
Pel, J.-W., Pollard, K., Sackett, P.~D., Sahu, K.~C., Vermaak, P., Watson, 
R., Williams, A., 
\& Sahu, M.~S.\ (1999c) Limb Darkening of a K Giant in the Galactic Bulge: PLANET Photometry of MACHO 97-BLG-28.  {\em ApJ,  522}, 1011-1021 

\refs Albrow, M.~D., An, J., Beaulieu, J.-P., Caldwell, J.~A.~R., DePoy, 
D.~L., Dominik, M., Gaudi, B.~S., Gould, A., Greenhill, J., Hill, K., Kane, 
S., Martin, R., Menzies, J., Pogge, R.~W., Pollard, K.~R., Sackett, P.~D., 
Sahu, K.~C., Vermaak, P., Watson, R., 
\& Williams, A.\ (2002) A Short, Nonplanetary, Microlensing Anomaly: Observations and Light-Curve Analysis of MACHO 99-BLG-47.  {\em ApJ,  572}, 1031-1040 

\refs Alcock, C., Akerlof, C.~W., Allsman, R.~A., Axelrod, T.~S., Bennett, 
D.~P., Chan, S., Cook, C.~H., Freeman, K.~C., Griest, K., Marshall, S.~L., 
Park, H.~S., Perlmutter, S., Peterson, B.~A., Pratt, M.~R., Quinn, P.~J., 
Rodgers, A.~W., Stubbs, C.~W., 
\& Sutherland, W.\ (1993) Possible Gravitational Microlensing of a Star in the Large Magellanic Cloud.  {\em Nature,  365}, 621-623

\refs Alcock, C., Allsman, R.~A., Alves, D., Axelrod, T.~S., Becker, A.~C., 
Bennett, D.~P., Cook, K.~H., Freeman, K.~C., Griest, K., Guern, J., Lehner, 
M.~J., Marshall, S.~L., Peterson, B.~A., Pratt, M.~R., Quinn, P.~J., Reiss, 
D., Rodgers, A.~W., Stubbs, C.~W., Sutherland, W., Welch, D.~L., 
\& The MACHO Collaboration (1996) Real-Time Detection and Multisite Observations of Gravitational Microlensing.  {\em ApJ,  463}, L67-L70

\refs Alcock, C., Allsman, R.~A., Alves, D., Axelrod, T.~S., Becker, A.~C., 
Bennett, D.~P., Cook, K.~H., Freeman, K.~C., Griest, K., Keane, M.~J., 
Lehner, M.~J., Marshall, S.~L., Minniti, D., Peterson, B.~A., Pratt, M.~R., 
Quinn, P.~J., Rodgers, A.~W., Stubbs, C.~W., Sutherland, W., Tomaney, 
A.~B., Vandehei, T., 
\& Welch, D.\ (1997) First Detection of a Gravitational Microlensing Candidate toward the Small Magellanic Cloud.  {\em ApJ,  491}, L11-L13

\refs Alcock, C., Allsman, R.~A., Alves, D.~R., Axelrod, T.~S., Becker, 
A.~C., Bennett, D.~P., Cook, K.~H., Dalal, N., Drake, A.~J., Freeman, 
K.~C., Geha, M., Griest, K., Lehner, M.~J., Marshall, S.~L., Minniti, D., 
Nelson, C.~A., Peterson, B.~A., Popowski, P., Pratt, M.~R., Quinn, P.~J., 
Stubbs, C.~W., Sutherland, W., Tomaney, A.~B., Vandehei, T., 
\& Welch, D.\ (2000) The MACHO Project: Microlensing Results from 5.7 Years of Large Magellanic Cloud Observations.  {\em ApJ,  542}, 281-307 

\refs Agol, E.\ (2002) Occultation and Microlensing.  {\em ApJ,  579}, 
430-436

\refs An, J.~H.\ (2005) Gravitational lens under perturbations: symmetry of 
perturbing potentials with invariant caustics.  {\em MNRAS,  356}, 
1409-1428 

\refs An, J.~H., 
\& Evans, N.~W.\ (2006) The Chang-Refsdal lens revisited.  {\em MNRAS,  369}, 317-334

\refs Aubourg, E., Bareyre, P., Brehin, S., Gros, M., Lachieze-Rey, M., 
Laurent, B., Lesquoy, E., Magneville, C., Milsztajn, A., Moscoso, L., 
Queinnec, F., Rich, J., Spiro, M., Vigroux, L., Zylberajch, S., Ansari, R., 
Cavalier, F., Moniez, M., Beaulieu, J.~P., Ferlet, R., Grison, P., 
Vidal-Madjar, A., Guibert, J., Moreau, O., Tajahmady, F., Maurice, E., 
Prevot, L., 
\& Gry, C.\ (1993) Evidence for Gravitational Microlensing by Dark Objects in the Galactic Halo.  {\em Nature,  365}, 623-625

\refs Beaulieu, J.-P., Bennett, D.~P., Fouqu{\'e}, P., Williams, A., 
Dominik, M., J{\o}rgensen, U.~G., Kubas, D., Cassan, A., Coutures, C., 
Greenhill, J., Hill, K., Menzies, J., Sackett, P.~D., Albrow, M., Brillant, 
S., Caldwell, J.~A.~R., Calitz, J.~J., Cook, K.~H., Corrales, E., Desort, 
M., Dieters, S., Dominis, D., Donatowicz, J., Hoffman, M., Kane, S., 
Marquette, J.-B., Martin, R., Meintjes, P., Pollard, K., Sahu, K., Vinter, 
C., Wambsganss, J., Woller, K., Horne, K., Steele, I., Bramich, D.~M., 
Burgdorf, M., Snodgrass, C., Bode, M., Udalski, A., Szyma{\'n}ski, M.~K., 
Kubiak, M., Wi{\c e}ckowski, T., Pietrzy{\'n}ski, G., Soszy{\'n}ski, I., 
Szewczyk, O., Wyrzykowski, {\L}., Paczy{\'n}ski, B., Abe, F., Bond, I.~A., 
Britton, T.~R., Gilmore, A.~C., Hearnshaw, J.~B., Itow, Y., Kamiya, K., 
Kilmartin, P.~M., Korpela, A.~V., Masuda, K., Matsubara, Y., Motomura, M., 
Muraki, Y., Nakamura, S., Okada, C., Ohnishi, K., Rattenbury, N.~J., Sako, 
T., Sato, S., Sasaki, M., Sekiguchi, T., Sullivan, D.~J., Tristram, P.~J., 
Yock, P.~C.~M., 
\& Yoshioka, T.\ (2006) Discovery of a cool planet of 5.5 Earth masses through gravitational microlensing.  {\em Nature,  439}, 437-440 

\refs Boutreux, T., 
\& Gould, A.\ (1996) Monte Carlo Simulations of MACHO Parallaxes from a Satellite.  {\em ApJ,  462}, 705- 

\refs Bennett, D.~P.\ (2004) The Detection of Terrestrial Planets via 
Gravitational Microlensing: Space vs. Ground-based Surveys.  {\em ASPC,  
321}, 59-67

\refs Bennett, D.~P.\ (2009a) Detection of Extrasolar Planets by 
Gravitational Microlensing. {\it Exoplanets: Detection, Formation, Properties, Habitability},
eds.\ J.\ Mason, (Berlin:Springer) (arXiv:0902.1761)

\refs Bennett, D.~P.\ (2009b) An Efficient Method for Modeling High 
Magnification Planetary Microlensing Events. {\em ApJ}, submitted (arXiv:0911.2703)

\refs Bennett, D.~P., Anderson, J., 
\& Gaudi, B.~S.\ (2007) Characterization of Gravitational Microlensing Planetary Host Stars.  {\em ApJ,  660}, 781-790 

\refs Bennett, D.~P., 
\& Rhie, S.~H.\ (1996) Detecting Earth-Mass Planets with Gravitational Microlensing.  {\em ApJ,  472}, 660-664

\refs Bennett, D.~P., 
\& Rhie, S.~H.\ (2002) Simulation of a Space-based Microlensing Survey for Terrestrial Extrasolar Planets.  {\em ApJ,  574}, 985-1003 

\refs Bennett, D.~P., Anderson, J., Bond, I.~A., Udalski, A., 
\& Gould, A.\ (2006) Identification of the OGLE-2003-BLG-235/MOA-2003-BLG-53 Planetary Host Star.  {\em ApJ,  647}, L171-L174 

\refs Bennett, D.~P., Anderson, J., 
\& Gaudi, B.~S.\ (2007a) Characterization of Gravitational Microlensing Planetary Host Stars.  {\em ApJ,  660}, 781-790 

\refs Bennett, D.~P., Anderson, J., Beaulieu, J.~P., Bond, I., Cheng, E., 
Cook, K., Friedman, S., Gaudi, B.~S., Gould, A., Jenkins, J., Kimble, R., 
Lin, D., Mather, J., Rich, M., Sahu, K., Sumi, T., Tenerelli, D., Udalski, 
A., 
\& Yoch, P.\ (2009) A Census of Explanets in Orbits Beyond 0.5 AU via Space-based Microlensing.  {\em Astro2010: The Astronomy and Astrophysics Decadal Survey, Science White Papers}, 18 

\refs Bennett, D.~P., Bond, I.~A., Udalski, A., Sumi, T., Abe, F., Fukui, 
A., Furusawa, K., Hearnshaw, J.~B., Holderness, S., Itow, Y., Kamiya, K., 
Korpela, A.~V., Kilmartin, P.~M., Lin, W., Ling, C.~H., Masuda, K., 
Matsubara, Y., Miyake, N., Muraki, Y., Nagaya, M., Okumura, T., Ohnishi, 
K., Perrott, Y.~C., Rattenbury, N.~J., Sako, T., Saito, T., Sato, S., 
Skuljan, L., Sullivan, D.~J., Sweatman, W.~L., Tristram, P.~J., Yock, 
P.~C.~M., Kubiak, M., Szyma{\'n}ski, M.~K., Pietrzy{\'n}ski, G., 
Soszy{\'n}ski, I., Szewczyk, O., Wyrzykowski, {\L}., Ulaczyk, K., Batista, 
V., Beaulieu, J.~P., Brillant, S., Cassan, A., Fouqu{\'e}, P., Kervella, 
P., Kubas, D., 
\& Marquette, J.~B.\ (2008) A Low-Mass Planet with a Possible Sub-Stellar-Mass Host in Microlensing Event MOA-2007-BLG-192.  {\em ApJ,  684}, 663-683 

\refs Bennett, D.~P., Rhie, S.~H., Nikolaev, S., Gaudi, B.~S., Udalski, A., 
Gould, A., Christie, G.~W., Maoz, D., Dong, S., McCormick, J., Szymanski, 
M.~K., Tristram, P.~J., Macintosh, B., Cook, K.~H., Kubiak, M., 
Pietrzynski, G., Soszynski, I., Szewczyk, O., Ulaczyk, K., Wyrzykowski, L., 
DePoy, D.~L., Han, C., Kaspi, S., Lee, C.~-., Mallia, F., Natusch, T., 
Park, B.~-., Pogge, R.~W., Polishook, D., Abe, F., Bond, I.~A., Botzler, 
C.~S., Fukui, A., Hearnshaw, J.~B., Itow, Y., Kamiya, K., Korpela, A.~V., 
Kilmartin, P.~M., Lin, W., Masuda, K., Matsubara, Y., Motomura, M., Muraki, 
Y., Nakamura, S., Okumura, T., Ohnishi, K., Perrott, Y.~C., Rattenbury, 
N.~J., Sako, T., Saito, T., Sato, S., Skuljan, L., Sullivan, D.~J., Sumi, 
T., Sweatman, W.~L., Yock, P.~C.~M., Albrow, M., Allan, A., Beaulieu, 
J.~-., Bramich, D.~M., Burgdorf, M.~J., Coutures, C., Dominik, M., Dieters, 
S., Fouque, P., Greenhill, J., Horne, K., Snodgrass, C., Steele, I., 
Tsapras, Y., Chaboyer, B., Crocker, A., 
\& Frank, S.\ (2009) Masses and Orbital Constraints for the OGLE-2006-BLG-109Lb,c Jupiter/Saturn Analog Planetary System. {\em ApJ}, submitted
(arXiv:0911.2706)

\refs Bolatto, A.~D., 
\& Falco, E.~E.\ (1994) The detectability of planetary companions of compact Galactic objects from their effects on microlensed light curves of distant stars.  {\em ApJ,  436}, 112-116 

\refs Bond, I.~A., Udalski, A., Jaroszy{\'n}ski, M., Rattenbury, N.~J., 
Paczy{\'n}ski, B., Soszy{\'n}ski, I., Wyrzykowski, L., Szyma{\'n}ski, 
M.~K., Kubiak, M., Szewczyk, O., {\.Z}ebru{\'n}, K., Pietrzy{\'n}ski, G., 
Abe, F., Bennett, D.~P., Eguchi, S., Furuta, Y., Hearnshaw, J.~B., Kamiya, 
K., Kilmartin, P.~M., Kurata, Y., Masuda, K., Matsubara, Y., Muraki, Y., 
Noda, S., Okajima, K., Sako, T., Sekiguchi, T., Sullivan, D.~J., Sumi, T., 
Tristram, P.~J., Yanagisawa, T., 
\& Yock, P.~C.~M.\ (2004) OGLE 2003-BLG-235/MOA 2003-BLG-53: A Planetary Microlensing Event.  {\em ApJ,  606}, L155-L158 

\refs Borucki, W.~J., Koch, D.~G., Lissauer, J.~J., Basri, G.~B., Caldwell, 
J.~F., Cochran, W.~D., Dunham, E.~W., Geary, J.~C., Latham, D.~W., 
Gilliland, R.~L., Caldwell, D.~A., Jenkins, J.~M., 
\& Kondo, Y.\ (2003) The Kepler mission: a wide-field-of-view photometer designed to determine the frequency of Earth-size planets around solar-like stars.  {\em SPIE,  4854}, 129-140 

\refs Bozza, V.\ (2000) Caustics in special multiple lenses.  {\em 
A\&A,  355}, 423-432 

\refs Bromley, B.~C.\ (1996) Finite-Size Gravitational Microlenses.  {\em 
ApJ,  467}, 537-539

\refs Calchi Novati, S., Paulin-Henriksson, S., An, J., Baillon, P., 
Belokurov, V., Carr, B.~J., Cr{\'e}z{\'e}, M., Evans, N.~W., 
Giraud-H{\'e}raud, Y., Gould, A., Hewett, P., Jetzer, P., Kaplan, J., 
Kerins, E., Smartt, S.~J., Stalin, C.~S., Tsapras, Y., 
\& Weston, M.~J.\ (2005) POINT-AGAPE pixel lensing survey of M 31. Evidence for a MACHO contribution to galactic halos.  {\em A\&A,  443}, 911-928 

\refs Cassan, A.\ (2008) An alternative parameterisation for binary-lens 
caustic-crossing events.  {\em A\&A,  491}, 587-595 

\refs Chang, K., 
\& Refsdal, S.\ (1979) Flux variations of QSO 0957+561 A, B and image splitting by stars near the light path.  {\em Nature,  282}, 561-564 

\refs Chang, K., 
\& Refsdal, S.\ (1984) Star disturbances in gravitational lens galaxies.  {\em A\&A,  132}, 168-178 

\refs Chung, S.-J., Han, C., Park, B.-G., Kim, D., Kang, S., Ryu, Y.-H., 
Kim, K.~M., Jeon, Y.-B., Lee, D.-W., Chang, K., Lee, W.-B., 
\& Kang, Y.~H.\ (2005) Properties of Central Caustics in Planetary Microlensing.  {\em ApJ,  630}, 535-542 

\refs Chung, S.-J., Kim, D., Darnley, M.~J., Duke, J.~P., Gould, A., Han, 
C., Jeon, Y.-B., Kerins, E., Newsam, A., 
\& Park, B.-G.\ (2006) The Possibility of Detecting Planets in the Andromeda Galaxy.  {\em ApJ,  650}, 432-437 

\refs Covone, G., de Ritis, R., Dominik, M., 
\& Marino, A.~A.\ (2000) Detecting planets around stars in nearby galaxies.  {\em A\&A,  357}, 816-822

\refs de Jong, J.~T.~A., Kuijken, K., Crotts, A.~P.~S., Sackett, P.~D., 
Sutherland, W.~J., Uglesich, R.~R., Baltz, E.~A., Cseresnjes, P., Gyuk, G., 
Widrow, L.~M., 
\& The MEGA collaboration (2004) First microlensing candidates from the MEGA survey of M 31.  {\em A\&A,  417}, 461-477 

\refs Derue, F., Afonso, C., Alard, C., Albert, J.-N., Andersen, J., 
Ansari, R., Aubourg, {\'E}., Bareyre, P., Bauer, F., Beaulieu, J.-P., 
Blanc, G., Bouquet, A., Char, S., Charlot, X., Couchot, F., Coutures, C., 
Ferlet, R., Fouqu{\'e}, P., Glicenstein, J.-F., Goldman, B., Gould, A., 
Graff, D., Gros, M., Ha{\"i}ssinski, J., Hamilton, J.-C., Hardin, D., de 
Kat, J., Kim, A., Lasserre, T., Le Guillou, L., Lesquoy, {\'E}., Loup, C., 
Magneville, C., Mansoux, B., Marquette, J.-B., Maurice, {\'E}., Milsztajn, 
A., Moniez, M., Palanque-Delabrouille, N., Perdereau, O., Pr{\'e}vot, L., 
Regnault, N., Rich, J., Spiro, M., Vidal-Madjar, A., Vigroux, L., 
\& Zylberajch, S.\ (2001) Observation of microlensing toward the galactic spiral arms. EROS II 3 year survey.  {\em A\&A,  373}, 126-138 

\refs Di Stefano, R., 
\& Perna, R.\ (1997) Identifying Microlensing by Binaries.  {\em ApJ,  488}, 55-63

\refs Di Stefano, R.\ (1999) Microlensing and the Search for 
Extraterrestrial Life.  {\em ApJ,  512}, 558-563 

\refs Di Stefano, R., 
\& Scalzo, R.~A.\ (1999a) A New Channel for the Detection of Planetary Systems through Microlensing. I. Isolated Events due to Planet Lenses.  {\em ApJ,  512}, 564-578 

\refs Di Stefano, R., 
\& Scalzo, R.~A.\ (1999b) A New Channel for the Detection of Planetary Systems through Microlensing. II. Repeating Events.  {\em ApJ,  512}, 579-600 

\refs Di Stefano, R.\ (2008) Mesolensing Explorations of Nearby Masses: 
From Planets to Black Holes.  {\em ApJ,  684}, 59-67 

\refs Di Stefano, R., 
\& Night, C.\ (2008) Discovery and Study of Nearby Habitable Planets with Mesolensing. {\em ApJ}, submitted (arXiv:0801.1510)

\refs Dominik, M.\ (1995) Improved Routines for the Inversion of the 
Gravitational Lens Equation for a Set of Source Points.  {\em 
A\&AS,  109}, 597-610

\refs Dominik, M., 
\& Hirshfeld, A.~C.\ (1996) Evidence for a binary lens in the MACHO LMC No. 1 microlensing event. {\em A\&A,  313}, 841-850 
 
\refs Dominik, M.\ (1998) Galactic microlensing with rotating binaries.  
{\em A\&A,  329}, 361-374 

\refs Dominik, M.\ (1999a) Ambiguities in FITS of observed binary lens 
galactic microlensing events.  {\em A\&A,  341}, 943-953 

\refs Dominik, M.\ (1999b) The binary gravitational lens and its extreme 
cases.  {\em A\&A,  349}, 108-125 

\refs Dominik, M.\ (2004a) Theory and practice of microlensing light curves 
around fold singularities.  {\em MNRAS,  353}, 69-86 

\refs Dominik, M.\ (2004b) Revealing stellar brightness profiles by means of 
microlensing fold caustics.  {\em MNRAS,  353}, 118-132 

\refs Dominik, M.\ (2006) Stochastic distributions of lens and source 
properties for observed galactic microlensing events.  {\em MNRAS,  367}, 
669-692 

\refs Dominik, M.\ (2007) Adaptive contouring - an efficient way to 
calculate microlensing light curves of extended sources.  {\em MNRAS,  
377}, 1679-1688 

\refs Dominik, M., Jorgensen, U.~G., Horne, K., Tsapras, Y., Street, R.~A., 
Wyrzykowski, L., Hessman, F.~V., Hundertmark, M., Rahvar, S., Wambsganss, 
J., Scarpetta, G., Bozza, V., Calchi Novati, S., Mancini, L., Masi, G., 
Teuber, J., Hinse, T.~C., Steele, I.~A., Burgdorf, M.~J., 
\& Kane, S.\ (2008) Inferring statistics of planet populations by means of automated microlensing searches.  
(arXiv:0808.0004)

\refs Dominik, M.\ (2009) Parameter degeneracies and (un)predictability of 
gravitational microlensing events.  {\em MNRAS,  393}, 816-821

\refs Dong, S., DePoy, D.~L., Gaudi, B.~S., Gould, A., Han, C., Park, 
B.-G., Pogge, R.~W., Udalski, A., Szewczyk, O., Kubiak, M., Szyma{\'n}ski, 
M.~K., Pietrzy{\'n}ski, G., Soszy{\'n}ski, I., Wyrzykowski, {\L}., Zebru{\'n}, K.\ (2006) Planetary Detection Efficiency of the Magnification 3000 Microlensing Event OGLE-2004-BLG-343.  {\em ApJ,  642}, 842-860 

\refs Dong, S., Udalski, A., Gould, A., Reach, W.~T., Christie, G.~W., 
Boden, A.~F., Bennett, D.~P., Fazio, G., Griest, K., Szyma{\'n}ski, M.~K., 
Kubiak, M., Soszy{\'n}ski, I., Pietrzy{\'n}ski, G., Szewczyk, O., 
Wyrzykowski, {\L}., Ulaczyk, K., Wieckowski, T., Paczy{\'n}ski, B., DePoy, 
D.~L., Pogge, R.~W., Preston, G.~W., Thompson, I.~B., 
\& Patten, B.~M.\ (2007) First Space-Based Microlens Parallax Measurement: Spitzer Observations of OGLE-2005-SMC-001.  {\em ApJ,  664}, 862-878 

\refs Dong, S., Gould, A., Udalski, A., Anderson, J., Christie, G.~W., 
Gaudi, B.~S., The OGLE Collaboration, Jaroszy{\'n}ski, M., Kubiak, M., 
Szyma{\'n}ski, M.~K., Pietrzy{\'n}ski, G., Soszy{\'n}ski, I., Szewczyk, O., 
Ulaczyk, K., Wyrzykowski, {\L}., The {$\mu$}FUN Collaboration, DePoy, 
D.~L., Fox, D.~B., Gal-Yam, A., Han, C., L{\'e}pine, S., McCormick, J., 
Ofek, E., Park, B.-G., Pogge, R.~W., The MOA Collaboration, Abe, F., 
Bennett, D.~P., Bond, I.~A., Britton, T.~R., Gilmore, A.~C., Hearnshaw, 
J.~B., Itow, Y., Kamiya, K., Kilmartin, P.~M., Korpela, A., Masuda, K., 
Matsubara, Y., Motomura, M., Muraki, Y., Nakamura, S., Ohnishi, K., Okada, 
C., Rattenbury, N., Saito, T., Sako, T., Sasaki, M., Sullivan, D., Sumi, 
T., Tristram, P.~J., Yanagisawa, T., Yock, P.~C.~M., Yoshoika, T., The 
PLANET/Robo Net Collaborations, Albrow, M.~D., Beaulieu, J.~P., Brillant, 
S., Calitz, H., Cassan, A., Cook, K.~H., Coutures, C., Dieters, S., 
Prester, D.~D., Donatowicz, J., Fouqu{\'e}, P., Greenhill, J., Hill, K., 
Hoffman, M., Horne, K., J{\o}rgensen, U.~G., Kane, S., Kubas, D., 
Marquette, J.~B., Martin, R., Meintjes, P., Menzies, J., Pollard, K.~R., 
Sahu, K.~C., Vinter, C., Wambsganss, J., Williams, A., Bode, M., Bramich, 
D.~M., Burgdorf, M., Snodgrass, C., Steele, I., Doublier, V., 
\& Foellmi, C.\ (2009a) OGLE-2005-BLG-071Lb, the Most Massive M Dwarf Planetary Companion? {\em ApJ,  695}, 970-987

\refs Dong, S., Bond, I.~A., Gould, A., Koz{\l}owski, S., Miyake, N., 
Gaudi, B.~S., Bennett, D.~P., Abe, F., Gilmore, A.~C., Fukui, A., Furusawa, 
K., Hearnshaw, J.~B., Itow, Y., Kamiya, K., Kilmartin, P.~M., Korpela, A., 
Lin, W., Ling, C.~H., Masuda, K., Matsubara, Y., Muraki, Y., Nagaya, M., 
Ohnishi, K., Okumura, T., Perrott, Y.~C., Rattenbury, N., Saito, T., Sako, 
T., Sato, S., Skuljan, L., Sullivan, D.~J., Sumi, T., Sweatman, W., 
Tristram, P.~J., Yock, P.~C.~M., The MOA Collaboration, Bolt, G., Christie, 
G.~W., DePoy, D.~L., Han, C., Janczak, J., Lee, C.-U., Mallia, F., 
McCormick, J., Monard, B., Maury, A., Natusch, T., Park, B.-G., Pogge, 
R.~W., Santallo, R., Stanek, K.~Z., The {$\mu$}FUN Collaboration, Udalski, 
A., Kubiak, M., Szyma{\'n}ski, M.~K., Pietrzy{\'n}ski, G., Soszy{\'n}ski, 
I., Szewczyk, O., Wyrzykowski, {\L}., Ulaczyk, K., 
\& The OGLE Collaboration (2009b) Microlensing Event MOA-2007-BLG-400: Exhuming the Buried Signature of a Cool, Jovian-Mass Planet.  {\em ApJ,  698}, 1826-1837

\refs Einstein, A.\ (1936) Lens-Like Action of a Star by the Deviation of 
Light in the Gravitational Field.  {\em Sci,  84}, 506-507 

\refs Fields, D.~L., Albrow, M.~D., An, J., Beaulieu, J.-P., Caldwell, 
J.~A.~R., DePoy, D.~L., Dominik, M., Gaudi, B.~S., Gould, A., Greenhill, 
J., Hill, K., J{\o}rgensen, U.~G., Kane, S., Martin, R., Menzies, J., 
Pogge, R.~W., Pollard, K.~R., Sackett, P.~D., Sahu, K.~C., Vermaak, P., 
Watson, R., Williams, A., Glicenstein, J.-F., 
\& Hauschildt, P.~H.\ (2003) High-Precision Limb-Darkening Measurement of a K3 Giant Using Microlensing.  {\em ApJ,  596}, 1305-1319 

\refs Fluke, C.~J., 
\& Webster, R.~L.\ (1999) Investigating the geometry of quasars with microlensing.  {\em MNRAS,  302}, 68-74 

\refs Ford, E.~B., 
\& Rasio, F.~A.\ (2008) Origins of Eccentric Extrasolar Planets: Testing the Planet-Planet Scattering Model.  {\em ApJ,  686}, 621-636 

\refs Gaudi, B.~S.\ (1998) Distinguishing Between Binary-Source and 
Planetary Microlensing Perturbations.  {\em ApJ,  506}, 533-539 

\refs Gaudi, B.~S., 
\& Gould, A.\ (1997a) Satellite Parallaxes of Lensing Events toward the Galactic Bulge.  {\em ApJ,  477}, 152-162 

\refs Gaudi, B.~S., 
\& Gould, A.\ (1997b) Planet Parameters in Microlensing Events.  {\em ApJ,  486}, 85-99

\refs Gaudi, B.~S., 
\& Gould, A.\ (1999) Spectrophotometric Resolution of Stellar Surfaces with Microlensing.  {\em ApJ,  513}, 619-625 

\refs Gaudi, B.~S., 
\& Han, C.\ (2004) The Many Possible Interpretations of Microlensing Event OGLE 2002-BLG-055.  {\em ApJ,  611}, 528-536 

\refs Gaudi et al.(1998) Microlensing by Multiple Planets in High-Magnification Events. {\em ApJ, 502}, L33-L37

\refs Gaudi, B.~S., 
\& Petters, A.~O.\ (2002) Gravitational Microlensing near Caustics. I. Folds.  {\em ApJ,  574}, 970-984 

\refs Gaudi, B.~S., 
\& Petters, A.~O.\ (2002) Gravitational Microlensing near Caustics. II. Cusps.  {\em ApJ,  580}, 468-489 

\refs Gaudi, B.~S., Albrow, M.~D., An, J., Beaulieu, J.-P., Caldwell, 
J.~A.~R., DePoy, D.~L., Dominik, M., Gould, A., Greenhill, J., Hill, K., 
Kane, S., Martin, R., Menzies, J., Naber, R.~M., Pel, J.-W., Pogge, R.~W., 
Pollard, K.~R., Sackett, P.~D., Sahu, K.~C., Vermaak, P., Vreeswijk, P.~M., 
Watson, R., 
\& Williams, A.\ (2002) Microlensing Constraints on the Frequency of Jupiter-Mass Companions: Analysis of 5 Years of PLANET Photometry.  {\em ApJ,  566}, 463-499 

\refs Gaudi, B.~S., Bennett, D.~P., Udalski, A., Gould, A., Christie, 
G.~W., Maoz, D., Dong, S., McCormick, J., Szyma{\'n}ski, M.~K., Tristram, 
P.~J., Nikolaev, S., Paczy{\'n}ski, B., Kubiak, M., Pietrzy{\'n}ski, G., 
Soszy{\'n}ski, I., Szewczyk, O., Ulaczyk, K., Wyrzykowski, {\L}., DePoy, ould
D.~L., Han, C., Kaspi, S., Lee, C.-U., Mallia, F., Natusch, T., Pogge, 
R.~W., Park, B.-G., Abe, F., Bond, I.~A., Botzler, C.~S., Fukui, A., 
Hearnshaw, J.~B., Itow, Y., Kamiya, K., Korpela, A.~V., Kilmartin, P.~M., 
Lin, W., Masuda, K., Matsubara, Y., Motomura, M., Muraki, Y., Nakamura, S., 
Okumura, T., Ohnishi, K., Rattenbury, N.~J., Sako, T., Saito, T., Sato, S., 
Skuljan, L., Sullivan, D.~J., Sumi, T., Sweatman, W.~L., Yock, P.~C.~M., 
Albrow, M.~D., Allan, A., Beaulieu, J.-P., Burgdorf, M.~J., Cook, K.~H., 
Coutures, C., Dominik, M., Dieters, S., Fouqu{\'e}, P., Greenhill, J., 
Horne, K., Steele, I., Tsapras, Y., Chaboyer, B., Crocker, A., Frank, S., 
\& Macintosh, B.\ (2008a) Discovery of a Jupiter/Saturn Analog with Gravitational Microlensing.  {\em Sci,  319}, 927-930

\refs Gaudi, B.~S., Patterson, J., Spiegel, D.~S., Krajci, T., Koff, R., 
Pojma{\'n}ski, G., Dong, S., Gould, A., Prieto, J.~L., Blake, C.~H., 
Roming, P.~W.~A., Bennett, D.~P., Bloom, J.~S., Boyd, D., Eyler, M.~E., de 
Ponthi{\`e}re, P., Mirabal, N., Morgan, C.~W., Remillard, R.~R., 
Vanmunster, T., Wagner, R.~M., 
\& Watson, L.~C.\ (2008b) Discovery of a Very Bright, Nearby Gravitational Microlensing Event.  {\em ApJ,  677}, 1268-1277 

\refs Goldreich, P., Lithwick, Y., 
\& Sari, R.\ (2004) Final Stages of Planet Formation.  {\em ApJ,  614}, 497-507

\refs Gould, A.\ (1992) Extending the MACHO search to about 10$^6$ solar 
masses.  {\em ApJ,  392}, 442-451

\refs Gould, A.\ (1994) Proper motions of MACHOs.  {\em ApJ,  421}, L71-L74 

\refs Gould, A.\ (1994) MACHO velocities from satellite-based parallaxes.  
{\em ApJ,  421}, L75-L78 

\refs Gould, A.\ (1995) MACHO parallaxes from a single satellite.  {\em 
ApJ,  441}, L21-L24 

\refs Gould, A.\ (1996) Microlensing and the Stellar Mass Function.  {\em 
PASP,  108}, 465-476

\refs Gould, A.\ (1999) Microlens Parallaxes with SIRTF.  {\em ApJ,  514}, 
869-877 

\refs Gould, A.\ (2000) Measuring the Remnant Mass Function of the Galactic 
Bulge.  {\em ApJ,  535}, 928-931 

\refs Gould, A.\ (2004) Resolution of the MACHO-LMC-5 Puzzle: The 
Jerk-Parallax Microlens Degeneracy.  {\em ApJ,  606}, 319-325 

\refs Gould, A.\ (2008) Hexadecapole Approximation in Planetary 
Microlensing.  {\em ApJ,  681}, 1593-1598 

\refs Gould, A.\ (2009) Wide Field Imager in Space for Dark Energy and 
Planets. {\em Astro2010: The Astronomy and Astrophysics Decadal Survey, Science White Papers}, 100
(arXiv:0902.2211)

\refs Gould, A., 
\& Gaucherel, C.\ (1997) Stokes's Theorem Applied to Microlensing of Finite Sources.  {\em ApJ,  477}, 580-584

\refs Gould, A., 
\& Loeb, A.\ (1992) Discovering planetary systems through gravitational microlenses.  {\em ApJ,  396}, 104-114 

\refs Gould, A., 
\& Han, C.\ (2000) Astrometric Resolution of Severely Degenerate Binary Microlensing Events.  {\em ApJ,  538}, 653-656 

\refs Gould, A., Miralda-Escude, J., 
\& Bahcall, J.~N.\ (1994) Microlensing Events: Thin Disk, Thick Disk, or Halo?  {\em ApJ,  423}, L105-108

\refs Gould, A., Udalski, A., An, D., Bennett, D.~P., Zhou, A.-Y.,
Dong, S., Rattenbury, N.~J., Gaudi, B.~S., Yock, P.~C.~M., Bond,
I.~A., Christie, G.~W., Horne, K., Anderson, J., Stanek, K.~Z., DePoy,
D.~L., Han, C., McCormick, J., Park, B.-G., Pogge, R.~W., Poindexter,
S.~D., Soszy{\'n}ski, I., Szyma{\'n}ski, M.~K., Kubiak, M.,
Pietrzy{\'n}ski, G., Szewczyk, O., Wyrzykowski, {\L}., Ulaczyk, K.,
Paczy{\'n}ski, B., Bramich, D.~M., Snodgrass, C., Steele, I.~A.,
Burgdorf, M.~J., Bode, M.~F., Botzler, C.~S., Mao, S., \& Swaving,
S.~C.\ (2006) Microlens OGLE-2005-BLG-169 Implies That Cool
Neptune-like Planets Are Common.  {\em ApJ, 644}, L37-L40

\refs Griest, K., 
\& Hu, W.\ (1992) Effect of binary sources on the search for massive astrophysical compact halo objects via microlensing.  {\em ApJ,  397}, 362-380

\refs Griest, K., 
\& Safizadeh, N.\ (1998) The Use of High-Magnification Microlensing Events in Discovering Extrasolar Planets.  {\em ApJ,  500}, 37-50

\refs Han, C.\ (1999) Analytic relations between the observed gravitational 
microlensing parameters with and without the effect of blending.  {\em 
MNRAS,  309}, 373-378 

\refs Han, C.\ (2006) Properties of Planetary Caustics in Gravitational 
Microlensing.  {\em ApJ,  638}, 1080-1085 

\refs Han, C.\ (2008) Microlensing Detections of Moons of Exoplanets.  {\em 
ApJ,  684}, 684-690

\refs Han, C.\ (2009) Distinguishing Between Planetary and Binary 
Interpretations of Microlensing Central Perturbations Under the Severe 
Finite-Source Effect.  {\em ApJ,  691}, L9-L12 

\refs Han, C., Gaudi, B.~S., An, J.~H., 
\& Gould, A.\ (2005) Microlensing Detection and Characterization of Wide-Separation Planets.  {\em ApJ,  618}, 962-972 

\refs Han, C., 
\& Gaudi, B.~S.\ (2008) A Characteristic Planetary Feature in Double-Peaked, High-Magnification Microlensing Events.  {\em ApJ,  689}, 53-58 

\refs Han, C., 
\& Gould, A.\ (1995) The Mass Spectrum of MACHOs from Parallax Measurements.  {\em ApJ,  447}, 53- 

\refs Han, C., 
\& Gould, A.\ (2003) Stellar Contribution to the Galactic Bulge Microlensing Optical Depth.  {\em ApJ,  592}, 172-175

\refs Han, C., 
\& Han, W.\ (2002) On the Feasibility of Detecting Satellites of Extrasolar Planets via Microlensing.  {\em ApJ,  580}, 490-493 

\refs Han, C., 
\& Park, M.-G.\ (2002) A New Channel to Search for Extra-Solar Systems with Multiple Planets via Gravitational Microlensing.  {\em JKAS,  35}, 35-40

\refs Hamadache, C., Le Guillou, L., Tisserand, P., Afonso, C., Albert, 
J.~N., Andersen, J., Ansari, R., Aubourg, {\'E}., Bareyre, P., Beaulieu, 
J.~P., Charlot, X., Coutures, C., Ferlet, R., Fouqu{\'e}, P., Glicenstein, 
J.~F., Goldman, B., Gould, A., Graff, D., Gros, M., Haissinski, J., de Kat, 
J., Lesquoy, {\'E}., Loup, C., Magneville, C., Marquette, J.~B., Maurice, 
{\'E}., Maury, A., Milsztajn, A., Moniez, M., Palanque-Delabrouille, N., 
Perdereau, O., Rahal, Y.~R., Rich, J., Spiro, M., Vidal-Madjar, A., 
Vigroux, L., 
\& Zylberajch, S.\ (2006) Galactic Bulge microlensing optical depth from EROS-2.  {\em A\&A,  454}, 185-199 

\refs Hardy, S.~J., 
\& Walker, M.~A.\ (1995) Parallax effects in binary microlensing events.  {\em MNRAS,  276}, L79-L82 

\refs Holtzman, J.~A., Watson, A.~M., Baum, W.~A., Grillmair, C.~J., Groth, 
E.~J., Light, R.~M., Lynds, R., 
\& O'Neil, E.~J., Jr.\ (1998) The Luminosity Function and Initial Mass Function in the Galactic Bulge.  {\em AJ,  115}, 1946-1957 

\refs Holz, D.~E., 
\& Wald, R.~M.\ (1996) Photon Statistics Limits for Earth-based Parallax Measurements of MACHO Events.  {\em ApJ,  471}, 64-67

\refs Horne, K., Snodgrass, C., 
\& Tsapras, Y.\ (2009) A metric and optimization scheme for microlens planet searches.  {\em MNRAS,  396}, 2087-2102

\refs Ida, S., \& Lin, D.~N.~C.\ (2004) Toward a Deterministic Model of Planetary Formation. II. The Formation and Retention of Gas Giant Planets around Stars with a Range of Metallicities.  {\em ApJ,  616}, 567-572

\refs Ida, S., 
\& Lin, D.~N.~C.\ (2005) Toward a Deterministic Model of Planetary Formation. III. Mass Distribution of Short-Period Planets around Stars of Various Masses.  {\em ApJ,  626}, 1045-1060 

\refs Ingrosso, G., Calchi Novati, S., De Paolis, F., Jetzer, P., Nucita, 
A.~A., 
\& Zakharov, A.~F.\ (2009) Pixel-lensing as a way to detect extrasolar planets in M31.  {\em MNRAS}, in press
(arXiv:0906.1050)

\refs Janczak, J., et al.\ (2009) Sub-Saturn Planet MOA-2008-BLG-310Lb: Likely To
Be in the Galactic Bulge. {\em ApJ}, submitted (arXiv:0908.0529)

\refs Jaroszy{\'n}ski, M., 
\& Mao, S.\ (2001) Predicting the second caustic crossing in binary microlensing events.  {\em MNRAS,  325}, 1546-1552 

\refs Johnson, J.~A., Gaudi, B.~S., Sumi, T., Bond, I.~A., 
\& Gould, A.\ (2008) A High-Resolution Spectrum of the Highly Magnified Bulge G Dwarf MOA-2006-BLG-099S.  {\em ApJ,  685}, 508-520 

\refs Juri{\'c}, M., 
\& Tremaine, S.\ (2008) Dynamical Origin of Extrasolar Planet Eccentricity Distribution.  {\em ApJ,  686}, 603-620 

\refs Kennedy, G.~M., Kenyon, S.~J., 
\& Bromley, B.~C.\ (2007) Planet formation around M-dwarfs: the moving snow line and super-Earths.  {\em Ap\&SS,  311}, 9-13 

\refs Kennedy, G.~M., 
\& Kenyon, S.~J.\ (2008) Planet Formation around Stars of Various Masses: The Snow Line and the Frequency of Giant Planets.  {\em ApJ,  673}, 502-512 

\refs Kervella, P., Th{\'e}venin, F., Di Folco, E., 
\& S{\'e}gransan, D.\ (2004) The angular sizes of dwarf stars and subgiants. Surface brightness relations calibrated by interferometry.  {\em A\&A,  426}, 297-307 

\refs Khavinson, D., \& Neumann, G.\ (2006) On the number of zeros of certain rational harmonic functions.
{\em Proc.\ Amer.\ Math.\ Soc., 134}, 1077-1085

\refs Kiraga, M., 
\& Paczynski, B.\ (1994) Gravitational microlensing of the Galactic bulge stars.  {\em ApJ,  430}, L101-L104 

\refs Laughlin, G., Bodenheimer, P., 
\& Adams, F.~C.\ (2004) The Core Accretion Model Predicts Few Jovian-Mass Planets Orbiting Red Dwarfs.  {\em ApJ,  612}, L73-L76 

\refs Lecar, M., Podolak, M., Sasselov, D., 
\& Chiang, E.\ (2006) On the Location of the Snow Line in a Protoplanetary Disk.  {\em ApJ,  640}, 1115-1118 

\refs Liebes, S.\ (1964) Gravitational Lenses. {\em PhRv,  133}, 835-844 

\refs Liebig, C., 
\& Wambsganss, J.\ (2009) Detectability of extrasolar moons as gravitational microlenses. {\em A\&A}, submitted (arXiv:0912.2076)

\refs Lissauer, J.~J.\ (1987) Timescales for planetary accretion and the 
structure of the protoplanetary disk.  {\em Icar,  69}, 249-265 

\refs Mao, S.\ (1992) Gravitational microlensing by a single star plus 
external shear.  {\em ApJ,  389}, 63-67 

\refs Mao, S.\ (2008) Introduction to Gravitational Microlensing\\
(arXiv:0811.0441)

\refs Mao, S., 
\& Di Stefano, R.\ (1995) Interpretation of gravitational microlensing by binary systems.  {\em ApJ,  440}, 22-27 

\refs Mao, S., 
\& Paczynski, B.\ (1991) Gravitational microlensing by double stars and planetary systems.  {\em ApJ,  374}, L37-L40 

\refs Minniti, D., Vandehei, T., Cook, K.~H., Griest, K., 
\& Alcock, C.\ (1998) Detection of Lithium in a Main Sequence Bulge Star Using Keck I as a 15M Diameter Telescope.  {\em ApJ,  499}, L175-L178

\refs Paczynski, B.\ (1986) Gravitational microlensing by the galactic 
halo.  {\em ApJ,  304}, 1-5 

\refs Paczynski, B.\ (1991) Gravitational microlensing of the Galactic 
bulge stars.  {\em ApJ,  371}, L63-L67 

\refs Paczynski, B.\ (1996) Gravitational Microlensing in the Local Group.  
{\em ARA\&A,  34}, 419-460 

\refs Paczynski, B., 
\& Stanek, K.~Z.\ (1998) Galactocentric Distance with the Optical Gravitational Lensing Experiment and HIPPARCOS Red Clump Stars.  {\em ApJ,  494}, L219-222

\refs Palanque-Delabrouille, N., Afonso, C., Albert, J.~N., Andersen, J., 
Ansari, R., Aubourg, E., Bareyre, P., Bauer, F., Beaulieu, J.~P., Bouquet, 
A., Char, S., Charlot, X., Couchot, F., Coutures, C., Derue, F., Ferlet, 
R., Glicenstein, J.~F., Goldman, B., Gould, A., Graff, D., Gros, M., 
Haissinski, J., Hamilton, J.~C., Hardin, D., de Kat, J., Lesquoy, E., Loup, 
C., Magneville, C., Mansoux, B., Marquette, J.~B., Maurice, E., Milsztajn, 
A., Moniez, M., Perdereau, O., Prevot, L., Renault, C., Rich, J., Spiro, 
M., Vidal-Madjar, A., Vigroux, L., Zylberajch, S., 
\& The EROS Collaboration (1998) Microlensing towards the Small Magellanic Cloud EROS 2 first year survey.  {\em A\&A,  332}, 1-9 

\refs Park, B.-G., Jeon, Y.-B., Lee, C.-U., 
\& Han, C.\ (2006) Microlensing Sensitivity to Earth-Mass Planets in the Habitable Zone.  {\em ApJ,  643}, 1233-1238 

\refs Paulin-Henriksson, S., Baillon, P., Bouquet, A., Carr, B.~J., 
Cr{\'e}z{\'e}, M., Evans, N.~W., Giraud-H{\'e}raud, Y., Gould, A., Hewett, 
P., Kaplan, J., Kerins, E., Lastennet, E., Le Du, Y., Melchior, A.-L., 
Smartt, S.~J., 
\& Valls-Gabaud, D.\ (2002) A Candidate M31/M32 Intergalactic Microlensing Event.  {\em ApJ,  576}, L121-L124 

\refs Peale, S.~J.\ (2001) Probability of Detecting a Planetary Companion 
during a Microlensing Event.  {\em ApJ,  552}, 889-911 

\refs Pejcha, O., 
\& Heyrovsk{\'y}, D.\ (2009) Extended-Source Effect and Chromaticity in Two-Point-Mass Microlensing.  {\em ApJ,  690}, 1772-1796 

\refs Petters, A.~O., Levine, H., 
\& Wambsganss, J.\ (2001), {\em Singularity theory and gravitational lensing}. Birkh{\" a}user, Boston. 

\refs Poindexter, S., Afonso, C., Bennett, D.~P., Glicenstein, J.-F., 
Gould, A., Szyma{\'n}ski, M.~K., 
\& Udalski, A.\ (2005) Systematic Analysis of 22 Microlensing Parallax Candidates.  {\em ApJ,  633}, 914-930

\refs Press, W.~H., Teukolsky, S.~A., Vetterling, W.~T., 
\& Flannery, B.~P.\ (1992), {\em Numerical recipes in FORTRAN. The art of scientific computing}. 
University Press, Cambridge.

\refs Rattenbury, N.~J., Bond, I.~A., Skuljan, J., 
\& Yock, P.~C.~M.\ (2002) Planetary microlensing at high magnification.  {\em MNRAS,  335}, 159-169 

\refs Refsdal, S.\ (1964) The gravitational lens effect.  {\em MNRAS,  
128}, 295-306

\refs Refsdal, S.\ (1966) On the possibility of determining the distances 
and masses of stars from the gravitational lens effect.  {\em MNRAS,  134}, 
315-319

\refs Renn, J., Sauer, T., 
\& Stachel, J.\ (1997) The origin of gravitational lensing: a postscript to Einstein's 1936 Science paper.  {\em Sci,  275}, 184-186 

\refs Rhie, S.~H.\ (1997) Infimum Microlensing Amplification of the Maximum 
Number of Images of n-Point Lens Systems.  {\em ApJ,  484}, 63-69

\refs Rhie, S.~H.\ (2003) How Cumbersome is a Tenth Order Polynomial?: The Case of Gravitational Triple Lens Equation.
(arXiv:astro-ph/0202294)

\refs Rhie, S.~H.\ (2003) n-point Gravitational Lenses with 5(n-1) Images.
(arXiv:astro-ph/0305166)

\refs Rhie, S.~H., 
\& Bennett, D.~P.\ (1999) Line Caustic Microlensing and Limb Darkening. (arXiv:astro-ph/9912050)

\refs Rhie, S.~H., Bennett, D.~P., Becker, A.~C., Peterson, B.~A., Fragile, 
P.~C., Johnson, B.~R., Quinn, J.~L., Crouch, A., Gray, J., King, L., 
Messenger, B., Thomson, S., Bond, I.~A., Abe, F., Carter, B.~S., Dodd, 
R.~J., Hearnshaw, J.~B., Honda, M., Jugaku, J., Kabe, S., Kilmartin, P.~M., 
Koribalski, B.~S., Masuda, K., Matsubara, Y., Muraki, Y., Nakamura, T., 
Nankivell, G.~R., Noda, S., Rattenbury, N.~J., Reid, M., Rumsey, N.~J., 
Saito, T., Sato, H., Sato, S., Sekiguchi, M., Sullivan, D.~J., Sumi, T., 
Watase, Y., Yanagisawa, T., Yock, P.~C.~M., 
\& Yoshizawa, M.\ (2000) On Planetary Companions to the MACHO 98-BLG-35 Microlens Star.  {\em ApJ,  533}, 378-391 

\refs Sackett, P.~D.\ (1999) Searching for Unseen Planets via Occultation 
and Microlensing. {\it Planets Outside the Solar System: Theory and Observations}, eds.\ J.-M. Mariotti and D. Alloin,
(Boston:Kluwer), 189-228 (arXiv:astro-ph/9811269)

\refs Sako, T., Sekiguchi, T., Sasaki, M., Okajima, K., Abe, F., Bond, 
I.~A., Hearnshaw, J.~B., Itow, Y., Kamiya, K., Kilmartin, P.~M., Masuda, 
K., Matsubara, Y., Muraki, Y., Rattenbury, N.~J., Sullivan, D.~J., Sumi, 
T., Tristram, P., Yanagisawa, T., 
\& Yock, P.~C.~M.\ (2008) MOA-cam3: a wide-field mosaic CCD camera for a gravitational microlensing survey in New Zealand.  {\em ExA,  22}, 51-66 

\refs Sauer, T. (2008) Nova Geminorum 1912 and the origin of the idea of gravitational lensing. {\em Arch. Hist. Exact Sci., 62},
1-22

\refs Schneider, P., 
\& Weiss, A.\ (1986) The two-point-mass lens - Detailed investigation of a special asymmetric gravitational lens.  {\em A\&A,  164}, 237-259

\refs Schneider, P., 
\& Weiss, A.\ (1992) The gravitational lens equation near cusps.  {\em A\&A,  260}, 1-2 

\refs Schneider, P., Ehlers, J., \& Falco, E.~E. (1992), 
{\em Gravitational Lenses}. Springer-Verlag, Berlin.  

\refs Smith, M.~C., Mao, S., 
\& Paczy{\'n}ski, B.\ (2003) Acceleration and parallax effects in gravitational microlensing.  {\em MNRAS,  339}, 925-936 

\refs Snodgrass, C., Horne, K., 
\& Tsapras, Y.\ (2004) The abundance of Galactic planets from OGLE-III 2002 microlensing data.  {\em MNRAS,  351}, 967-975 

\refs Sumi, T., Abe, F., Bond, I.~A., Dodd, R.~J., Hearnshaw, J.~B., Honda, 
M., Honma, M., Kan-ya, Y., Kilmartin, P.~M., Masuda, K., Matsubara, Y., 
Muraki, Y., Nakamura, T., Nishi, R., Noda, S., Ohnishi, K., Petterson, 
O.~K.~L., Rattenbury, N.~J., Reid, M., Saito, T., Saito, Y., Sato, H., 
Sekiguchi, M., Skuljan, J., Sullivan, D.~J., Takeuti, M., Tristram, P.~J., 
Wilkinson, S., Yanagisawa, T., 
\& Yock, P.~C.~M.\ (2003) Microlensing Optical Depth toward the Galactic Bulge from Microlensing Observations in Astrophysics Group Observations during 2000 with Difference Image Analysis.  {\em ApJ,  591}, 204-227 

\refs Sumi, T., Bennett, D.~P., Bond, I.~A., Udalski, A., Batista, V., 
Dominik, M., Fouqu{\'e}, P., Kubas, D., Gould, A., Macintosh, B., Cook, K., 
Dong, S., Skuljan, L., Cassan, A., The MOA Collaboration: F.~Abe, Botzler, 
C.~S., Fukui, A., Furusawa, K., Hearnshaw, J.~B., Itow, Y., Kamiya, K., 
Kilmartin, P.~M., Korpela, A., Lin, W., Ling, C.~H., Masuda, K., Matsubara, 
Y., Miyake, N., Muraki, Y., Nagaya, M., Nagayama, T., Ohnishi, K., Okumura, 
T., Perrott, Y.~C., Rattenbury, N., Saito, T., Sako, T., Sullivan, D.~J., 
Sweatman, W.~L., P., Yock, P.~C.~M., The PLANET Collaboration: 
J.~P.~Beaulieu, Cole, A., Coutures, C., Duran, M.~F., Greenhill, J., 
Jablonski, F., Marboeuf, U., Martioli, E., Pedretti, E., Pejcha, O., Rojo, 
P., Albrow, M.~D., Brillant, S., Bode, M., Bramich, D.~M., Burgdorf, M.~J., 
Caldwell, J.~A.~R., Calitz, H., Corrales, E., Dieters, S., Dominis Prester, 
D., Donatowicz, J., Hill, K., Hoffman, M., Horne, K., J, U.~G., Kains, N., 
Kane, S., Marquette, J.~B., Martin, R., Meintjes, P., Menzies, J., Pollard, 
K.~R., Sahu, K.~C., Snodgrass, C., Steele, I., Street, R., Tsapras, Y., 
Wambsganss, J., Williams, A., Zub, M., The OGLE Collaboration: M.~K.~Szyma, 
Kubiak, M., Pietrzy, G., Soszy, I., Szewczyk, O., Ulaczyk, K., The 
\$$\backslash$mu\$FUN Collaboration: W.~Allen, Christie, G.~W., DePoy, 
D.~L., Gaudi, B.~S., Han, C., Janczak, J., Lee, C.~-., McCormick, J., 
Mallia, F., Monard, B., Natusch, T., Park, B.-G., Pogge, R.~W., 
\& Santallo, R.\ (2009) A Cold Neptune-Mass Planet OGLE-2007-BLG-368Lb: Cold Neptunes Are Common. {\em ApJ}, submitted (arXiv:0912.1171)

\refs Thomas, C.~L., Griest, K., Popowski, P., Cook, K.~H., Drake, A.~J., 
Minniti, D., Myer, D.~G., Alcock, C., Allsman, R.~A., Alves, D.~R., 
Axelrod, T.~S., Becker, A.~C., Bennett, D.~P., Freeman, K.~C., Geha, M., 
Lehner, M.~J., Marshall, S.~L., Nelson, C.~A., Peterson, B.~A., Quinn, 
P.~J., Stubbs, C.~W., Sutherland, W., Vandehei, T., 
\& Welch, D.~L.\ (2005) Galactic Bulge Microlensing Events from the MACHO Collaboration.  {\em ApJ,  631}, 906-934 

\refs Tsapras, Y., Street, R., Horne, K., Snodgrass, C., Dominik, M., 
Allan, A., Steele, I., Bramich, D.~M., Saunders, E.~S., Rattenbury, N., 
Mottram, C., Fraser, S., Clay, N., Burgdorf, M., Bode, M., Lister, T.~A., 
Hawkins, E., Beaulieu, J.~P., Fouqu{\'e}, P., Albrow, M., Menzies, J., 
Cassan, A., 
\& Dominis-Prester, D.\ (2009) RoboNet-II: Follow-up observations of microlensing events with a robotic network of telescopes.  {\em AN,  330}, 4-11

\refs Udalski, A., Szymanski, M., Kaluzny, J., Kubiak, M., Krzeminski,
W., Mateo, M., Preston, G.~W., \& Paczynski, B.\ (1993) The optical
gravitational lensing experiment. Discovery of the first candidate
microlensing event in the direction of the Galactic Bulge.  {\em AcA,
43}, 289-294

\refs Udalski, A., Szymanski, M., Kaluzny, J., Kubiak, M., Mateo, M.,
Krzeminski, W., \& Paczynski, B.\ (1994) The Optical Gravitational
Lensing Experiment. The Early Warning System: Real Time Microlensing.
{\em AcA, 44}, 227-234

\refs Udalski, A., Zebrun, K., Szymanski, M., Kubiak, M., Pietrzynski,
G., Soszynski, I., \& Wozniak, P.\ (2000) The Optical Gravitational
Lensing Experiment. Catalog of Microlensing Events in the Galactic
Bulge.  {\em AcA, 50}, 1-65

\refs Udalski, A.\ (2003) The Optical Gravitational Lensing Experiment. 
Real Time Data Analysis Systems in the OGLE-III Survey.  {\em AcA,  53}, 
291-305 

\refs Udalski, A., Jaroszy{\'n}ski, M., Paczy{\'n}ski, B., Kubiak, M., 
Szyma{\'n}ski, M.~K., Soszy{\'n}ski, I., Pietrzy{\'n}ski, G., Ulaczyk, K., 
Szewczyk, O., Wyrzykowski, {\L}., Christie, G.~W., DePoy, D.~L., Dong, S., 
Gal-Yam, A., Gaudi, B.~S., Gould, A., Han, C., L{\'e}pine, S., McCormick, 
J., Park, B.-G., Pogge, R.~W., Bennett, D.~P., Bond, I.~A., Muraki, Y., 
Tristram, P.~J., Yock, P.~C.~M., Beaulieu, J.-P., Bramich, D.~M., Dieters, 
S.~W., Greenhill, J., Hill, K., Horne, K., 
\& Kubas, D.\ (2005) A Jovian-Mass Planet in Microlensing Event OGLE-2005-BLG-071.  {\em ApJ,  628}, L109-L112 

\refs Uglesich, R.~R., Crotts, A.~P.~S., Baltz, E.~A., de Jong, J., Boyle, 
R.~P., 
\& Corbally, C.~J.\ (2004) Evidence of Halo Microlensing in M31.  {\em ApJ,  612}, 877-893 

\refs van Belle, G.~T.\ (1999) Predicting Stellar Angular Sizes.  {\em 
PASP,  111}, 1515-1523

\refs Vermaak, P.\ (2000) The effects of resolved sources and blending on 
the detection of planets via gravitational microlensing.  {\em MNRAS,  
319}, 1011-1019 

\refs Vermaak, P.\ (2003) Rapid analysis of binary lens gravitational 
microlensing light curves.  {\em MNRAS,  344}, 651-656

\refs Wambsganss, J.\ (1997) Discovering Galactic planets by gravitational 
microlensing: magnification patterns and light curves.  {\em MNRAS,  284}, 
172-188 

\refs Witt, H.~J.\ (1990) Investigaion of high amplification events in 
light curves of gravitationally lensed quasars.  {\em A\&A,  236}, 311-322 

\refs Witt, H.~J.\ (1995) The Effect of the Stellar Size on Microlensing at 
the Baade Window.  {\em ApJ,  449}, 42-46 

\refs Witt, H.~J., 
\& Mao, S.\ (1994) Can lensed stars be regarded as pointlike for microlensing by MACHOs?.  {\em ApJ,  430}, 505-510 

\refs Witt, H.~J., 
\& Mao, S.\ (1995) On the Minimum Magnification Between Caustic Crossings for Microlensing by Binary and Multiple Stars.  {\em ApJ,  447}, L105-L108

\refs Wozniak, P., 
\& Paczynski, B.\ (1997) Microlensing of Blended Stellar Images.  {\em ApJ,  487}, 55-60

\refs Yoo, J., DePoy, D.~L., Gal-Yam, A., Gaudi, B.~S., Gould, A., Han, C., 
Lipkin, Y., Maoz, D., Ofek, E.~O., Park, B.-G., Pogge, R.~W., Udalski, A., 
Soszy{\'n}ski, I., Wyrzykowski, {\L}., Kubiak, M., Szyma{\'n}ski, M., 
Pietrzy{\'n}ski, G., Szewczyk, O., Zebru{\'n}, K.\ (2004) OGLE-2003-BLG-262: Finite-Source Effects from a Point-Mass Lens.  {\em ApJ,  603}, 139-151 

\refs Zakharov, A.~F.\ (1995) On the magnification of gravitational lens 
images near cusps..  {\em A\&A,  293}, 1-4 

\refs Zakharov, A.~F.\ (1999) On the some properties of gravitational lens 
equation near cusps.  {\em A\&AT,  18}, 17-25 

\end{document}